\documentclass[preprint,showpacs,preprintnumbers,amsmath,amssymb]{revtex4-1}

\usepackage{graphicx}% Include figure files
\usepackage{dcolumn}% Align table columns on decimal point
\usepackage{bm}% bold math
\usepackage{amsmath,empheq}
\usepackage{bigints}
\usepackage{float}
\usepackage{epsfig}
\usepackage{epstopdf}
\usepackage{eqnarray}
\usepackage{mathrsfs}
\usepackage{natbib}
\usepackage{mathtools}
\usepackage{fdsymbol}
\usepackage{oplotsymbl}
\usepackage{pbox}
\usepackage{array}
\usepackage{makecell}

\usepackage[toc,page]{appendix}
\begin{document}
\raggedbottom
\title{Chimera states in coupled map lattices: Spatiotemporally intermittent behavior and an equivalent cellular automaton}
\author{Joydeep Singha}
\email{joydeep@physics.iitm.ac.in}
\affiliation{Department of Physics, Indian Institute of Technology Madras, Chennai, 600036, India.}
\author{Neelima Gupte}
\email{gupte@physics.iitm.ac.in}
\affiliation{Department of Physics, Indian Institute of Technology Madras, Chennai, 600036, India.}

\begin{abstract}
We study the existence of chimera states, i.e. mixed states, in a globally coupled sine circle map lattice, with different strengths of inter-group and intra-group coupling. We find that at specific values of the parameters of the CML, a completely random initial condition evolves to chimera states, having a phase synchronised and a phase desynchronised group, where the space time variation of the phases of the maps in the desynchronised group shows structures similar to spatiotemporally intermittent regions. Using the complex order parameter we obtain a phase diagram that identifies  the region in the parameter space which supports chimera states of this type, as well as other types of phase configurations. An equivalent cellular automaton is obtained which shows space time behaviour similar to the CML. The chimera regions show coexisting deterministic and probabilistic behaviour in the subgroup probabilities which show a transition to purely probabilistic behaviour at the boundaries of the region. We also derive a mean field equation for this cellular automaton and compare its solutions with the corresponding phase configurations obtained in the parameter space, and show that its behaviour matches with the behaviour seen for the CML.
\end{abstract}
\maketitle

\section{Introduction}%\label{introduction}
The chimera phase pattern is a remarkable spatiotemporal property found in spatially extended dynamical systems such as coupled phase oscillators \cite{kuramoto2002, strogatz2004, strogatz2005, strogatz2008, strogatz2010, sethia2008, sheeba2009, sheeba2010, omelchenko2008, laing2009, wang2011, showalter2012, showalter2013, showalter2018, shashi2013, bountis2014, panaggio2016, tareda2016, dai2018, wu2018, maistrenko2014, jaros2015, xie2014, dudkowski2014, hovel2015, yao2015, xie2015} and was recently discovered in coupled map lattice models  \cite{nayak2011, neelima2016, hagerstrom2012, li2018}. In the context of dynamical systems, the `chimera'  state is defined to be a state with the characteristic stable coexistence of a synchronous group of oscillators together with a desynchronised group of oscillators. Similar dynamical behaviour was found in early studies of unihemispheric sleep \cite{rattenborg2000} and the asynchronous eye closure \cite{mathews2006} of sea mammals, birds and reptiles. This kind of  spatio-temporal behaviour was first shown to exist in non-locally coupled complex Ginzberg-Landau oscillators \cite{kuramoto2002} by Kuramoto and Battogtokh. Later, this mixed state was discovered and analysed in a variety of systems such as rings of phase oscillators \cite{strogatz2004, strogatz2005, strogatz2008, strogatz2010}, delay-coupled rings of phase oscillators \cite{sethia2008} and bipartite oscillator populations \cite{sheeba2009, sheeba2010}, Stuart-Landau oscillators \cite{omelchenko2008}, networks of Kuramoto oscillators \cite{laing2009, wang2011}, coupled chemical oscillators \cite{showalter2012, showalter2013, showalter2018}, and mechanical oscillator networks \cite{shashi2013}. 

Here, we study the existence of chimera states in a coupled map lattice which is a discrete analog of coupled phase oscillators both in space and time. The chimera phase state as well as other other mixed states were reported in specific systems of coupled map lattices in both theoretical \cite{nayak2011, neelima2016} models and  experimental systems \cite{hagerstrom2012, li2018}. The CML, used here, is of the form used in Refs. \cite{nayak2011, neelima2016} and consists of two populations of globally coupled identical sine circle maps where the strength of the coupling within each population and that between the maps belonging to distinct populations take different values. The emergence of chimera states in models having two species of identical dynamical units, has been explored earlier in different coupled oscillator models such as two populations of phase oscillators \cite{strogatz2008, bountis2014, panaggio2016, tareda2016, dai2018}, and for Fitzhugh-Nagumo oscillators \cite{wu2018}. The existence of chimera states in globally coupled systems has also been  reported for systems of Stuart-Landau oscillators and for the complex Ginzberg-Landau equation \cite{lennart2015, schmidt2015, konrad2014}.

\par We note that different types of chimera states with interesting spatio-temporal behaviours have been studied in various contexts. These include multiheaded chimera states \cite{maistrenko2014, jaros2015}, travelling chimera states\cite{xie2014, dudkowski2014}, multi-chimera states \cite{hovel2015, yao2015}, twisted chimera states \cite{xie2015}, and amplitude chimera states. It was also shown earlier that the specific CML which we study here can support another kind of mixed state, namely the splay-chimera state where the  coexistence of a phase synchronised group of maps and a phase desynchronised group of maps consisting of splay phase configurations was reported \cite{neelima2016}. In this paper, we report the existence of yet another kind of chimera state for this system, where the evolution of random initial conditions in certain regions of the parameter space results in a new class of chimera solutions where the space time variation of the desynchronised group shows spatiotemporally intermittent behaviour. In addition to the chimera states described here, this system supports various other kinds of phase configurations viz. globally synchronised states, two phase clustered states, fully phase desynchronised states, etc. We show that the transition between these phase configurations upon the change of the parameters can be identified from  the complex order parameters which take unique values for each of these states. We thus obtain the phase diagram of the coupled map lattice and identify the regimes which support chimera states of this type, and regimes which support other phase configurations. Our analysis focusses on the chimera region of the phase diagram  and its neighbourhood. 
We note that chimeras with co-existing coherent and incoherent regions have been seen in oscillator systems such as systems of globally coupled systems of Kuramoto oscillators with delayed feedback \cite{azamat2014} and Stuart-Landau oscillators \cite{konrad2014, gregory2010} and Ginzberg-Landau oscillators \cite{krischer2015, lennart2015} and locally coupled system of oscillators \cite{clerc2016, clerc2017}.

\par We note that despite the recent identification of  different types of chimera states, there has been little work done towards the theoretical analysis of their spatiotemporal properties. Early work on chimeras identified the chimera state in oscillator systems. The analysis carried out for such systems consisted of setting up a mean field framework for such systems, and the identification of their fixed points \cite{strogatz2004}. Subsequent analysis of such systems has been on similar lines \cite{strogatz2004, strogatz2005, strogatz2008, strogatz2010, sethia2008, sheeba2009, sheeba2010, omelchenko2008, laing2009, wang2011}.
The  analysis of the STI chimeras above has also been in the oscillator context. 
Thus,  the analysis of chimera states in a large number of nonlinear systems would benefit by the identification of  general methods, independent of the underlying system, in particular by the incorporation of CML techniques.  We note that considerable insight into the spatio-temporal behaviour of CML-s has been obtained by the construction of equivalent cellular automata models \cite{chate1988, bohr2003, zahera2010}.  A  cellular automaton model equivalent to coupled phase oscillators which shows chimera states for specific ranges of local coupling has been recently constructed \cite{garcia2016}. 
  
Here, we analyse the spatio-temporal behaviour of the chimera states of our model by constructing an equivalent cellular automaton for our model, specifically in the regime where spatio-temporally intermittent chimeras are seen. Using the global coupling topology of the CML, we identify appropriate conditional probabilities, which specify the transition between the laminar and burst states as the system evolves in time, and calculate these probabilities numerically from the space time variation of the phases of the maps of the CML. The identification of the correct probabilities is non-trivial as probabilities which reflect the global coupled structure of the CML have to be identified. The resulting cellular automaton as a global interaction structure and can be used to study different globally coupled oscillator models. The transition to STI chimera states seen in this specific model is signalled by a transition of the CML probabilities characteristic of the synchronized part of the chimera from probabilistic to deterministic behavior. Thus, the CA analysis can pick up the transition in the CML. 

\par We also show that a mean field equation for the fraction  of the laminar/turbulent sites can be derived  in terms of these transition probabilities, and show that the fixed point of this mean field equation gives the values of the fraction of the sites that are laminar or burst sites. These fractions match accurately with those calculated numerically from the space time variation of the CML. Using the solution of the mean field equations, we obtain the phase diagram of the CA in terms of the control parameters of the CML. We compare this phase diagram with that obtained using the group-wise order parameters and the global order parameter, and show that the two phase diagrams match correctly. Moreover, our construction  identifies and differentiates between the regions of the parameter space which correspond to stable chimera states with, and without  defects in the phase synchronised group and without defects in the phase synchronised group,  and find their dependence on the nonlinearity and the coupling. The domains of different dynamical behaviours in the phase diagram are also mapped to the parameter space of the mean field equation of the CA. 

\par Our paper is organised in the following manner: Section \ref{model} discusses the coupled sine circle map lattice model under study.  In section III, we introduce the complex order parameters and obtain a phase diagram using their calculated values. We also discuss here the variety of phase configurations that can be found when the system is evolved using random initial conditions. In section \ref{chim} we discuss the behavior of the chimera consisting of a  phase synchronised group and desynchronised group with spatio-temporally intermittent regions and the method of identifying and labelling the laminar and burst sites. We define and discuss the transition probabilities and calculate them in section \ref{CA}. Section \ref{density} derives the mean field equation and compares the fraction  of laminar sites obtained from the fixed point solution of this equation with those obtained from direct numerical calculation and also performs the stability analysis. Section \ref{transition} compares the phase diagrams and discuss the transition between states in terms of the transition probabilities of the CA. Section \ref{conclusion} summarises our conclusions.

\section{The model}\label{model}
\begin{figure}[H]
\centering \includegraphics[width = 11cm, height = 6cm]{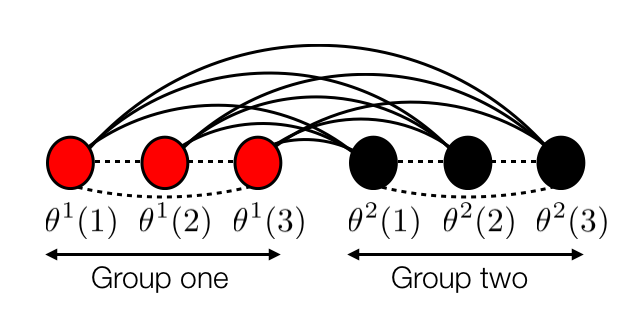}
 \caption{\label{fig: topology}\footnotesize (color online) The schematic of the system of Eq. \ref{sinecml} with 3 maps in each group. The intergroup coupling is shown by solid lines and the intra-group coupling is denoted by dotted lines.}
\end{figure}

In this paper, we study a lattice of coupled sine circle maps, where the maps are distributed into two groups, and the maps are globally coupled, but with two distinct values for the intragroup and intergroup coupling. A single sine circle map is given by,
\begin{equation}
\theta_{n + 1} = \theta_{n} + \Omega -\frac{K}{2\pi}\sin(2\pi\theta_{n}) \mod{1}
\label{sinecir}
\end{equation}
where $\theta$ is the phase of the map, $0<\theta<1$ and $n$ is the time step. The parameter $\Omega$ denotes the frequency ratio in the absence of nonlinearity and $K$ determines the strength of nonlinearity. A single sine circle map shows Arnold tongues organised by frequency locking and quasi-periodic behaviours. It shows universality in the mode locking structure prior to both the period doubling route to chaos and quasi-periodic route to chaos depending on the value of $\Omega$. The evolution equation for the coupled sine circle map lattice considered here is given by,
\begin{equation}
\theta_{n+1}^{\sigma}(i) = \theta_{n}^{\sigma}(i)  + \Omega - \frac{K}{2\pi} \sin(2\pi \theta_{n}^{\sigma}(i)) + \sum\limits_{\sigma' = 1}^{2} \frac{\epsilon_{\sigma \sigma'}}{N_{\sigma'}} \left[ \sum\limits_{j = 1}^{N_{\sigma'}}(\theta_{n}^{\sigma'}(j) + \Omega - \frac{K}{2\pi} \sin(2\pi\theta_{n}^{\sigma'}(j)))\right] 
\mod 1
\label{sinecml}
\end{equation}
The  equation above defines the evolution of the $i$th map in the group $\sigma$, where $\sigma$ takes values $1,2$, and $N$ is the number of maps in each of the groups. We also define the coupling parameters to be $\epsilon_{11} = \epsilon_{22} = \epsilon_{1}$ and  $\epsilon_{12} = \epsilon_{21} = \epsilon_{2}$ with the constraint $\epsilon_1 + \epsilon_2 = 1$. Therefore,  our model consists of two groups of identical sine circle maps where $N_{\sigma}$ is the number of maps in the group $\sigma$. Here we take $N_1 = N_2 = N$. Each map in a given group is coupled to all the maps in its own group by the parameter $\epsilon_1$ whereas it is coupled to the maps in the other group by the parameter $\epsilon_2$. Thus the system in equation \ref{sinecml} is controlled by three independent parameters, $K, \Omega, \epsilon_1$. A schematic of the CML of Eq. \ref{sinecml} with three lattice sites in each group is shown in Fig. \ref{fig: topology}. 

\begin{figure}[H]
\centering \begin{tabular}{ccc}
  \includegraphics[scale = 0.45]{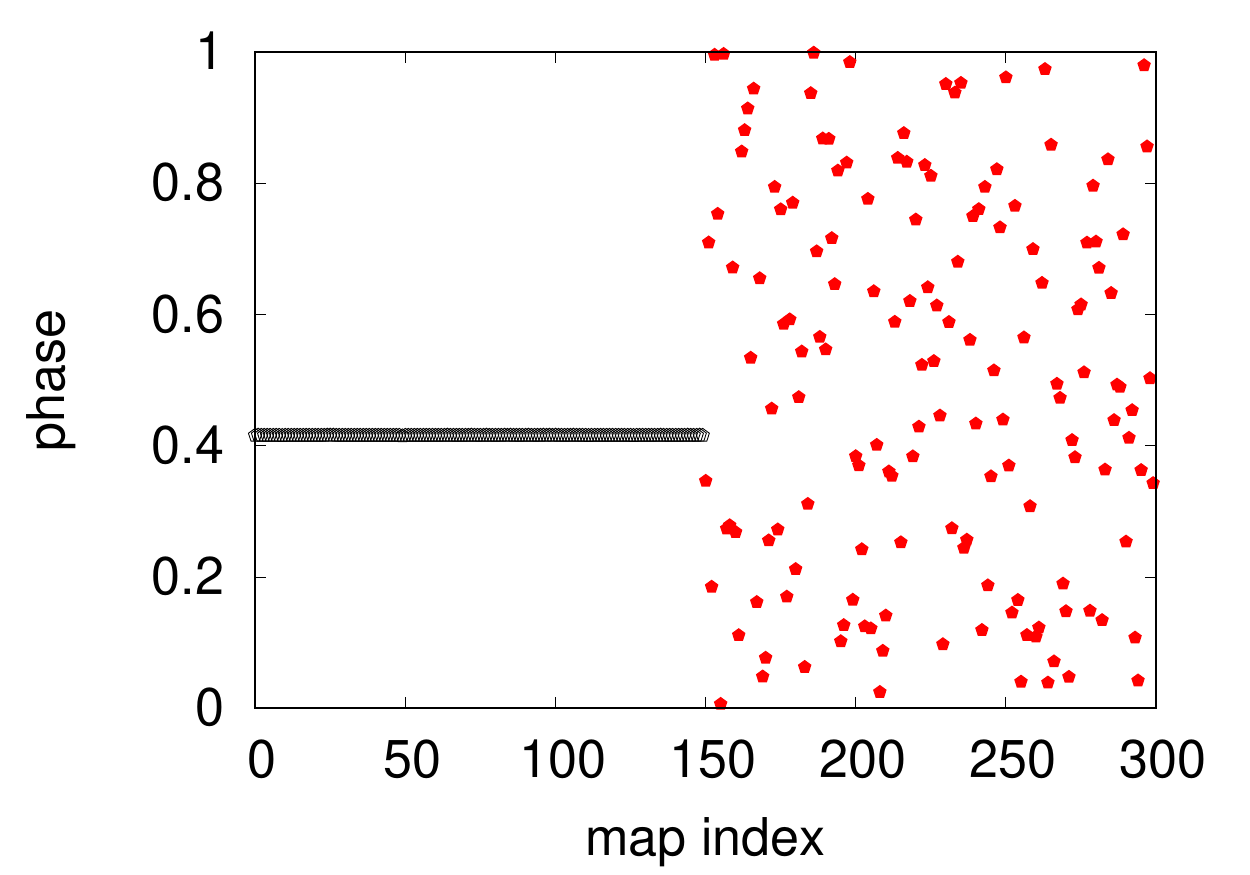}&
  \hspace{-0.8cm}
  \includegraphics[scale = 0.5]{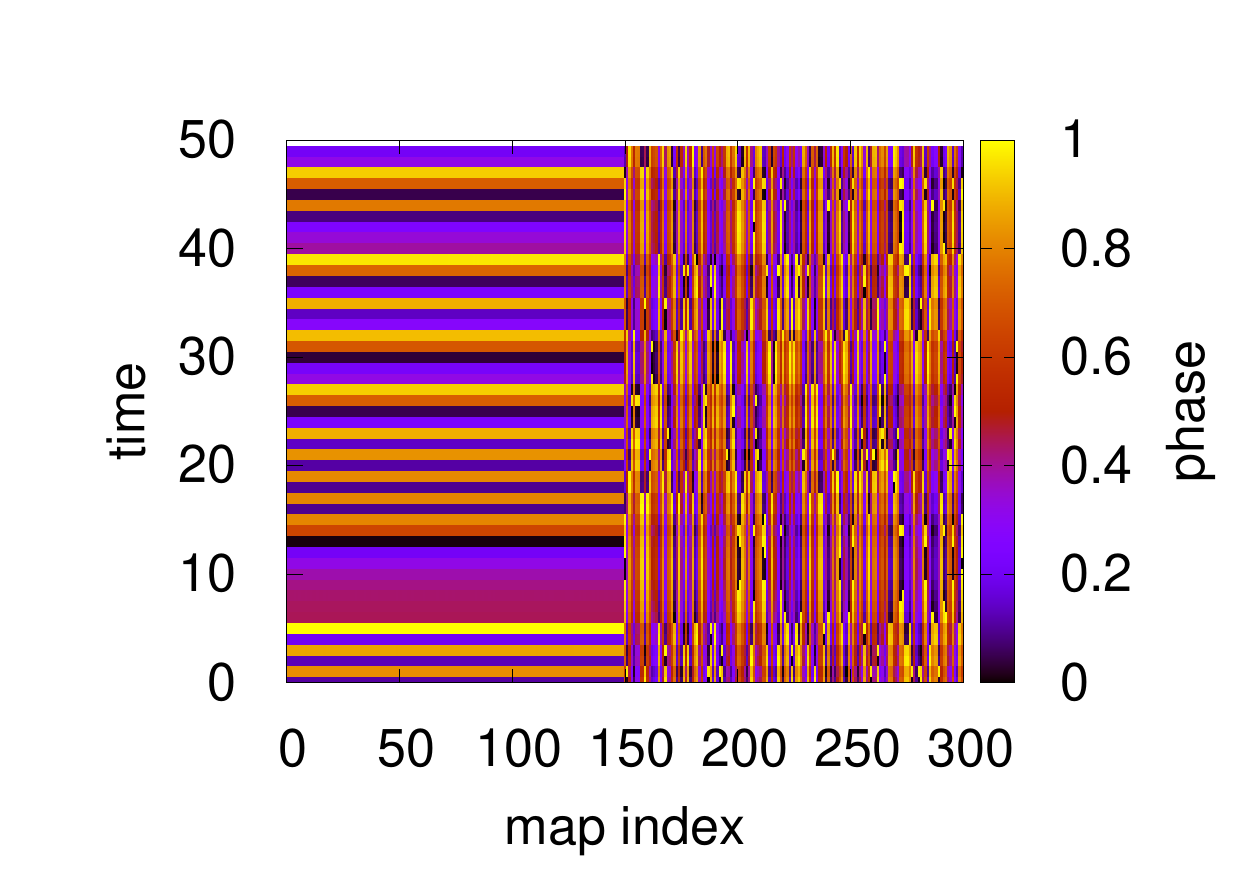}&
  \hspace{-0.5cm}
  \includegraphics[scale = 0.45]{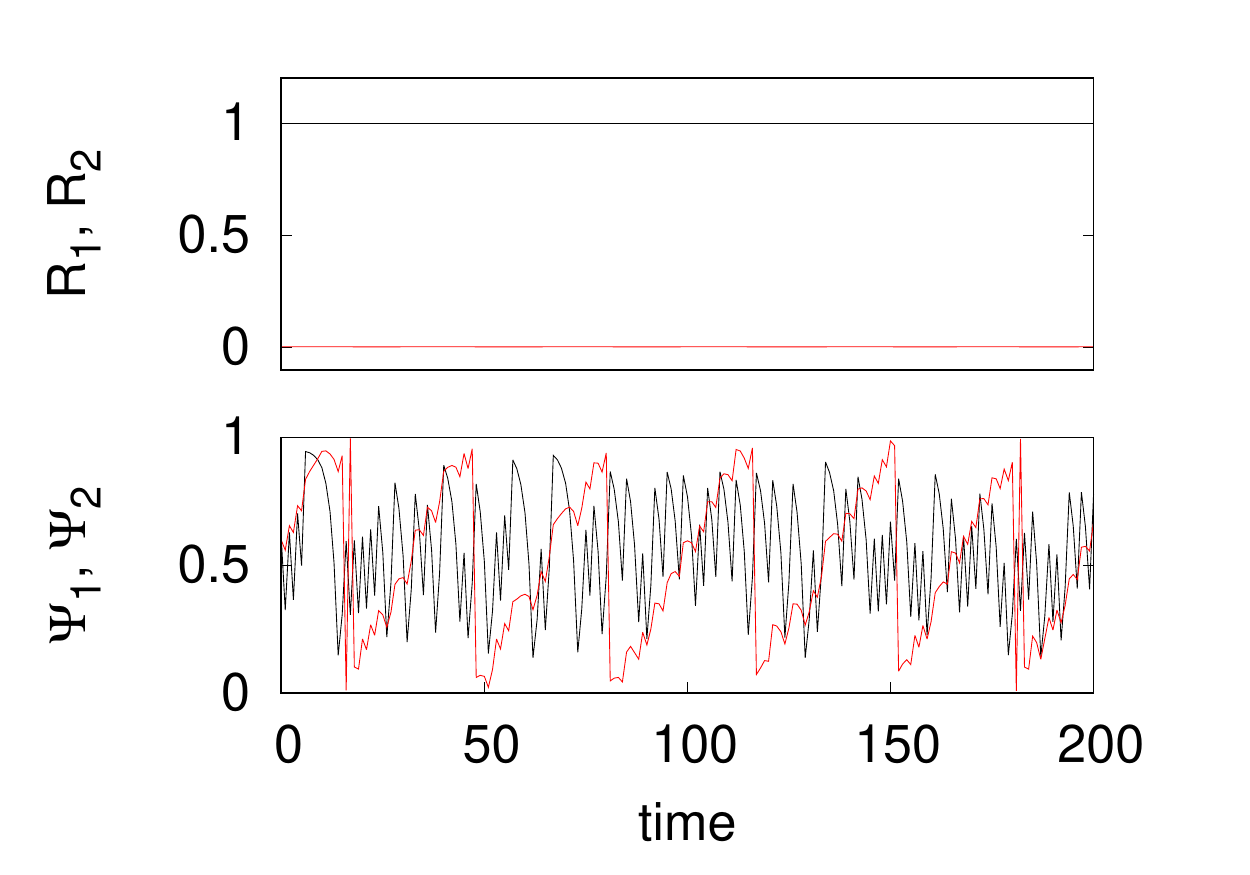}\\
   (a) & (b) & (c)\\
  \includegraphics[scale = 0.45]{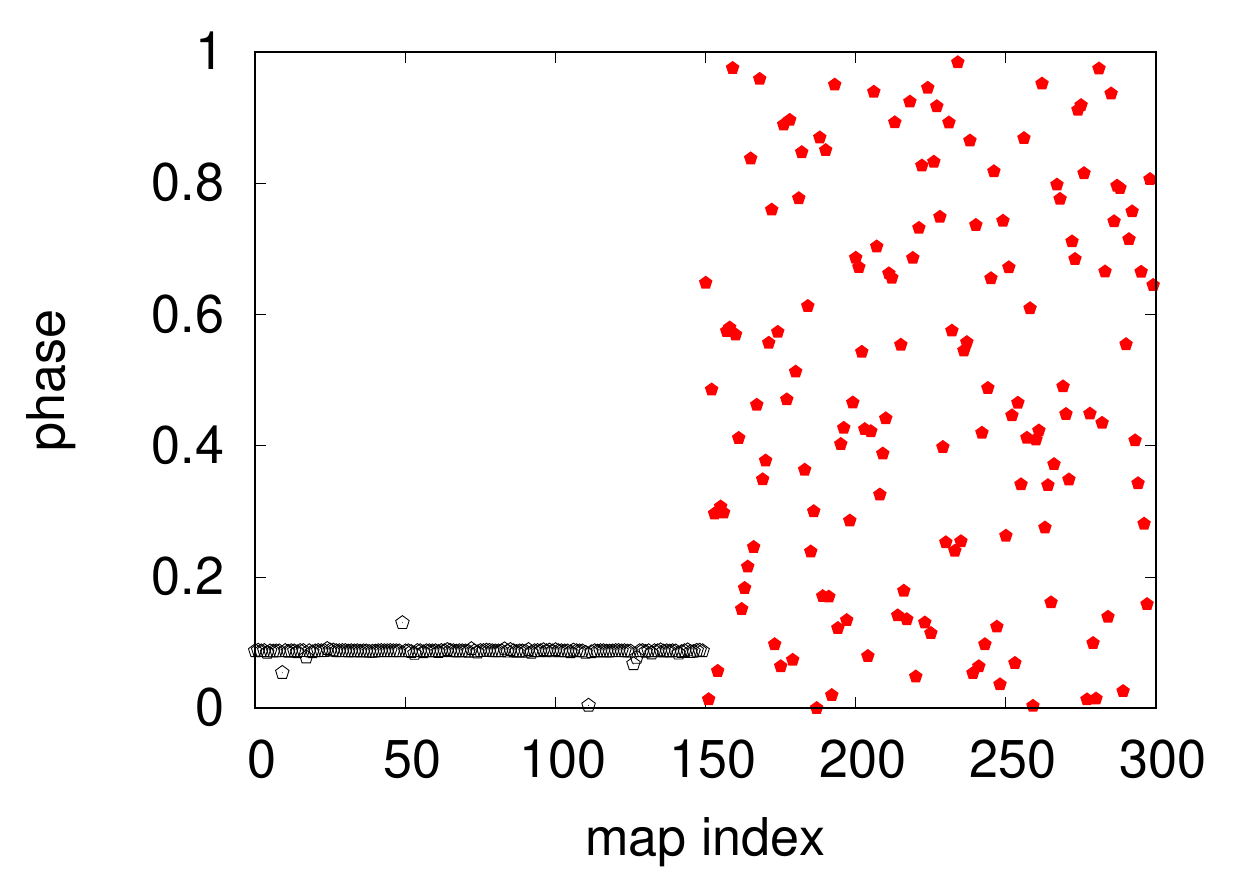}&
  \hspace{-0.8cm}
  \includegraphics[scale = 0.5]{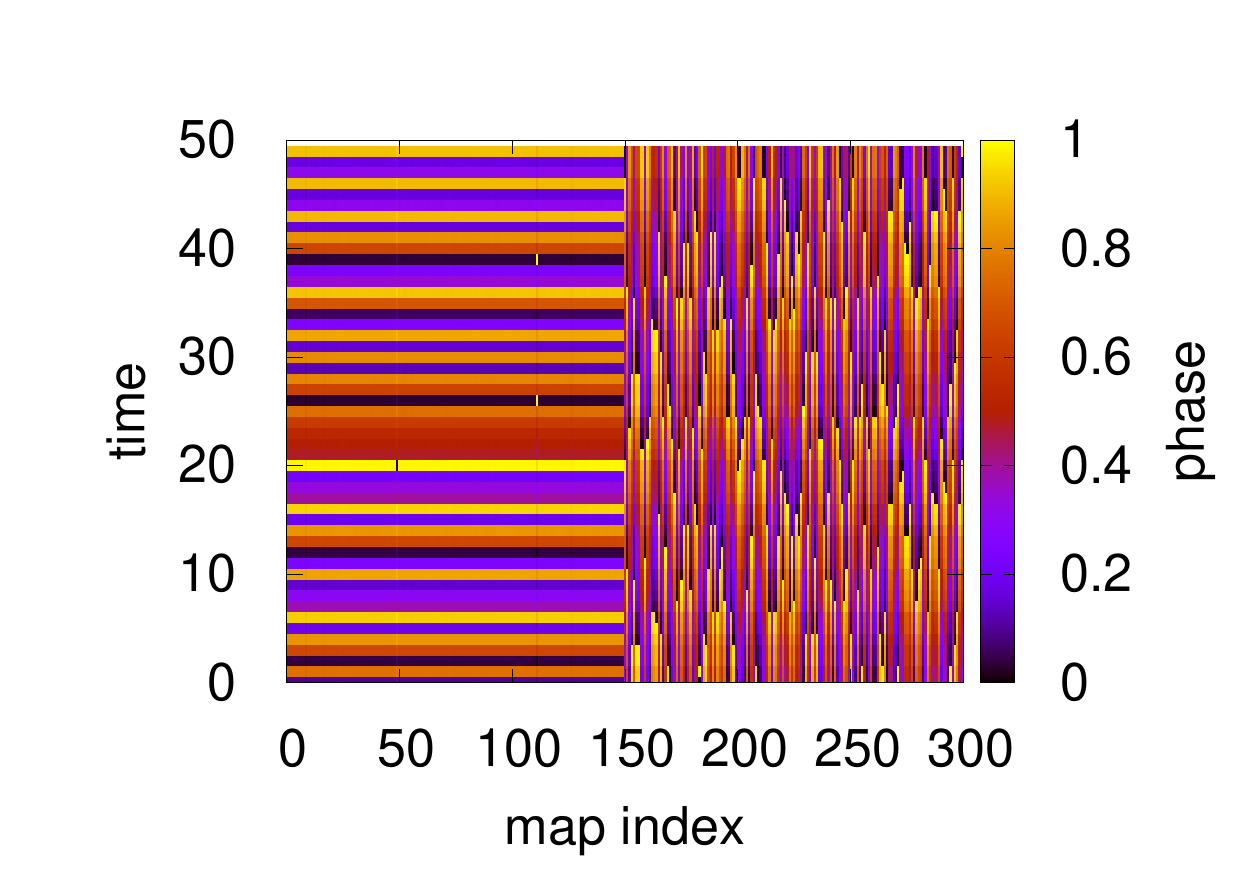}&
  \hspace{-0.5cm}
  \includegraphics[scale = 0.45]{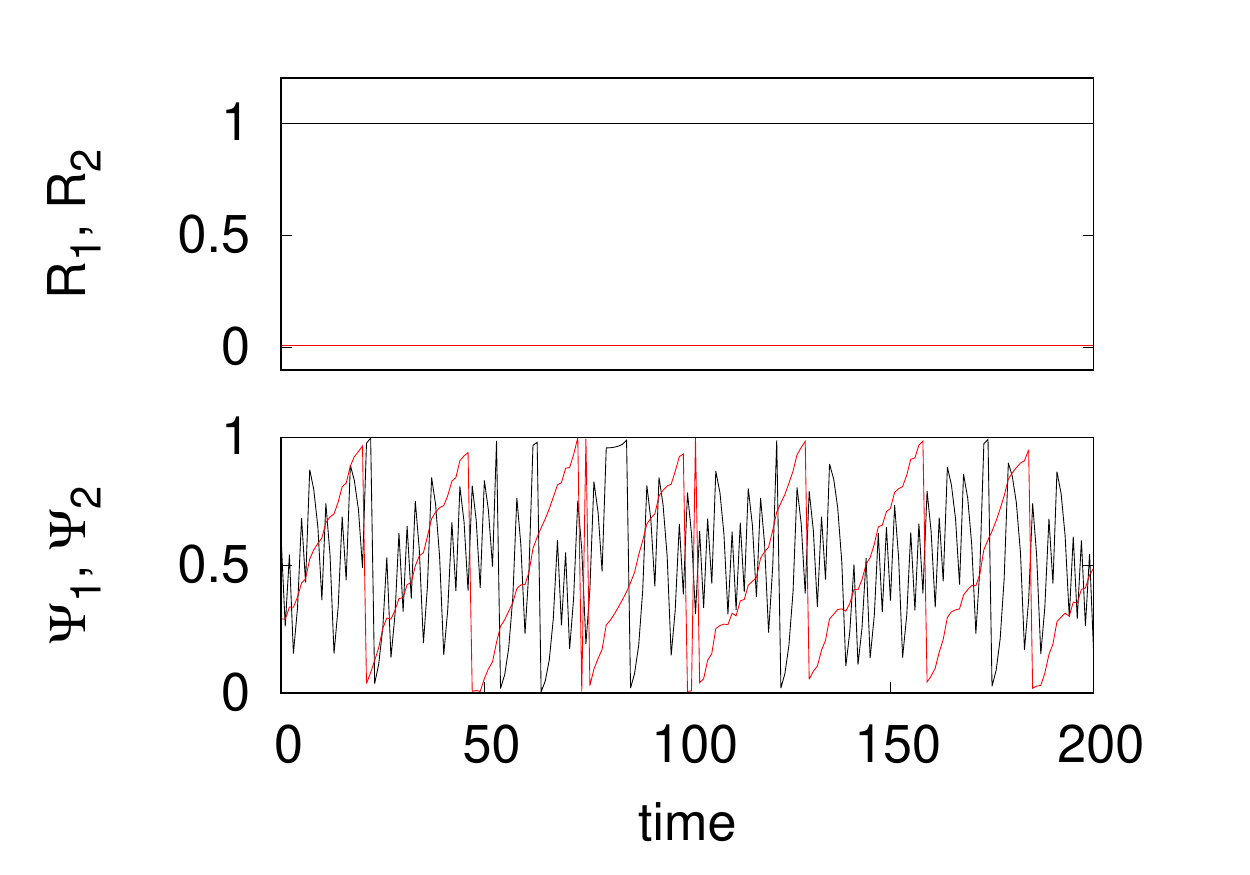}\\ 
  (d) & (e) & (f)\\
 \end{tabular}
\caption{\label{fig: initial}\footnotesize (color online) (a) The snapshot of the chimera state with a purely synchronised group is shown. The parameters are $K = 10^{-5}, \Omega = 0.27, \epsilon_1 = 0.82, N = 150$ (b) The space time plot of the chimera state without the defects in the synchronised group. (c) The temporal variation of $R^{1}, R^{2}, \Psi^{1}, \Psi^{2}$ for the chimera states with complete phase synchronisation in group one. (d) A snapshot of the chimera state with defects in the synchronised group. The parameters are $K = 10^{-5}, \Omega = 0.27, \epsilon_1 = 0.93, N = 150$. The sites between 1 to 150 belong to group one and the sites between 151 to 300 belong to group two. (e) The space time plot of the chimera state shown in (a). (f) The temporal variation of order parameter $R^{1}, R^{2}$ and average phases, $\Psi^{1}, \Psi^{2}$ (defined in Eq. \ref{ord_ang_1} and \ref{ord_ang_2} for group one and two respectively) for the chimera states with defects in the synchronised group shown in (d). In both (c) and (f), $R^1$ and $\Psi^1$ are shown in black whereas $R^{2}$ and $\Psi^{2}$ are denoted in red.}
\end{figure}

As said above the system under consideration is a multivariable system with maps that are coupled globally with two groups which differ in their intergroup and intragroup coupling. As a consequence of this, different initial conditions generally evolve to distinct attractors with different spatiotemporal properties; e.g. an initial condition where an identical phase is assigned to each site will always evolve to a globally synchronised state. In \cite{nayak2011} it was shown that an initial condition, where all the phases of the maps in one group are identical while the maps in the other group are set to random phases between zero and one, evolves to chimera states, clustered chimera states, clustered states etc. at different region in the parameter space. Another initial condition with a system wide splay phase configuration was shown to evolve to a splay phase state, and to splay chimera states depending on the parameters \cite{neelima2016}. Initial conditions such as these break the symmetry between the groups. In this paper, we explore this CML using a very general initial condition where the phases of each of the maps in both of the groups are randomly distributed between zero and one.
We report that at certain parameter values, the fully random initial condition evolves to a chimera state which consists of a spatially phase synchronised group and a spatially and temporally phase desynchronised group (Fig. \ref{fig: initial}). At particular values of $K, \Omega, \epsilon_1$ and $N$ we find a chimera phase state with a purely synchronised subgroup where all maps in group one belong to a phase synchronised cluster (see Fig. \ref{fig: initial}(a)) whereas at other parameters we observe chimera states, where the spatially phase synchronised subgroup has defects, as the phases of a small fraction of circle maps do not belong to the synchronised cluster (Fig. \ref{fig: initial}.(d)). We also see in Fig. \ref{fig: initial}.(b) and (e) that the space time variation of the desynchronised group in both type of chimera states shows spatiotemporally intermittent structures, as synchronised islands in the shape of cones can be observed within the desynchronised phases. 

\section{Phase diagram}\label{PD}
We note that the system is controlled by the parameters $K, \Omega, \epsilon_1$. Apart from this set of parameters, the system dynamics also depends on the size, $2N$ of the system and the initial condition. We fix the size of the system at $N = 150$ and vary the parameters to look for the chimera phase configuration. To identify the chimera states as seen in Fig. \ref{fig: initial} we use the order parameters, $R_{n}^{1}, R_{n}^{2}, R_{n}$ and the average phase, $\Psi^{1}_{n}, \Psi^{2}_{n}, \Psi_{n}$ defined respectively for each of the groups at time step $n$ as, 
\begin{equation}
 R_{n}^{1}\exp{\left( i2\pi\Psi^{1}_{n} \right)} = \frac{1}{N}\sum\limits_{j = 1}^{N} \exp{\left( i2\pi\theta^{1}_{n}(j) \right)}
 \label{ord_ang_1}
\end{equation}
\begin{equation}
R_{n}^{2}\exp{\left( i2\pi\Psi^{2}_{n} \right)} = \frac{1}{N}\sum\limits_{j = 1}^{N} \exp{\left( i2\pi\theta^{2}_{n}(j) \right)}
\label{ord_ang_2}
\end{equation}
\begin{equation}
R_n \exp{\left( i2\pi\Psi_{n} \right)} = \frac{1}{2N}\sum\limits_{\sigma = 1}^{2}\sum\limits_{j = 1}^{N}\exp{\left( i2\pi\theta_{n}^{\sigma}(j) \right)}
\label{global}
\end{equation}

It is clear that $R_{n}^{1}$, $R_{n}^{2}$ becomes one when the phases of the maps in the corresponding group are spatially synchronised at time step $n$. In that case, the phases at which the groups synchronise are given by $\Psi^{1}_{n}, \Psi^{2}_{n}$ respectively. Similarly their values become approximately zero when the phases are uniformly distributed between zero and one. Similar conclusions can be drawn for $R_n, \Psi_n$ if the whole system is phase synchronised or desynchronised. If all the maps are fully phase synchronised at a time step, then $\Psi^{1}_{n}$ and $\Psi^{2}_{n}$ become equal at that time step, while $R_{n}^{1}, R_{n}^{2}, R_{n}$ become one. These properties of these quantities enable us to look for the chimera states of the types shown in Figs. \ref{fig: initial}.(c) and (d), as we vary the parameters $K, \Omega, \epsilon_1$.

\par
\begin{figure}[H]
\centering \begin{tabular}{c}
 \includegraphics[scale = 0.55]{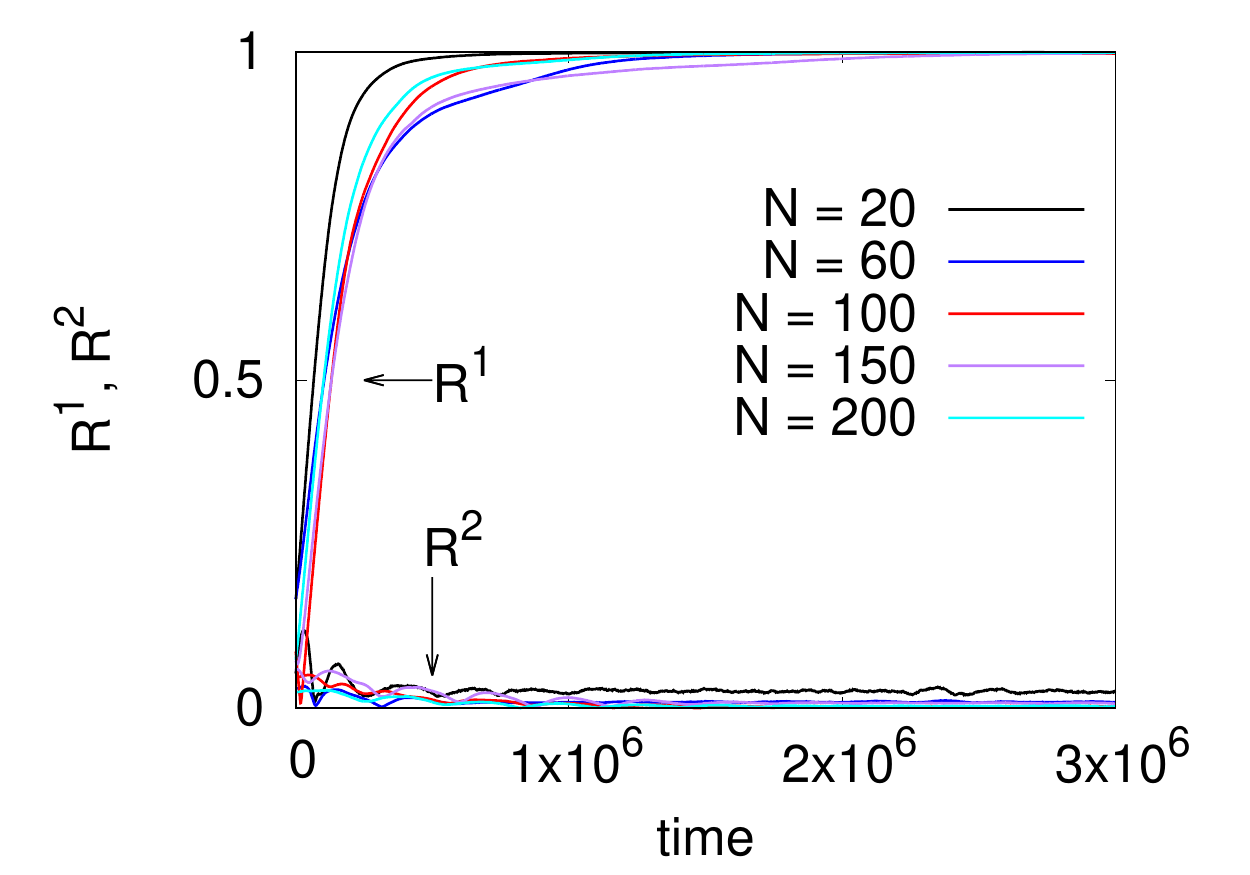}\\
 (a)\\
 \end{tabular}
 \centering \begin{tabular}{cc}
 \includegraphics[scale = 0.6]{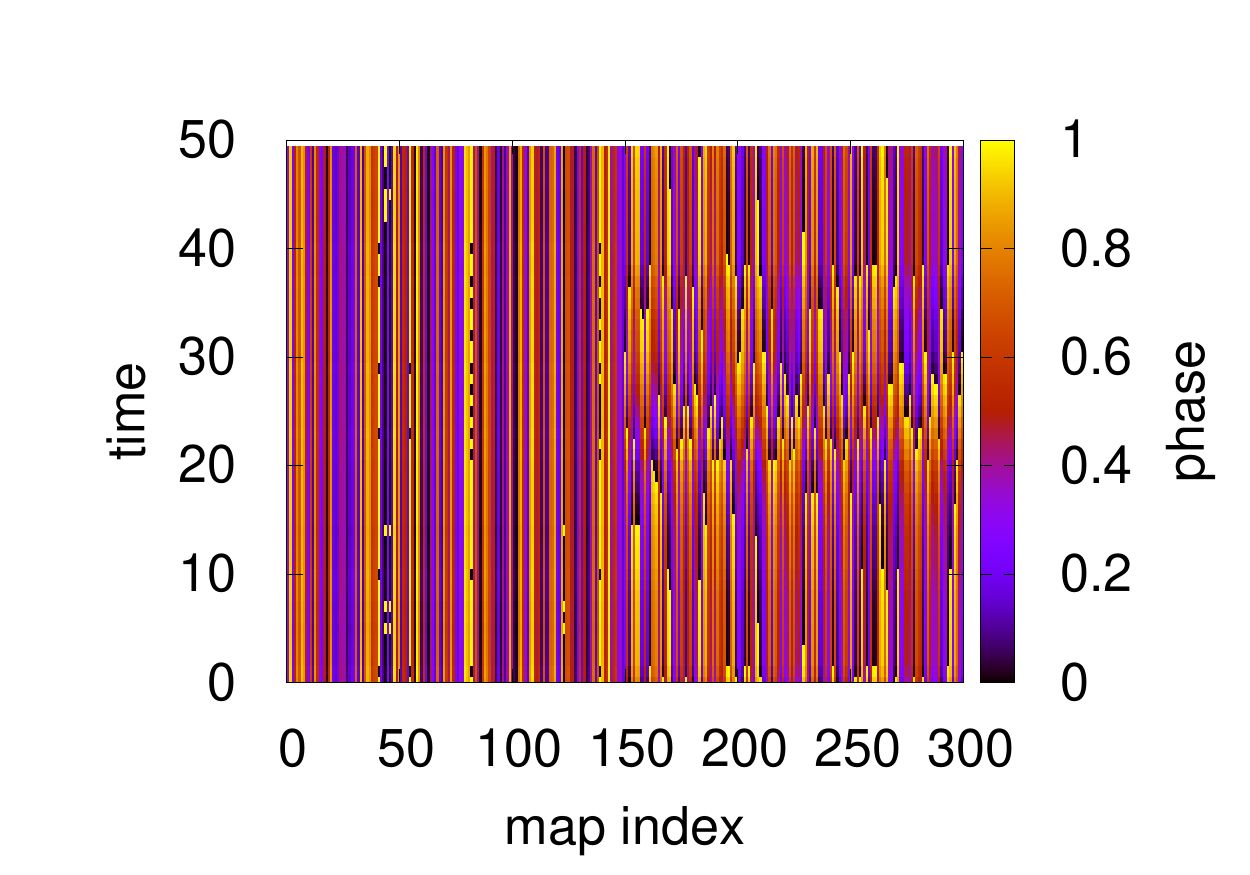}&
 \includegraphics[scale = 0.6]{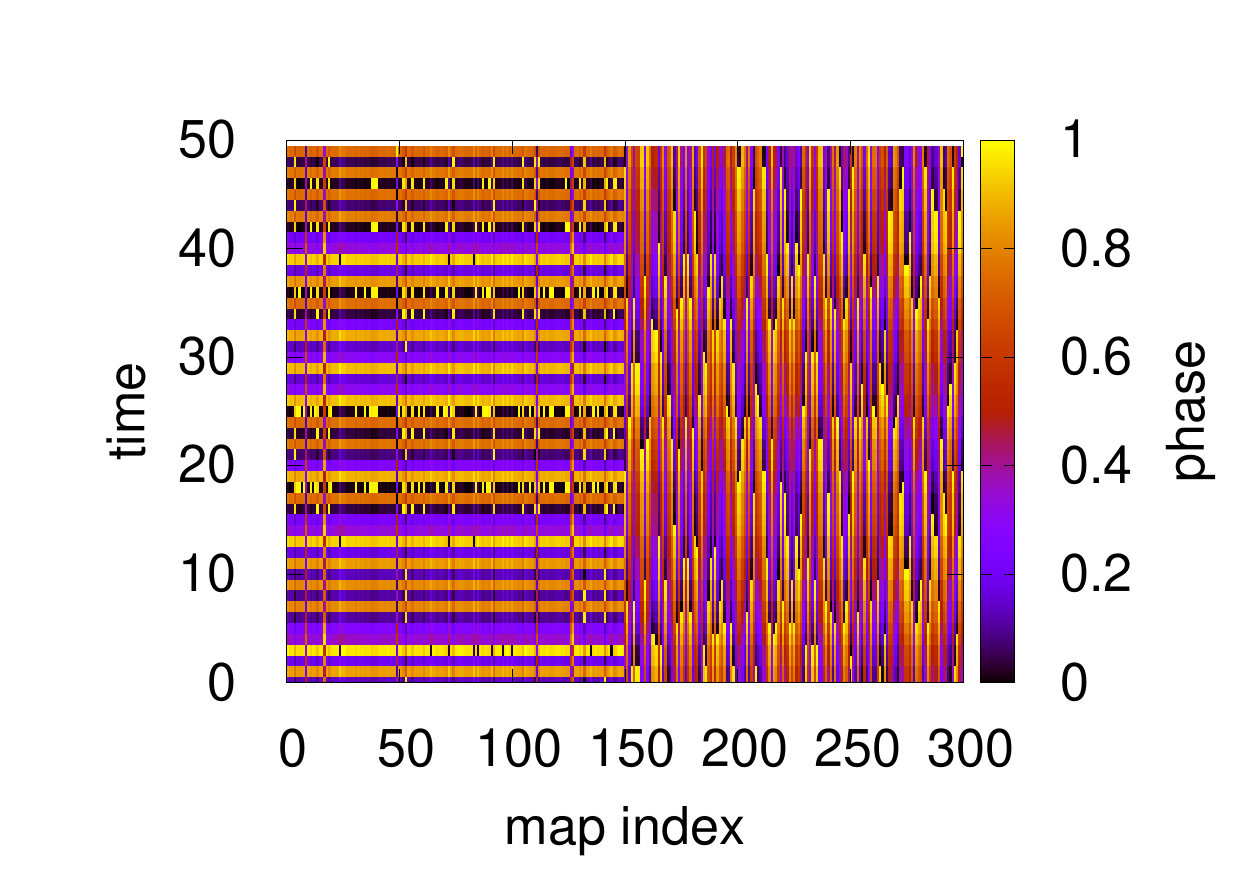}\\
 (b) & (c)\\
 \end{tabular}
\caption{\label{fig: trans} \footnotesize (color online) (a) The variation of the order parameters, $R, R_{n}^{1}, R_{n}^{2}$ with time for $N = 20, 60, 100, 150$ and $200$. The space time plot of the system after (b) 20000 time steps and (c) 500000 time steps for $N = 150$. For the above plots we use the parameters $K = 10^{-5}, \Omega = 0.27, \epsilon_1 = 0.93$.}
\end{figure}

 It is clear that the minimum number of time steps required for the system to settle into chimera states of interest here is a function of the system size. Figure \ref{fig: trans}.(a) shows the variation of the order parameters $R_{n}^{1}, R_{n}^{2}, R_{n}$ with time for different system sizes, $N = 20, 60, 100, 150, 200$, as the Eq. \ref{sinecml} settles from a completely random initial condition to the chimera state shown in Fig. \ref{fig: initial}. As expected, the transient time for systems of smaller sizes is relatively less than that required by larger systems. Overall we see that subgroup order parameter $R^{1}_{n}$ increases to values above 0.8 after three hundred thousand time steps, and slowly tends to one approximately after three million time steps while the subgroup order parameter $R_{n}^{2}$ becomes zero. Such values of the group wise order parameters imply chimera phase configurations. Thus our system takes at least three million time steps to settle into the chimera states of the type shown in Fig. \ref{fig: initial}. The space time variation of the phases of the maps at intermediate time steps show that the CML is in mixed configurations which are different (see Fig. \ref{fig: trans}.(b), (c)) from the chimera states under consideration. Therefore we always evolve the system for  $3\times 10^6$ iterations or more, in all our subsequent numerical calculations.

\begin{figure}[H]
\centering \begin{tabular}{cc}
  \includegraphics[scale = 0.65]{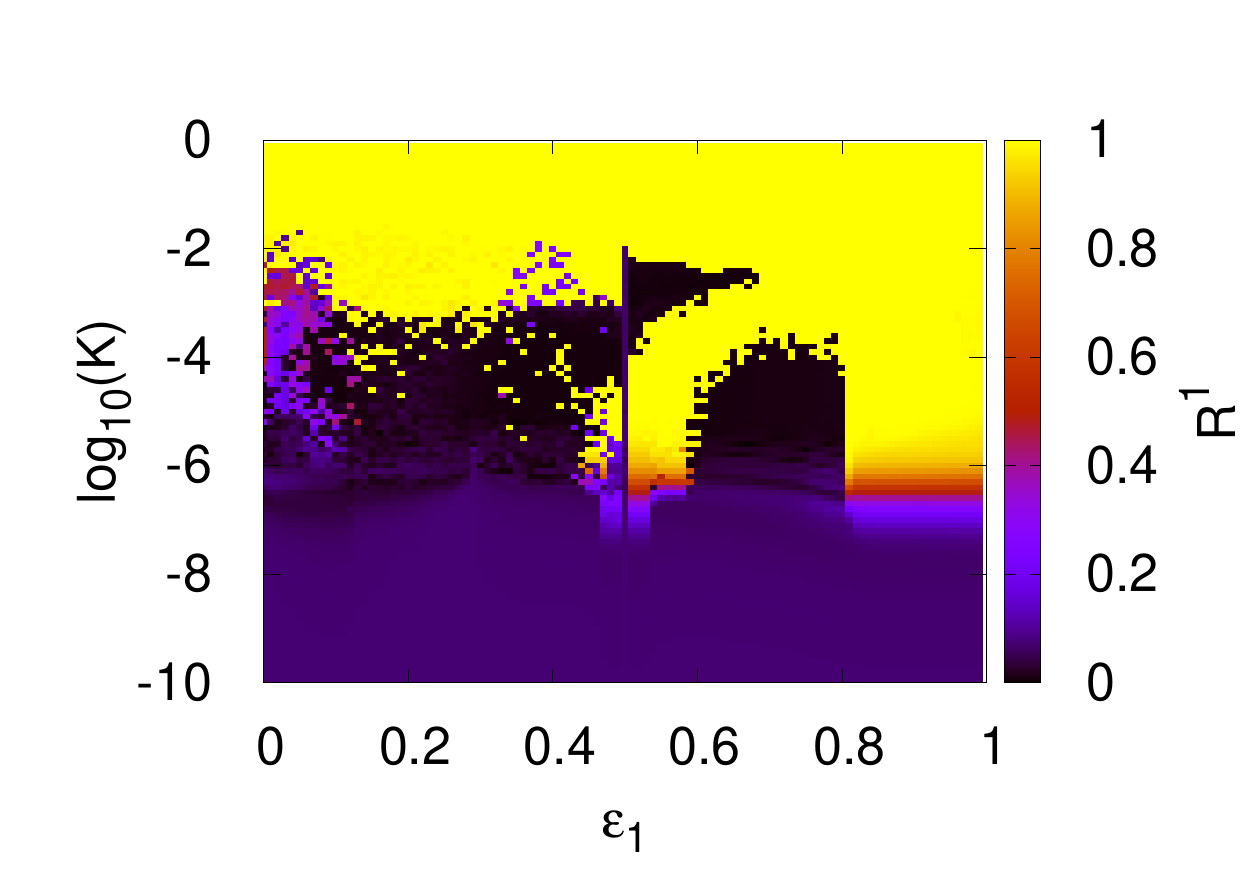}&
  \includegraphics[scale = 0.65]{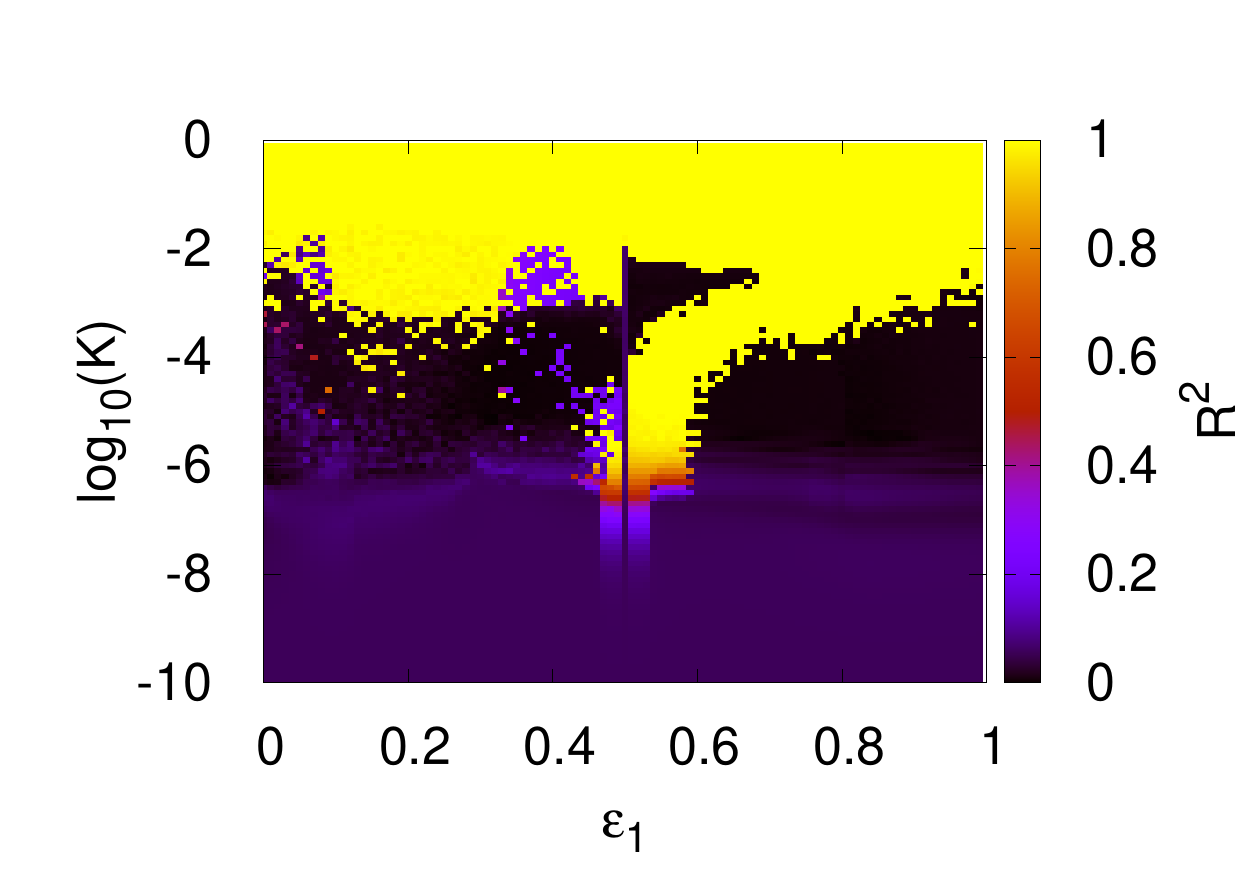}\\
  (a) & (b)\\
  \end{tabular}
  \centering \begin{tabular}{c}
  \includegraphics[scale = 0.65]{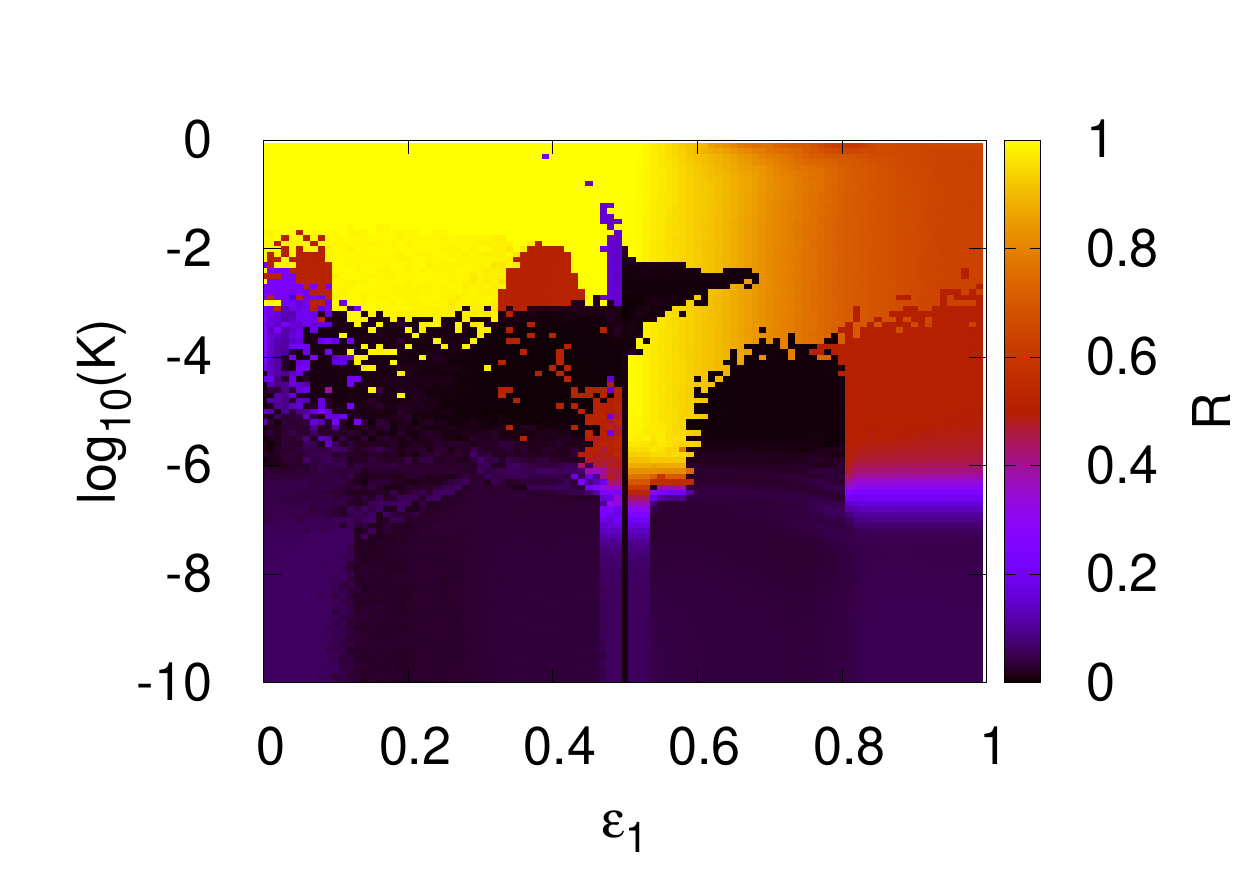}\\
  (c)\\
  \end{tabular}
  \caption{\label{fig: order_1} \footnotesize (color online) The order parameters (a)$R^{1}$, (b)$R^{2}$ and (c)$R$ are plotted for the values  $N = 150, \Omega = 0.27$. The color code for the values of the order parameter is indicated in the vertical bar in each plot. At each parameter value we use a random initial condition and iterate the system initially for $4 \times 10^6$ time steps, after which the order parameters $(R^1, R^2, R)$ are calculated and averaged over $10^5$ time steps. The region where chimera states are seen is identified  by the order parameter values $R^1 \approx 1$ and $R^2 \approx 0$.}
  \end{figure}

We obtain a phase diagram for $\Omega = 0.27$ and vary the parameters $K, \epsilon_1$ in the range $10^{-8} < K < 1$ and $0 < \epsilon_1 < 1$. At each values of these parameters we use a fixed set of initial phase values which are randomly distributed between zero and one. We calculate $R_{n}^{1}$, $R_{n}^{2}$, $R_n$ for $10^5$ time steps and calculate the average after the system of Eq.\ref{sinecml} is iterated for three million time steps. Figure \ref{fig: order_1} show the values of $R^{1}$, $R^{2}$ and $R$ respectively with the variation of $K, \epsilon_1$ at $\Omega = 0.27$.

\begin{figure}[H]
\centering \begin{tabular}{cc}
\includegraphics[scale = 0.65]{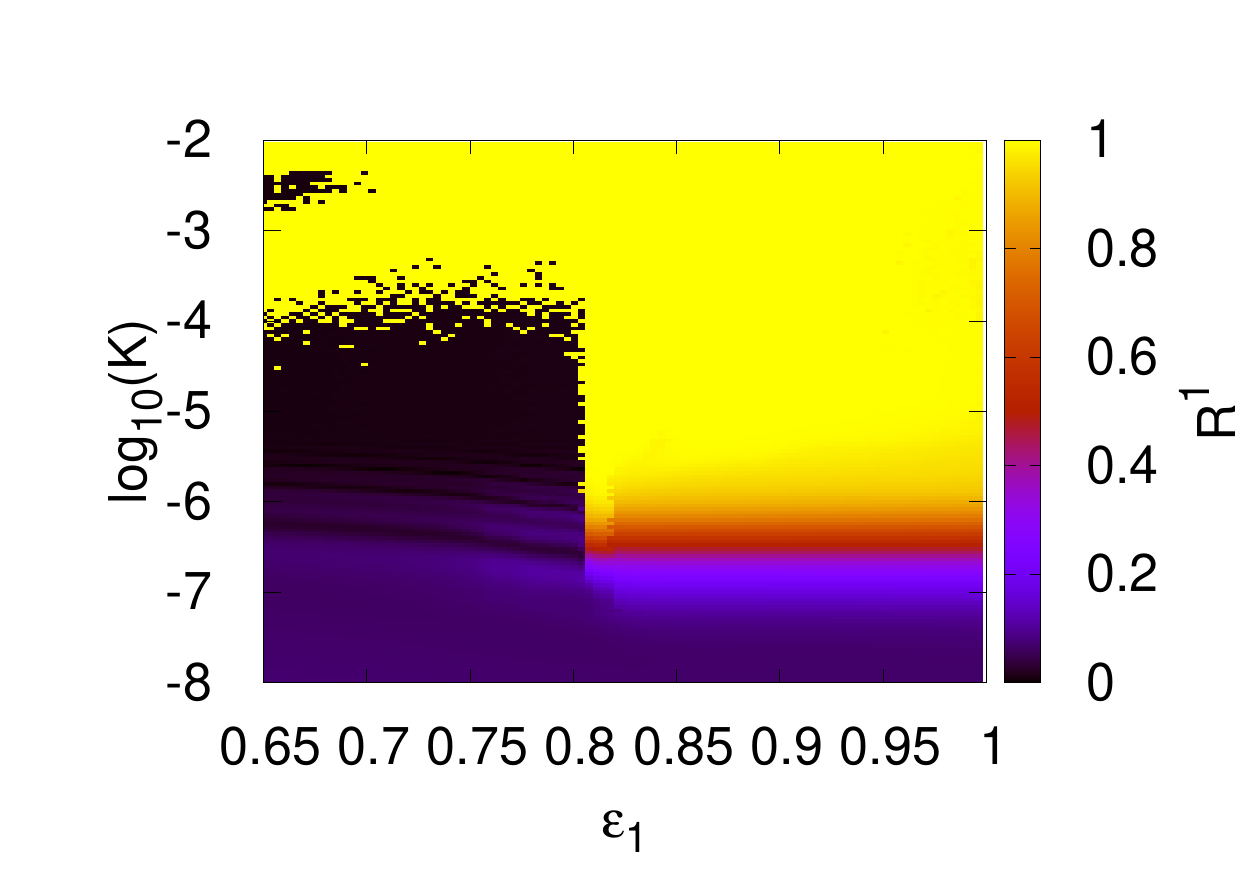}&
\includegraphics[scale = 0.65]{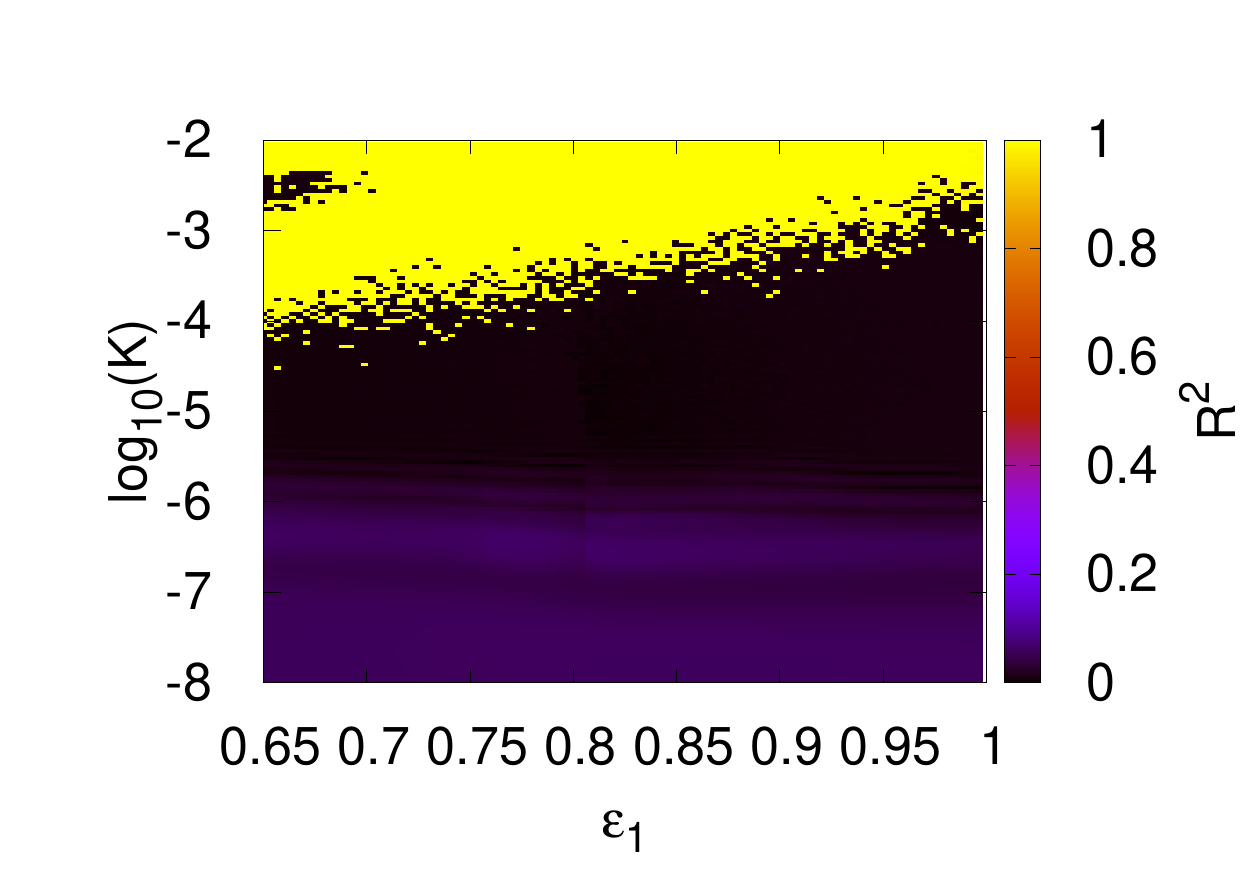}\\
(a) & (b)\\
\end{tabular}
\caption{\label{fig: order_zoom}\footnotesize (color online) The order parameters (a) $R^1$ corresponding to group $1$, and (b) $R^2$ corresponding to group $2$  are calculated and plotted between the parameter region $0.65 < \epsilon_1 < 1.0$, $10^{-8} < K < 10^{-2}$ for $\Omega = 0.27$. At each of the parameter values we use a random initial condition and iterate the system for $4 \times 10^6$ time steps after which we calculate the order parameters $(R^1, R^2)$ and average them over $10^5$ time steps. Chimera states are found in the region where the subgroup order parameters take values  $R^1 \approx 1, R^2 \approx 0$. } 
\end{figure}

 These show the existence of the chimera states (Fig. \ref{fig: initial}.(c)) in a region in $K, \epsilon_1$ space approximately given by $0.8 < \epsilon_1 < 1, 10^{-3} < K < 10^{-6}$ surrounded by other phase configurations around it. A magnified version of this phase diagram around this region is shown in Fig. \ref{fig: order_zoom}. Five types of distinct phase configurations can be found in the phase diagram of Fig. \ref{fig: order_zoom}. These are chimera states, two clustered states, globally synchronised states and fully desynchronised states. The details of these dynamical states are as follows, 
\par
\begin{figure}[H]
  \centering \begin{tabular}{ccc}
 \hspace{-.5cm}
 \includegraphics[scale = 0.43]{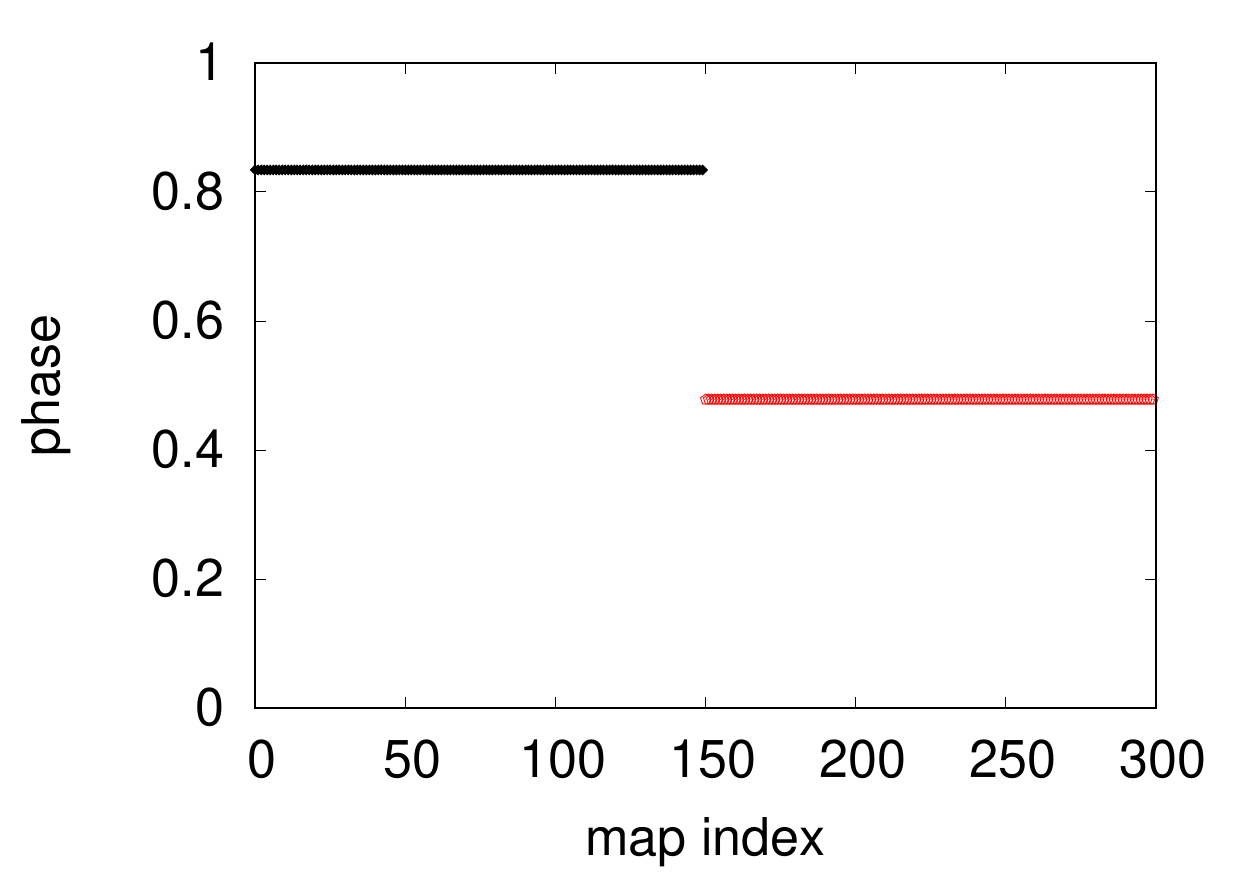}&
 \hspace{-0.5cm}\includegraphics[scale = 0.44]{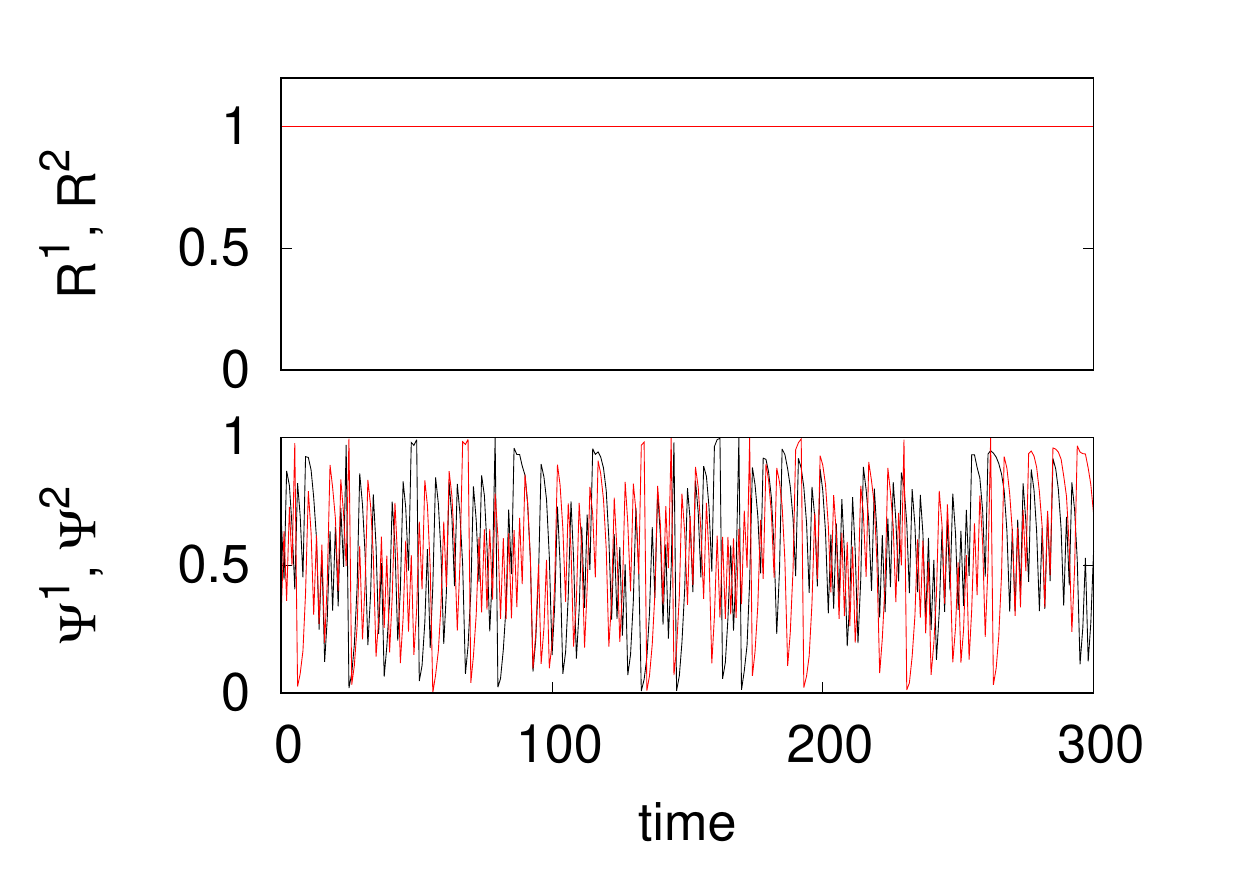}&
 \hspace{-0.5cm}\includegraphics[scale = 0.43]{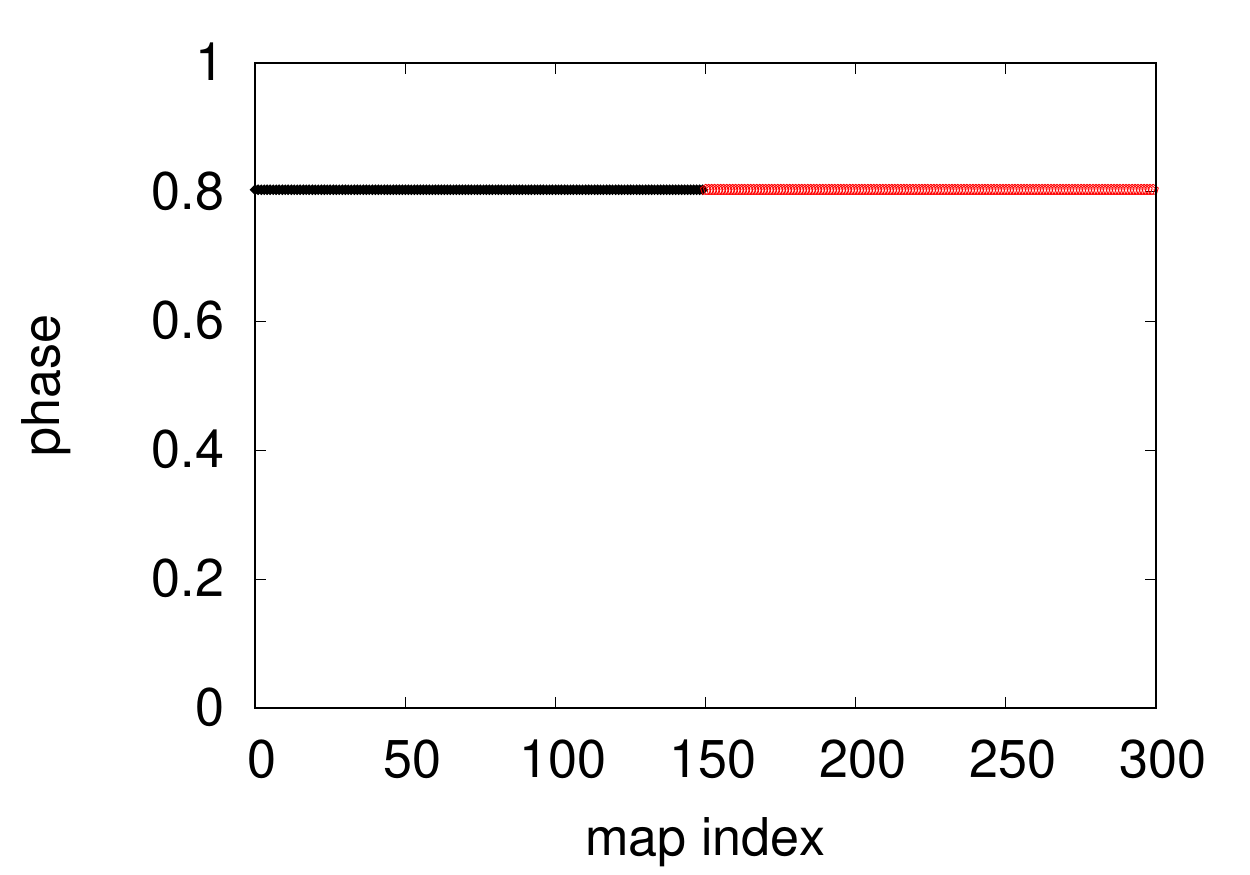}\\
  (a) & (b) & (c)\\
  \hspace{-.5cm}
  \includegraphics[scale = 0.44]{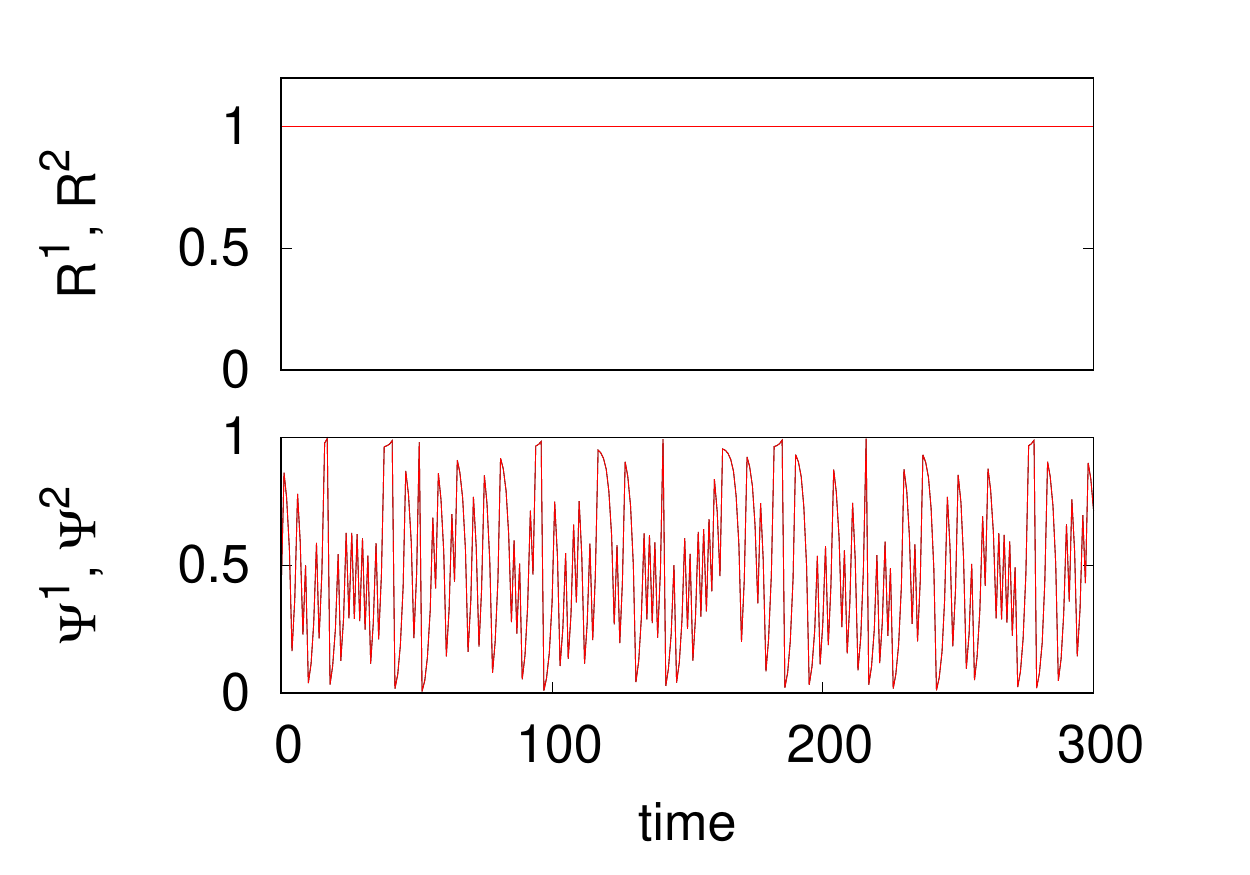}&
  \hspace{-0.5cm}\includegraphics[scale = 0.43]{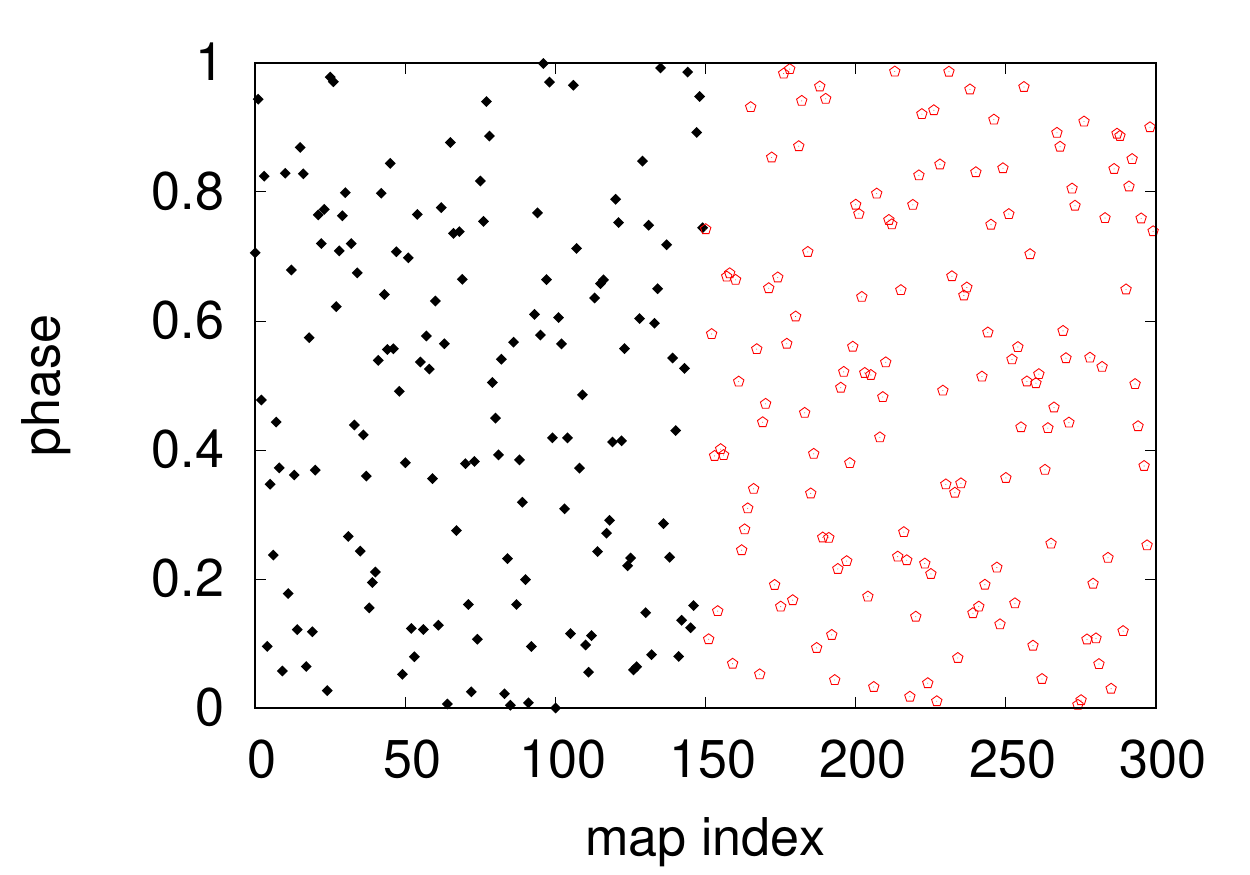}&
  \hspace{-0.5cm}\includegraphics[scale = 0.44]{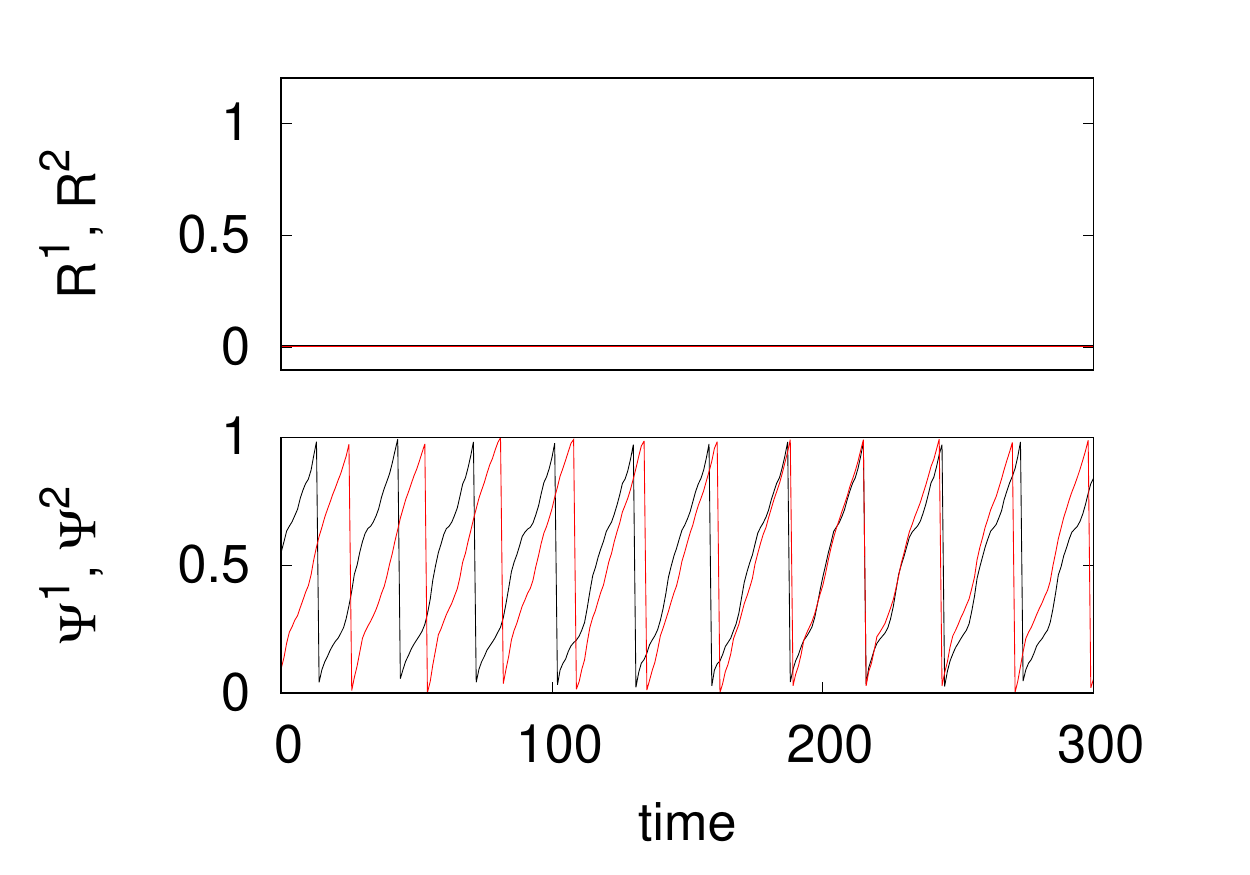}\\
  (d) & (e) & (f)\\
 \end{tabular}
\caption{\label{fig: states}\footnotesize (color online) (a) A snapshot of the two cluster phase configuration at $K = 10^{-2}, \epsilon_1 = 0.93$. is shown. (b) The variation of $R^{1}, R^{2}, \Psi^{1}, \Psi^{2}$. is shown for the same parameters. (c) A snapshot of the fully synchronised state at $K = 10^{-2}, \epsilon_1 = 0.45$ and (d) the variation of $R^{1}, R^{2}, \Psi^{1}, \Psi^{2}$ with time for the same parameters is shown. (e) A snapshot of a fully de-synchronised state at $K = 10^{-5}, \epsilon_1 = 0.75$ and the order parameters and the average phases are shown in (f). Other parameters viz. $\Omega, N$ were kept fixed at $0.27$ and $150$ respectively. All of the above phase configurations were obtained for the same set of initial conditions.}
\end{figure}

\begin{enumerate}
\item \textbf{Case 1 and case 2 : Chimera states (Fig. \ref{fig: initial})}: We obtain a chimera state when either $R^{1}$ or $R^{2}$ is one and the value of the other quantity is near zero. We get this condition at several of the parameter values for $\Omega = 0.27$. In particular when $-6 < \log_{10}K < -4$ and $0.8 < \epsilon_1 < 1.0$, at some parameters we find, case 1 : $R^{1} = 1$ and $R^{2} \approx 0$ (see Fig. \ref{fig: initial}.(c)) which indicates the chimera states with pure synchronisation in the synchronised group. Case 2 corresponds to chimera states with defects in the synchronised group for which we find $R^{1} \lesssim 1$ and $R^{2} \approx 0$ (Fig. \ref{fig: initial}.(f)). The temporal variation of $R^1, R^2$ also shows this behaviour. The variation of $\Psi_1$ and $\Psi_2$ with time shows that the variation of the average phases of the phase synchronised and desynchronised group are qualitatively different (see Figs. \ref{fig: initial}.c and f).

\item \textbf{Case 3 : Fully desynchronised states (Figs. \ref{fig: states}.(e), (f))}: These are found at those parameter values where $R^{1}$, $R^{2}$, $R$ are approximately zero. At these parameter values, all the maps in both the groups are temporally and spatially phase desynchronised. The temporal variation of $\Psi^{1}, \Psi^{2}$ suggest that the average phase of both the groups are approximately periodic. They are observed approximately for $\log_{10}K < -6$ and in the region $\log_{10}K < -4$ for $\epsilon_1 < 0.8$.

\item \textbf{Case 4 : Two clustered states (Fig. \ref{fig: states}.(a))}: We find that $R^{1} = 1$ and $R^{2} = 1$ in the parameter region approximately given by $-4 < \log_{10}K < 0$ and $0.5 < \epsilon_1 < 1$. The phases of the maps in each of the groups are such that they are spatially phase synchronised as suggested by the temporal variation of $R^1, R^2$ while the phases at which they synchronise are not equal as suggested by $\Psi^1, \Psi^2$ (see Fig. \ref{fig: states}.(b)). Figure \ref{fig: states}.(b) also suggests that each of these phase clusters do not synchronise to a temporally fixed phase value as can be seen from the variation of the average phases $\Psi^{1}, \Psi^{2}$ (see Fig. \ref{fig: states}.b). 

\item \textbf{Case 5 : Globally synchronised states (Fig. \ref{fig: states}.(c))}: These are characterised by the order parameter values when all three quantities, $R^{1}, R^{2}, R$ are approximately one. They can be seen mostly above $K \approx 10^{-3}$ for $\epsilon_1$ below $0.8$. The temporal variation of the average phases of each of the groups, $\Psi^{1}, \Psi^{2}$ in Fig. \ref{fig: states}.(d) suggests that all the maps are spatially phase synchronised at all time steps although the phase at which they synchronise is not a temporal fixed point similar to the temporal variation of the two clustered state.

\end{enumerate}

\par
\begin{figure}[H]
\centering \begin{tabular}{cc}
 \includegraphics[scale = 0.55]{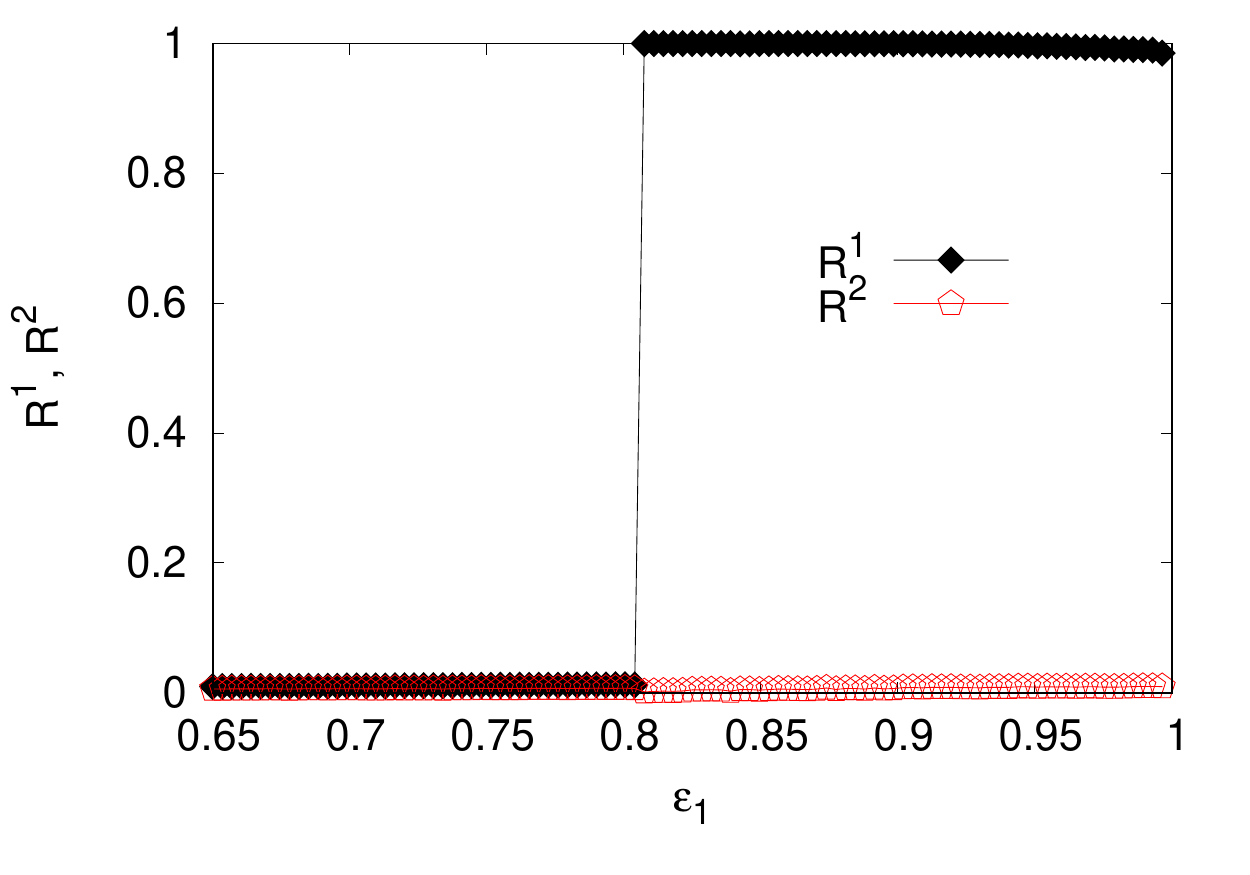}&
 \includegraphics[scale = 0.55]{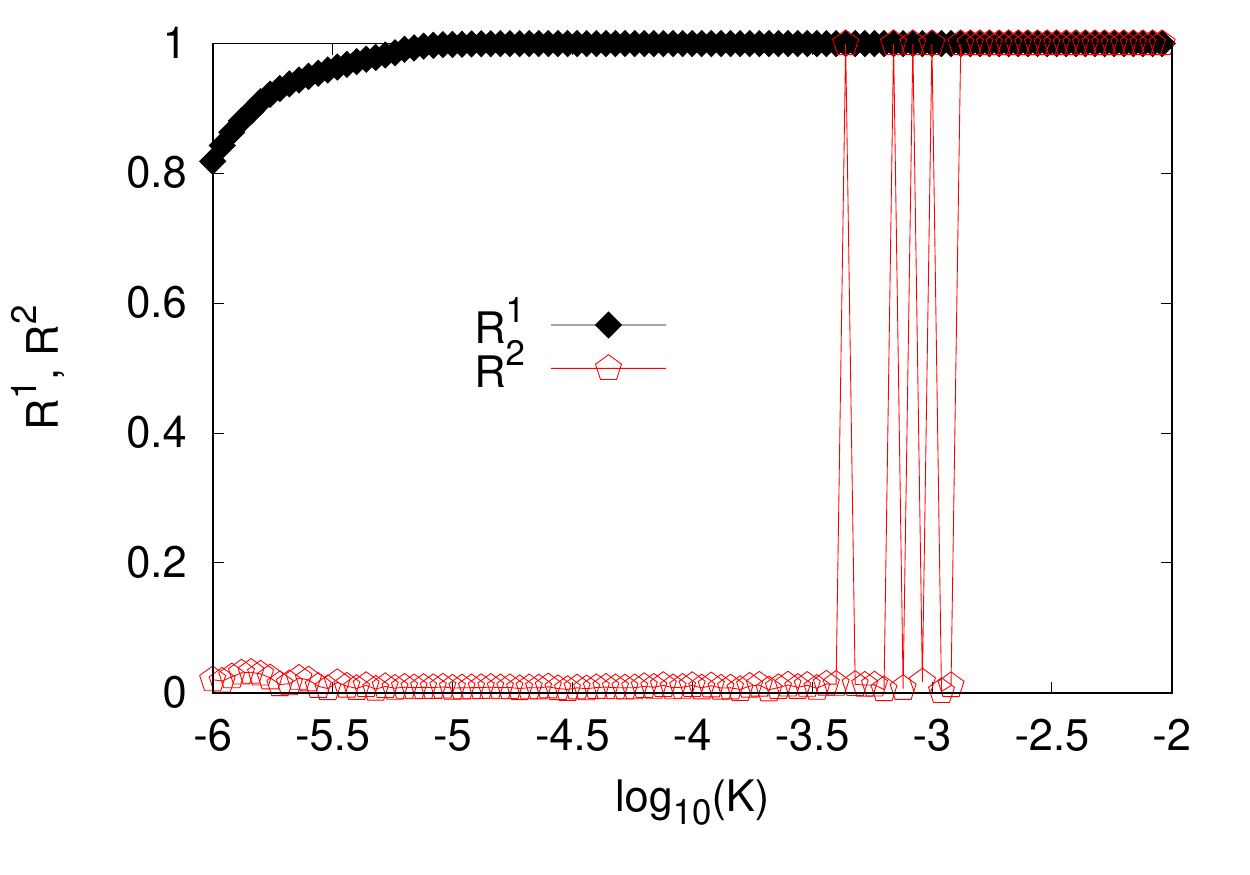}\\
 (a) & (b)\\
 \end{tabular}
 \caption{\label{fig: density_1}\footnotesize (color online) The variation of the group wise order parameters is plotted for (a) the parameter values  $K = 10^{-5}, \Omega = 0.27, N = 150$ as $\epsilon_1$ varies between $0.65$ and one. (b)The order parameters $R^1$ and $R^2$ are plotted as $K$ varies between $10^{-6}$ and $10^{-2}$ for the parameter $\epsilon_1 =0.93$ with $\Omega = 0.27$.}
\end{figure}

 In this paper we are mainly interested in the region of the parameter space where the chimera states are seen and its transition to other phase configurations which are shown in Figs. \ref{fig: states}. In particular we are interested in the region between $-8 < \log_{10}K < -2$ and $0.65 < \epsilon_1 < 1.0$. The variation of the order parameters  $R^1, R^2$ in this region are shown in Fig. \ref{fig: order_zoom}. Figure \ref{fig: order_zoom} shows that the fully desynchronised states seen in the region $-5.5 < \log_{10}K < -4$ and $0.65 < \epsilon_1 < 0.8$ transform to chimera states at $\epsilon_1 = 0.8$. The global phase desynchronised state seen between $-8 < \log_{10}K < -5.5$ and $0.8 < \epsilon_1 < 1$ transforms to chimera states as $\log_{10}K$ increases beyond $-5.5$. Between the parameter values $-4 < \log_{10}K < -3$ and $0.65 < \epsilon_1 < 1$ the chimera states transform to two clustered states. The transitions between these phase configurations due to the variation of the parameters  $K, \epsilon_1$ is better understood from the variation of the order parameters $R^1, R^2$ at different cross sections of the phase diagram in the Fig. \ref{fig: order_zoom}. Figure \ref{fig: density_1}.(a) shows the variation of $R^{1}_{n}$ and $R^{2}_{n}$ with values of  $\epsilon_1$ lying in the range between $0.65$ and one for $K = 10^{-5}$. It can be seen that both the subgroup order parameters  $R^1, R^2$ take values near zero when $\epsilon_1$ is less than 0.8. These values of the order parameters suggest that the system is in a fully phase desynchronised phase configuration for this range of $\epsilon_1$ and $K$ values. When the parameter $\epsilon_1> 0.8$ we see that $R^{1} = 1$ for group one and $R^{2} \approx 0$ zero for group two indicating a chimera phase configuration. When $\epsilon_1 \rightarrow 1$ we see from Fig. \ref{fig: density_1}.(a) that $R^{1} \lesssim 1$ while $R^2 \approx 0$. This indicates that some of the circle maps from the group one have phases that do not belong to the synchronised cluster at these values of $\epsilon_1$. As discussed earlier, this indicates the presence of a chimera phase state with defects in the synchronised group. We take another cross section of this phase diagram at the parameter $\epsilon_1 = 0.93$ in Fig. \ref{fig: density_1}.(b) that shows the variation $R^1$ and $R^2$ with $\log_{10}K$ as it increases from $-6$ to $-2$. We find that the subgroup order parameters take values $R^1 = 1$ and $R^2 \approx 0$ near $\log_{10}K \approx -5.7$ implying the existence of the chimera phase configuration till $\log_{10}K \approx -3.3$. When $\log_{10}K$ lies  between $-3.36$ and $-2.9$ we observe an interchange between these two states for a small variation of $K$. We find $R^1$ and $R^2$ both become one when $\log_{10}K > -2.7$. 
In the next section we discuss the properties of the chimera states as shown in Fig. \ref{fig: initial} and  construct the equivalent cellular automaton for the CML in this regime. 

\vspace{-0.5cm}

\section{Chimera states with STI like structures in the desynchronised group}\label{chim}
We have seen in the previous section that the chimera states with spatiotemporally intermittent behaviour are seen in a large region in the $K, \epsilon_1$ parameter space with $\Omega = 0.27$. We can observe from the space time plots and the temporal variation of $R^1$ (see Fig. \ref{fig: initial}) of this chimera state that the maps in the synchronised group are spatially phase synchronised but the phase at which they synchronise is not a temporal fixed point as shown by the variation of $\Psi^{1}$ (Figs. \ref{fig: initial}.c and f). The variation of $R^2, \Psi^2$ in Figs. \ref{fig: initial}.c and \ref{fig: initial}.(f) maps in the desynchronised group can be seen to be spatially and temporally desynchronised. We calculate the Lyapunov exponents for the space time variation of the chimera states found at $K = 10^{-5}, \Omega = 0.27, \epsilon_1 = 0.93$ and $N = 150$. We find that the largest Lyapunov exponent is 0.692 while the second largest LE is 0.621 whereas the rest of the exponents are zero. This shows that the temporal behaviour of the chimera phase state is hyper-chaotic. We also find the return map at a site by randomly choosing a typical site from each of the groups. We observe that there is a distinct difference between the return map of a site from group one and group two. The return maps for groups one and two show non-banded and banded structures respectively. The space time behaviour of the phases of the circle maps in the desynchronised group suggest the existence of synchronised islands having identical phases inside clusters of spatiotemporally phase desynchronised sites. This implies that there exist intermittent laminar and burst regions in the space time variation of the phases of the maps in group two. In the case of the synchronised group, we can clearly observe that all the sites are always in a spatially laminar stage at each time step. We discuss the method for identifying these laminar and burst sites for our system in the next section.
 
\begin{figure}[H]
\centering \begin{tabular}{cc}
 \includegraphics[scale = 0.55]{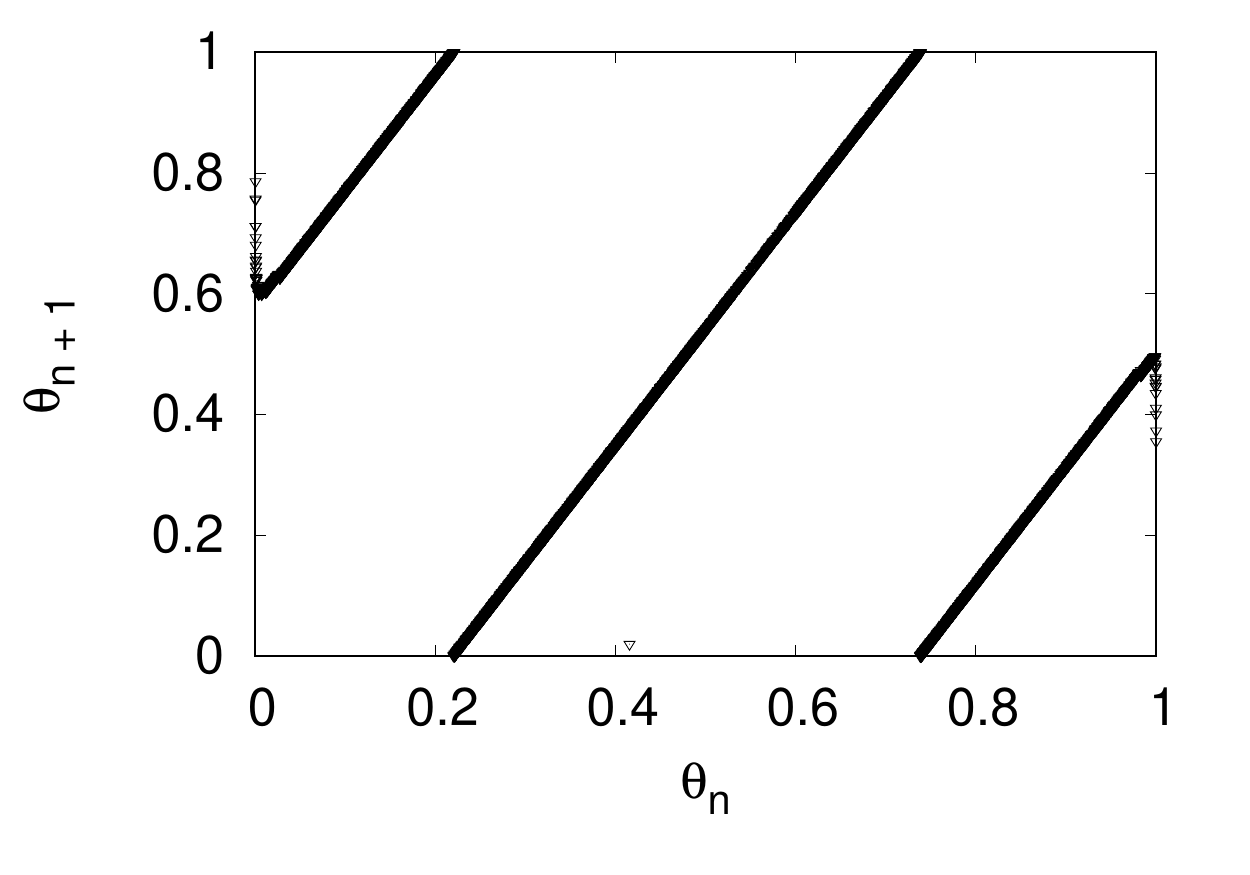}&
 \includegraphics[scale = 0.55]{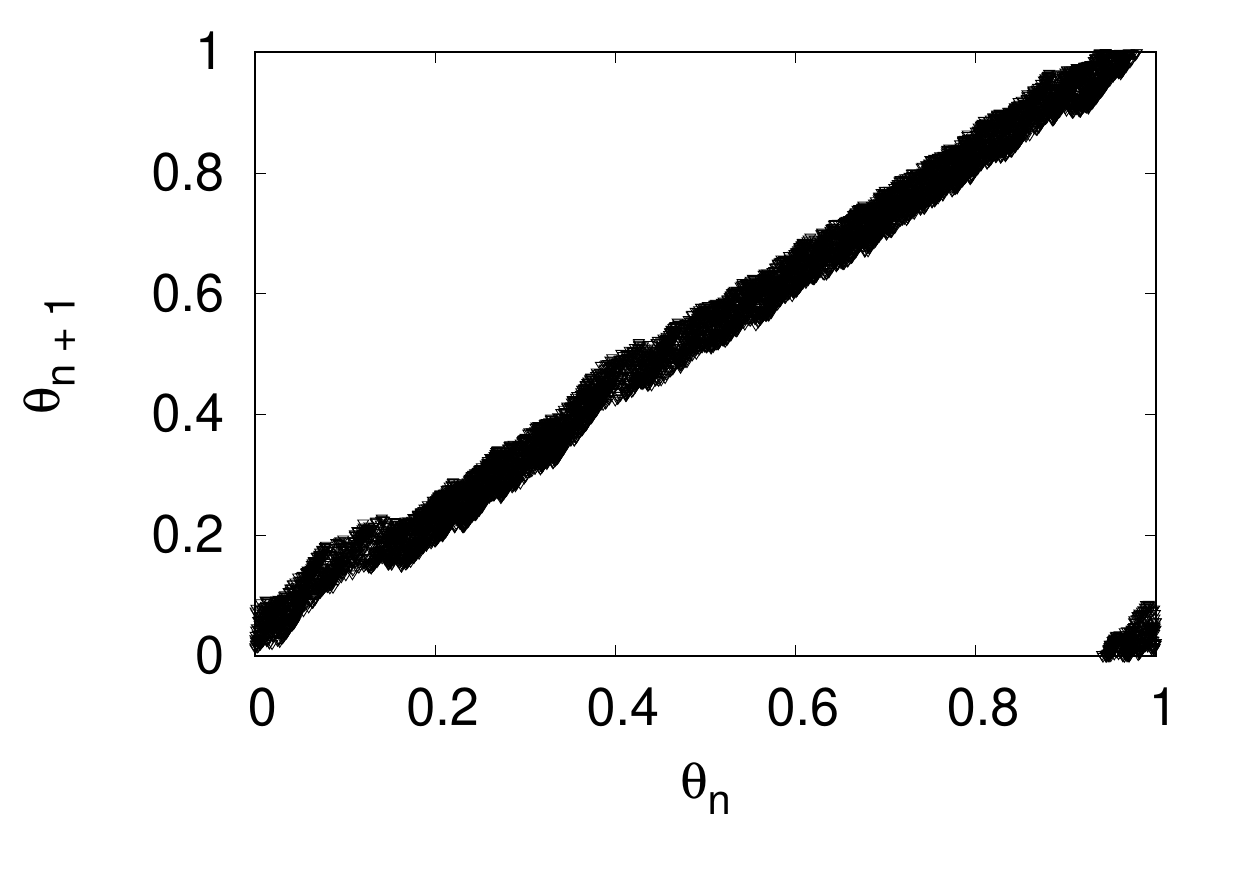}\\
 (a) & (b)
\end{tabular}
\caption{\label{fig: return} \footnotesize (color online) The return map of site 10 from (a) group one and (b) group two when the CML is already evolved in to the chimera state at parameters $K = 10^{-5}, \Omega = 0.27, \epsilon_1 = 0.93, N = 150$.}
\end{figure}

\subsection{Identifying the laminar and burst sites}\label{laminar_burst} We consider any two sites $(i, j)$ as laminar sites when the phases of the circle maps at these sites are such that the quantity $\Delta_{ij} = \left|\frac{1}{2}\left| \exp (2\pi i \theta^{\sigma}(i)) + \exp(2\pi i \theta^{\sigma'}(j))\right| - 1\right|$ is less than an assigned cutoff value set by the parameter $\delta$. The quantity $\Delta_{ij}$, which can also be considered as a two site order parameter (compare with the definition of $R^1, R^2, R$ of the group-wise and global order parameter given in Eq. \ref{global}), is used instead of directly computing the phase difference because $\Delta_{ij}$ takes into account the fact that equation \ref{sinecml} has a modulo one operation. It is necessary to take account of  the global coupling topology to identify the laminar and burst sites. 

\begin{figure}[H]
\centering \begin{tabular}{cc}
\includegraphics[scale = 0.278]{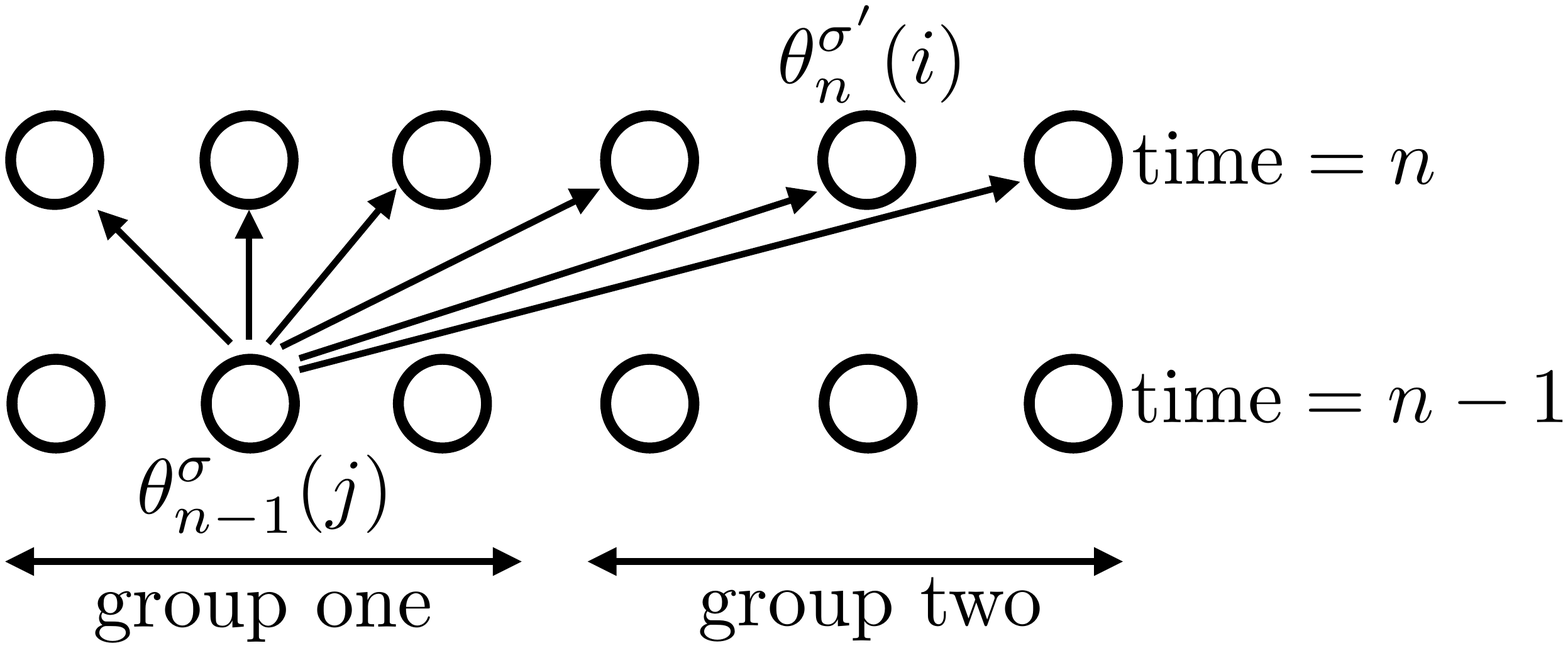}&
\includegraphics[scale = 0.278]{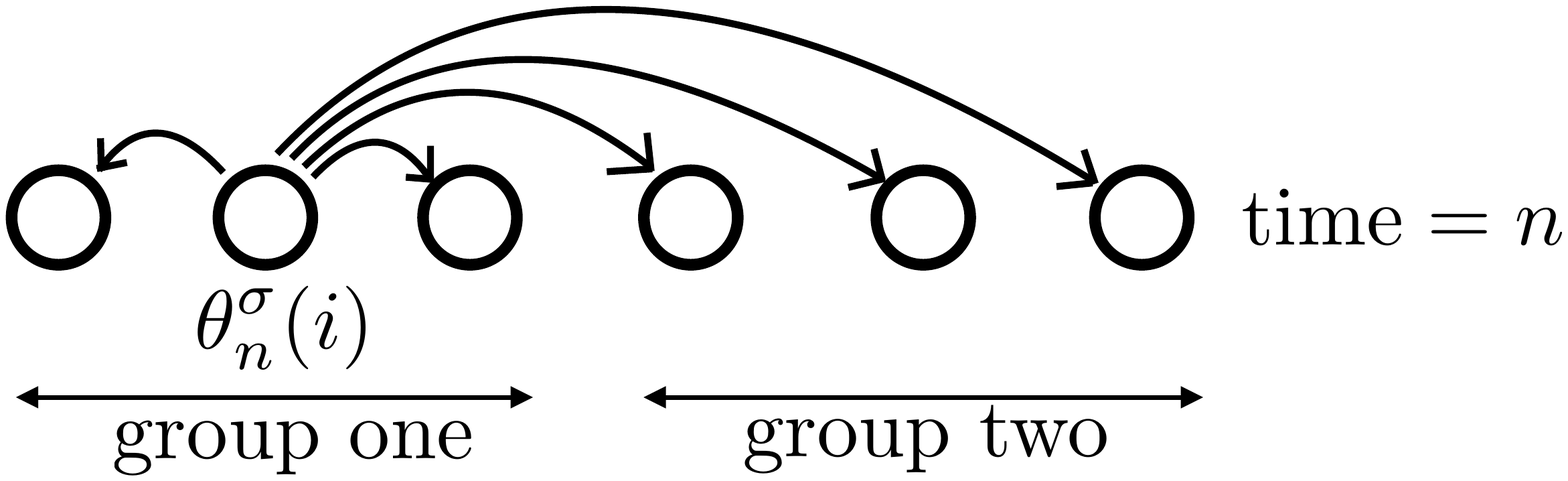}\\
(a) & (b)\\
\end{tabular}
\caption{\label{fig: infec}\footnotesize (color online) A schematic of the method for identifying the laminar and burst sites. Each of these arrows in the above diagrams indicate the pair of sites between which the condition $\Delta_{ij}$ is checked} 
\end{figure}

We identify the laminar and burst sites in the spatiotemporal variation of the phases of the CML in two steps which we describe here :
\begin{enumerate}
\item We consider the phases of the CML at two consecutive time steps, $n$ and $n - 1$. The phase of the map at site $i$ in group $\sigma$ at time step $n$, denoted as $\theta_{n}^{\sigma}(i)$. We choose two sites each from time steps $n - 1$ and $n$ that can belong to any of the groups and they are denoted by $\theta_{n - 1}^{\sigma}(j)$ and $\theta_{n}^{\sigma'}(i)$. We now check if $\Delta_{ij} < \delta$ for all $i = 1, 2, \cdots, N$ for both $\sigma' = 1, 2$ and label those lattice sites as laminar, if the corresponding phase, $\theta_{n}^{\sigma'}(i)$ satisfies the condition. We also label the lattice site at $\theta_{n - 1}^{\sigma}(j)$ as laminar if at least one such $i$ is found for which $\Delta_{ij} < \delta$ (see diagram \ref{fig: infec}.(a) for reference). We repeat this method for $j = 1, 2, \cdots, N$ for $\sigma = 1, 2$. We thus check if there is any temporal infection between the sites at time step $n - 1$ and time step $n$. Once the laminar sites at time step $n$ are identified by this method we check if there is any spatial infection between sites. We describe this in next step.

\item Now, we calculate $\Delta_{ij} = \left|\frac{1}{2}\left| \exp(i2\pi \theta_{n}^{\sigma'}(j)) + \exp(i2\pi \theta_{n}^{\sigma}(i))\right| - 1\right|$ for all $j = 1, 2, \cdots, N$ when $\sigma'  \neq \sigma$ and for $j = 1, 2, \cdots, N$ except $j \neq i$ when $\sigma' = \sigma$ and we check the condition $\Delta_{ij} < \delta$. A simple schematic is shown in Fig. \ref{fig: infec}.(b) for clarification. We label $\theta_{n}^{\sigma}(i)$ as a laminar site at time step $n$ if the condition is satisfied at least once. 
\end{enumerate}

After checking the phases of the maps at all sites at time step $n$ for temporal and spatial infections for laminarity in a similar fashion, we move on to the phases of the maps in the next time step. This identification is used to construct an equivalent cellular automaton. 

\section{Construction of the equivalent cellular automaton}
\label{CA}
We mentioned earlier that equivalent cellular automatons have been constructed for CMLs with nearest neighbour coupling in \cite{chate1988,bohr2003, zahera2006}. A cellular automaton having the range of nonlocal coupling as a parameter was shown to support chimera states recently \cite{garcia2016}. In this section we now proceed to construct an equivalent cellular automaton which mimics the dynamics of the laminar and burst sites as defined in the previous section during the evolution of the coupled map lattice in Eq. \ref{sinecml} after the system stabilises into an attractor within the numerical accuracy. We define the CA on a lattice of size $2N$ consisting of two groups with $N$ sites in each of them. The evolution equation of the CML (equation \ref{sinecml}) tells us that the global coupling terms in the equation appear in the form of a summation of the individual contribution of each of the maps multiplied by the appropriate coupling strength depending on the groups that the maps belong to. We infer from this form of coupling that the probability that a site $i$ behaves like a laminar or burst site at time step $n$, depends on the total number of laminar and burst sites in each of the groups rather than on their positions in the system. Hence the conditional probabilities which are required to construct the equivalent cellular automaton depend on the total number of sites in each of the groups, where the state variables take the value one and zero.

\begin{figure}[H]
\centering \begin{tabular}{cc}
 \includegraphics[width = 9.5cm, height = 7.5cm]{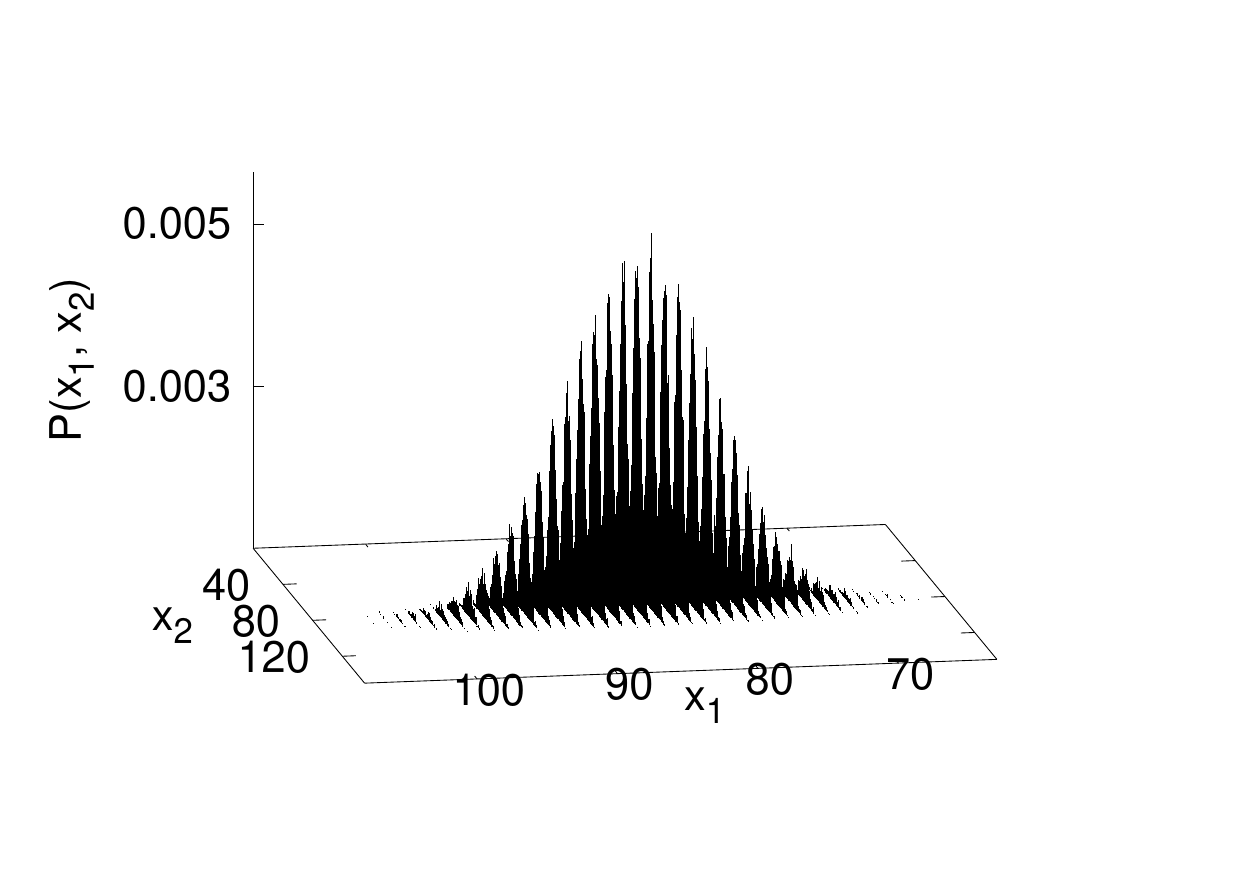}&
 \includegraphics[width = 8cm, height = 7.5cm]{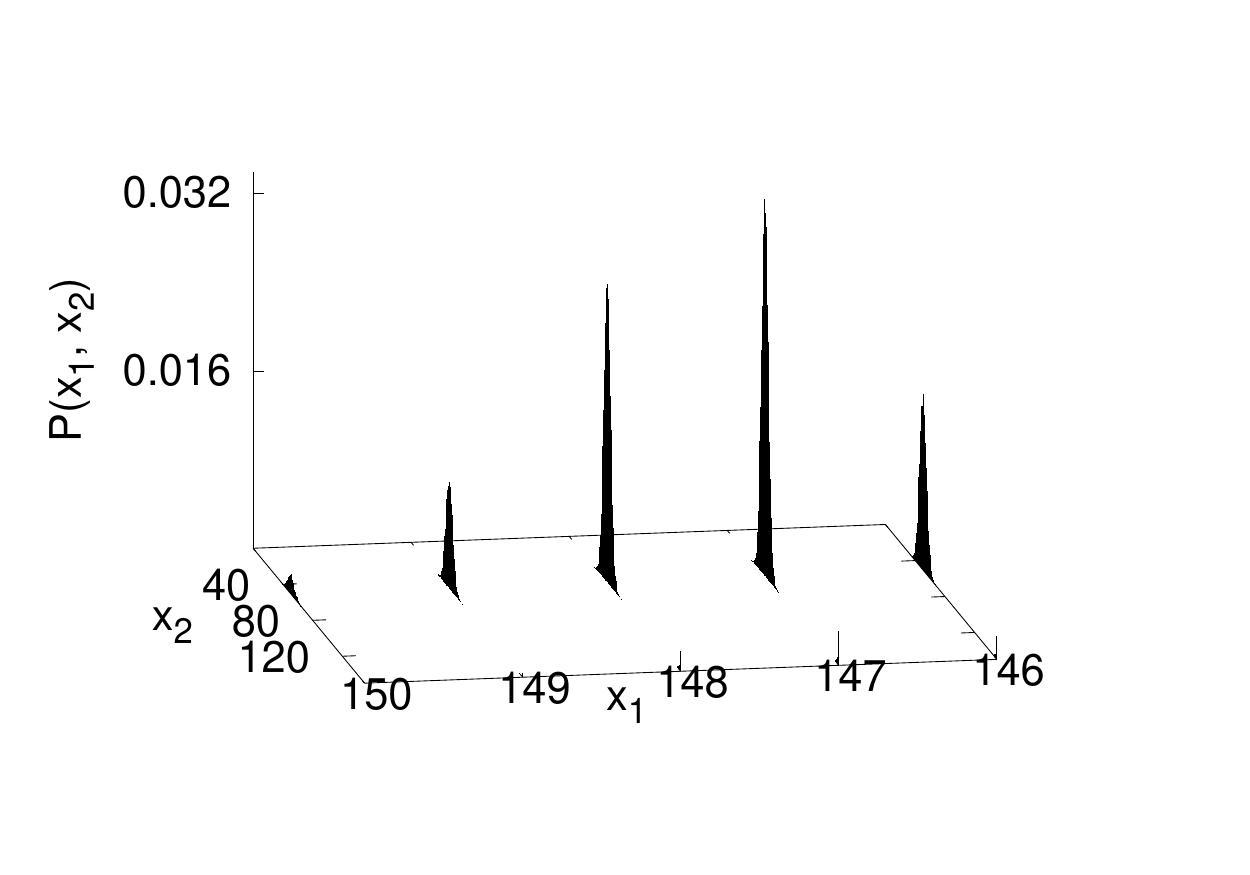}\\
  (a) & (b)\\
 \end{tabular}
\caption{\label{fig: prob_ones}\footnotesize (color online) The probability $P(x_1, x_2)$ is calculated for the laminar and burst sites for (a) fully phase desynchronised state (case 3), (b) chimera phase state with defects in the synchronised group (case 2). In (a) we see that $P(x_1, x_2)$ is nonzero approximately in the range $60 \leq x_2 \leq 110$, $70 \leq x_1 \leq 100$ while in (b) the probabilities are nonzero for $40 \leq x_2 \leq 60$ for each values of $x_1 = 146, 147, 148, 149, 150$.}
\end{figure}

\begin{figure}[H]
 \centering \begin{tabular}{cc}
 \includegraphics[width = 9.5cm, height = 7.5cm]{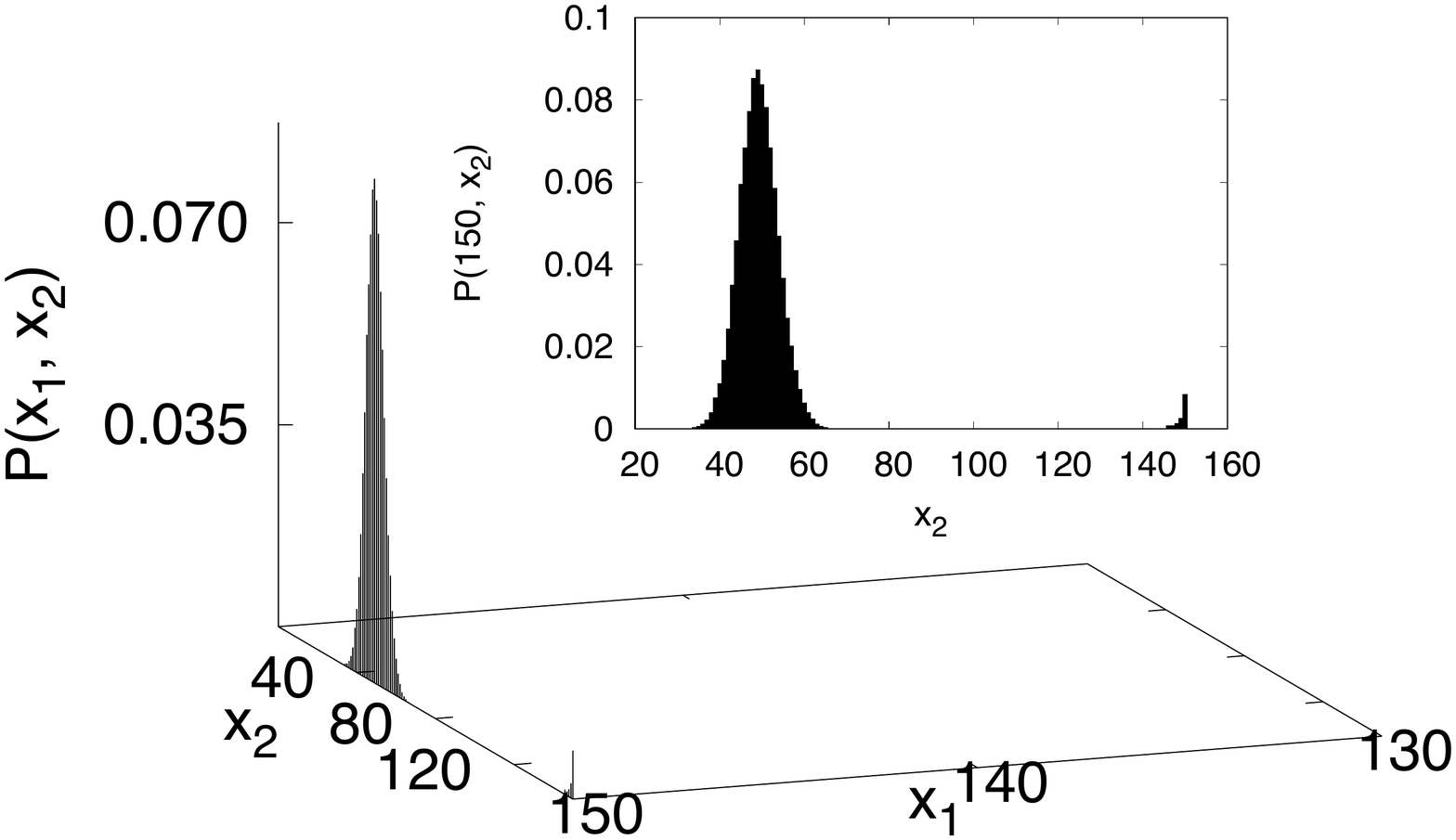}&
 \includegraphics[width = 6.5cm, height = 6cm]{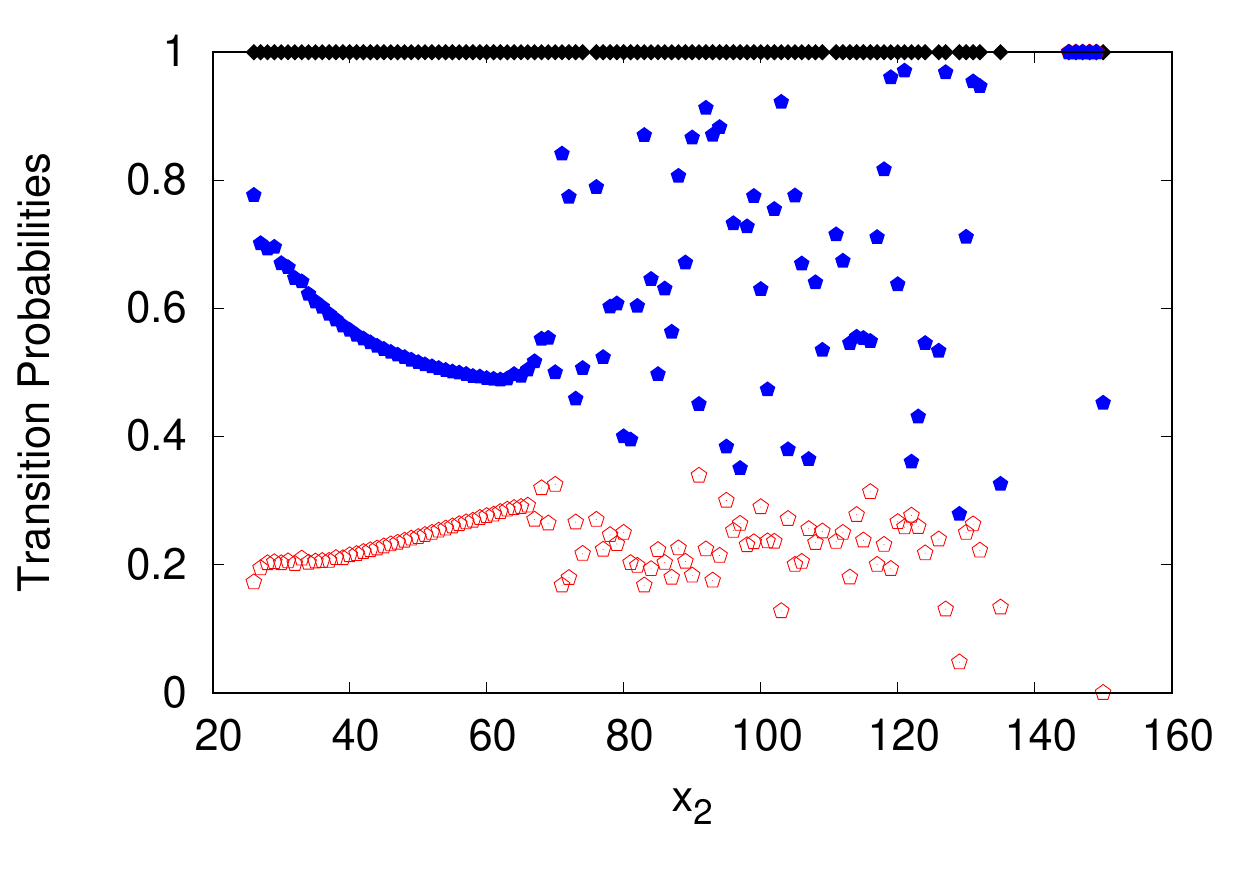}\\
 (a) & (b)\\
 \end{tabular}
\caption{\label{fig: prob_chim_pure} \footnotesize (color online) (a) The probability $P(x_1, x_2)$ is calculated for the laminar and burst sites for the dynamics of chimera state with purely phase synchronised part (case 1). Hence,  $x_1$ is always 150 for a purely synchronised group in the chimera state but $x_2$ can be between 35 and 65.(b) Transition probabilities $P^{x_1, x_2}(1|1)$ (black points) for the purely phase synchronised group, $P^{x_1, x_2}(1|0)$ (red points) and $p^{x_1, x_2}(1|1)$ (blue points) for the phase desynchronised group. The parameters in the CML are $K = 10^{-5}, \Omega = 0.27, \epsilon_1 = 0.82, N = 150$.}
\end{figure}

Let us assign a state variable $s_{n}^{\sigma}(i)$ to each site $i$ in group $\sigma$ at time step $n$ and let $s_{n}^{\sigma}(i)$ take the value $0$ for a burst site otherwise it is assigned the value $1$ for a laminar site. Let $x_1$ and $x_2$ denote the total number of laminar sites which have $s_{n}^{\sigma}(i) = 1$ in groups one and two respectively at any time step $n$. The total number of possible combinations of $x_1, x_2$ are $N^{2}$.  We calculate the probability $P(x_1, x_2)$ of the occurrence of a given combination of $x_1, x_2$ at any time step for the configurations observed here, viz the fully phase desynchronised state, the chimera state with phase synchronised group with defects and the chimera state with phase synchronised group without defects. Figure \ref{fig: prob_ones}.(a) shows the probabilities $P(x_1, x_2)$ which are numerically calculated from the CML and plotted against the values of $x_1, x_2$ for the fully phase desynchronised state in case 3. It is clear from the plot that significant values of $P(x_1, x_2)$ occur in a small subset of the available space of $x_1$ and $x_2$, $60 \leq x_2 \leq 110$, $70 \leq x_1 \leq 100$. Such significant values of $P(x_1, x_2)$ occur even less frequently for the case of chimera states with defects in the phase synchronised group (case 2) (see Fig. \ref{fig: prob_ones}.(b)). Here the significant nonzero values lie between $40 \leq x_2 \leq 60$ for each values of $x_1 = 146, 147, 148, 149$ and $150$. For the case of the chimera state with pure synchronisation in group one (case 1), the only possible value of $x_1$ is 150 and $35 \leq x_2 \leq 65$ (Fig. \ref{fig: prob_chim_pure}.(a)).

We define the conditional probability for the dynamics of $s_{n}^{\sigma}(i)$ as, $P^{x_1, x_2}(s_{n + 1}^{\sigma}(i)|(s_{n}^{\sigma}(i))$ which is the transition probability that a lattice site $i$ chosen at random in group $\sigma$ at time step $n$ having value $s_{n}^{\sigma}(i)$ transforms to $s_{n + 1}^{\sigma}(i)$ at time step $n + 1$, given that there are $x_1$ and $x_2$ laminar sites in groups one and two respectively at time step $n$. So there exist four possibilities for each combination of $x_1$ and $x_2$, which are $P^{x_1, x_2}(0|0)$, $P^{x_1, x_2}(1|0)$, $P^{x_1, x_2}(0|1)$, $P^{x_1, x_2}(1|1)$ for the dynamics of sites in each group. It is clear from the definition of these transition probabilities, that at each possible combination of $x_1$ and $x_2$ allowed by the dynamics of the CML, they will always satisfy,

\begin{equation}
P^{x_1, x_2}(0|0)  + P^{x_1, x_2}(1|0) = P^{x_1, x_2}(0|1)  + P^{x_1, x_2}(1|1) = 1
\end{equation}

We calculate $P^{x_1, x_2}(1|0)$ and $P^{x_1, x_2}(1|1)$ numerically from the CML configurations. These are calculated separately  for all the groups, i.e.  for the chimera state with a purely phase synchronised group (case 1), chimera states with synchronised group having defects (case 2), and fully phase desynchronised states (case 3).  For case 1, i.e.  the chimera state with a purely synchronised group one, all of the sites in the synchronised group are spatially laminar at every time step. Hence $P^{x_1, x_2}(1|1)$ for the purely phase synchronised group of maps turns out to be one (see Fig. \ref{fig: prob_chim_pure}(b)), while $P^{x_1, x_2}(1|0)$ cannot be calculated, as a site with initial value zero cannot be found. We also find that $P^{x_1, x_2}(1|1)$ is greater than $P^{x_1, x_2}(1|0)$ for the phase desynchronised group (Fig. \ref{fig: prob_chim_pure} (b)) in this case. 
\par
\begin{figure}[H]
\centering \begin{tabular}{cc}
  \includegraphics[scale = 0.58]{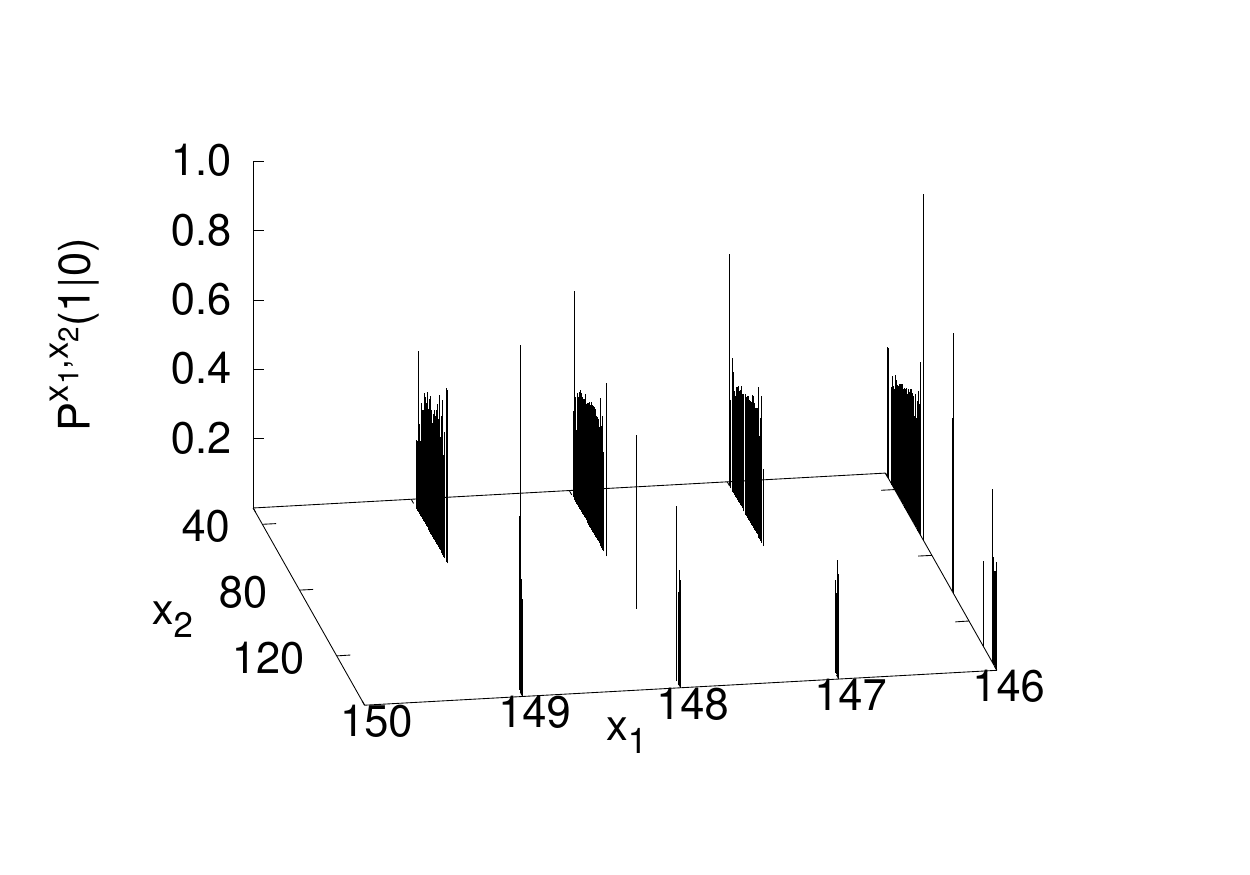}&
  \includegraphics[scale = 0.58]{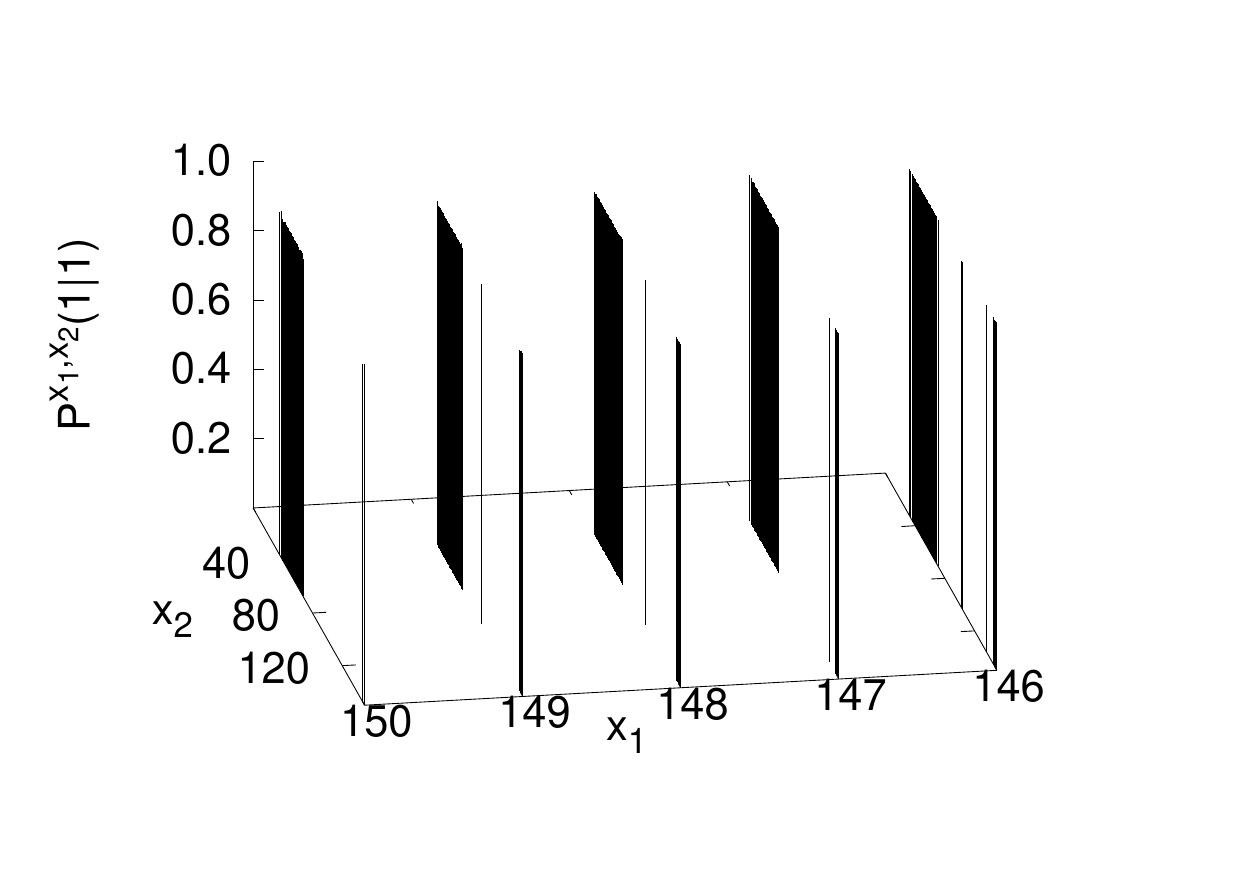}\\
  (a) & (b)\\
  \includegraphics[scale = 0.58]{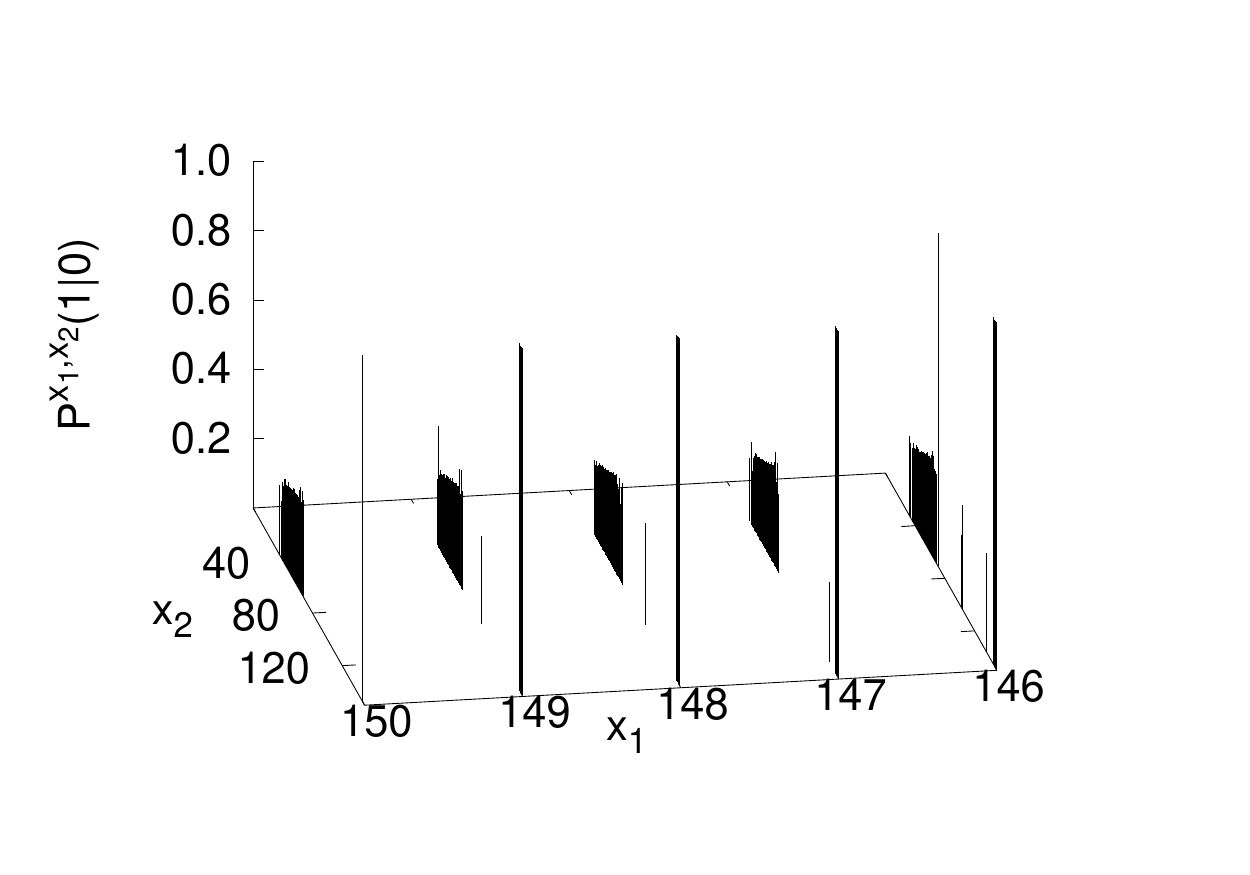}&
  \includegraphics[scale = 0.58]{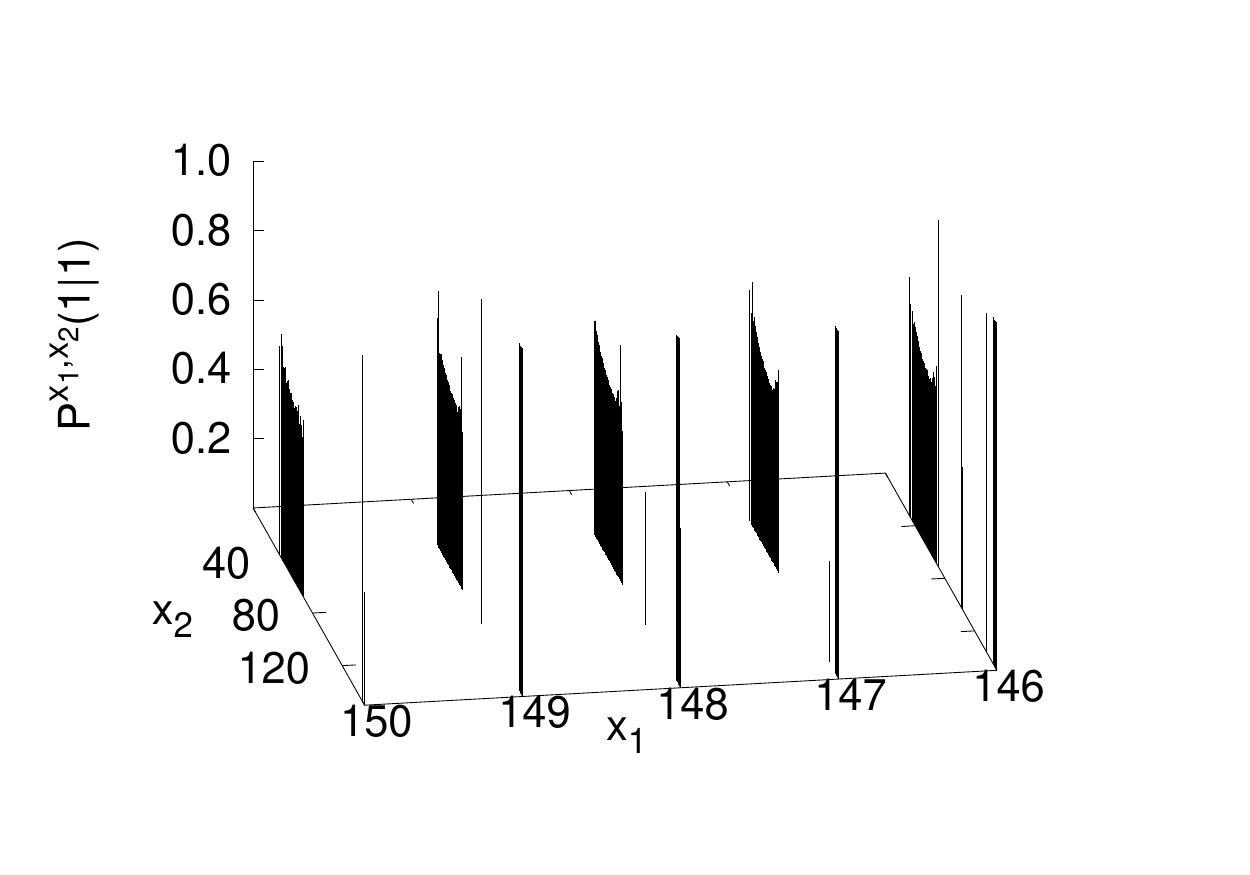}\\
  (c) & (d)\\
 \end{tabular}
\caption{\label{fig: prob_chim_defect} \footnotesize (color online) The transition probabilities are calculated for the laminar and burst sites for the chimera state with a phase desynchronised group of maps and phase synchronised group of maps with defects. The quantities which are plotted here are (a) $P^{x_1, x_2}(1|0)$ and (b) $P^{x_1, x_2}(1|1)$ for group one and (c) $P^{x_1, x_2}(1|0)$, (d) $P^{x_1, x_2}(1|1)$ for group two. It can be seen that $P^{x_1, x_2}(1|1)$ have higher values than $P^{x_1, x_2}(1|0)$ in the case of both the groups. However the values of $P^{x_1, x_2}(1|1)$ is higher in the case of group one compared to group two. The parameters are $K = 10^{-5}, \Omega = 0.27, \epsilon_1 = 0.93, N = 150$.}
\end{figure}

\begin{figure}[H]
\centering \begin{tabular}{cc}
 \includegraphics[width = 7cm, height = 7cm]{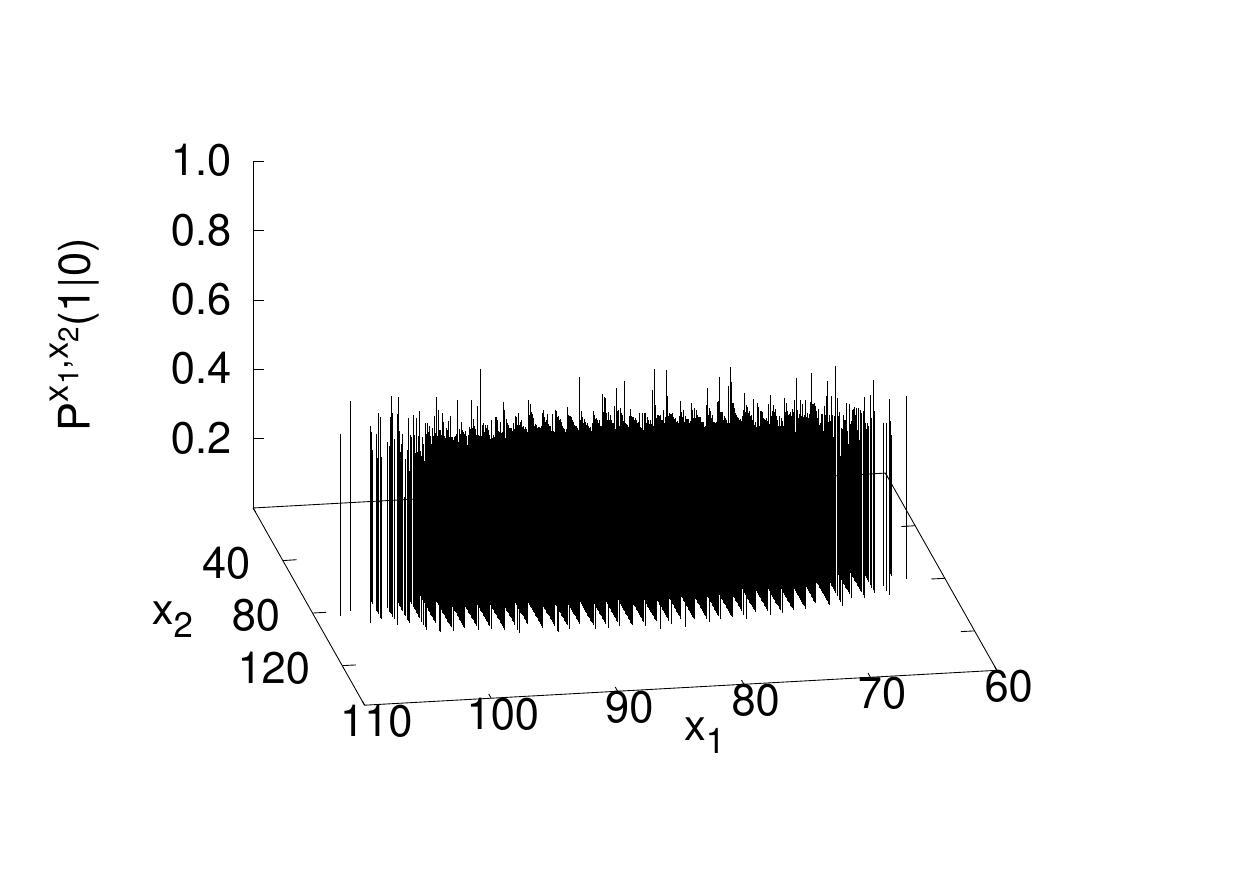}&
 \includegraphics[width = 7cm, height = 7cm]{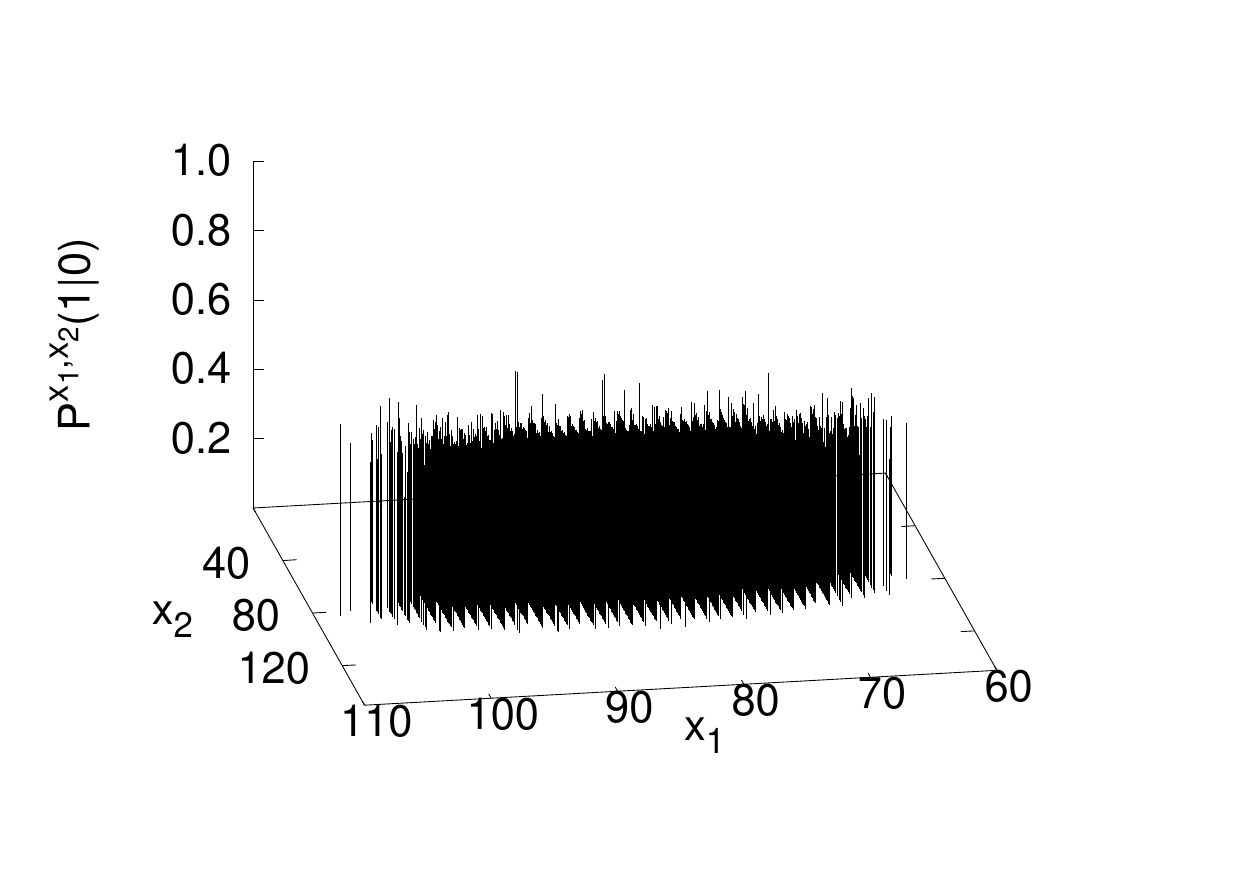}\\
 (a) & (b)\\
 \end{tabular}
 \centering \begin{tabular}{c}
  \includegraphics[width = 7cm, height = 5.5cm]{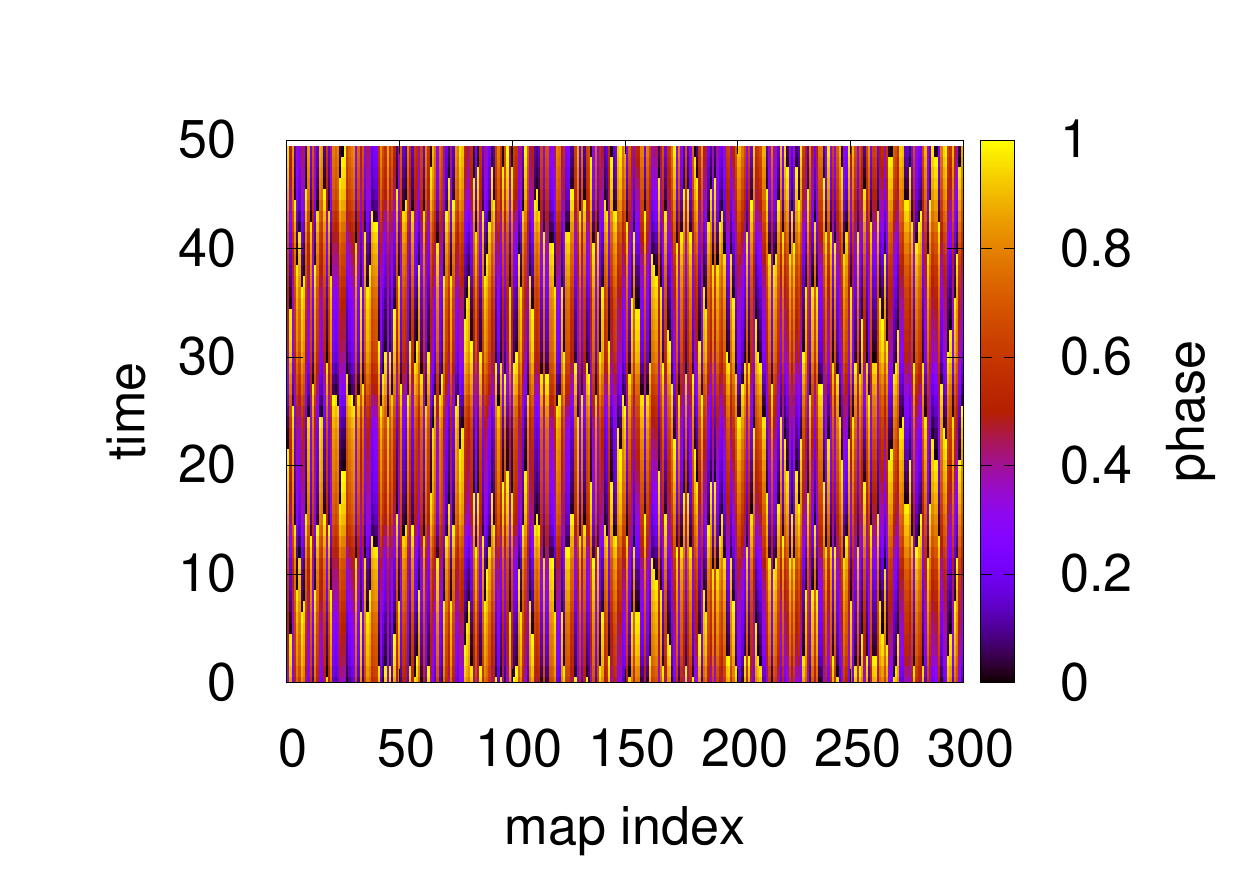}\\
  (c)\\
  \end{tabular}
  \caption{\label{fig: prob_full_desync} \footnotesize (color online) The transition probabilities are calculated for the fully phase desynchronised state shown in the space time plot in the figure. The system of Eq. \ref{sinecml} evolves from a random initial phase configuration for four million  iterates, and the subsequent two million iterations are used to calculate the conditional probabilities. Figure (a) and (b) plot  $P^{x_1, x_2}(1|0)$ for (a) group one and (b) group two respectively. It can be seen that this quantity has similar values for both the groups. Similarly $P^{x_1, x_2}(1|1)$ also has equal values in groups one and two (not shown here). (c) The space time plot of the fully phase desynchronised state is shown. It is evident that these values of transition probabilities show similar behavior due to the identical dynamics of the maps in  both the groups, as 
confirmed by the space time plot. The parameters used in the CML are $K = 10^{-5}, \Omega = 0.27, \epsilon_1 = 0.75, N = 150$. }
\end{figure}

 For the chimera state with defects in the phase synchronised group, i.e.  case 2, we observe that $P^{x_1, x_2}(1|1)$ is greater than $P^{x_1, x_2}(1|0)$ (see Figs. \ref{fig: prob_chim_defect} (a) and (b)). In particular $P^{x_1, x_2}(1|1) \approx 1$ for the significant values of $P(x_1, x_2)$ (see Fig. \ref{fig: prob_chim_defect} (b)). Both $P^{x_1, x_2}(1|1)$ and $P^{x_1, x_2}(1|0)$ are less than the conditional probabilities calculated for the synchronised group (Figs. \ref{fig: prob_chim_defect} (c) and (d)). We observe that $P^{x_1, x_2}(1|0)$ and $P^{x_1, x_2}(1|1)$ have comparable values for both the groups for which $P(x_1, x_2)$ is nonzero in the case of the fully phase desynchronised group in case 3 (see Fig. \ref{fig: prob_full_desync}). This is expected since in this case the dynamics of the maps from groups one and two are similar as compared to case 1 and case 2. In the next section we calculate the fraction of laminar sites using the values of $P(x_1, x_2)$ and $P^{x_1, x_2}(s_{n + 1}^{\sigma}(i)|(s_{n}^{\sigma}(i))$. 

\subsection{Calculation of fraction of laminar sites}\label{density}
 Let $m_{\sigma}$ be the fraction of sites in the CA where $s_{i}$ takes the value one during the evolution of the phases of the maps in CML. If the CML of equation \ref{sinecml} settles into one of the attractors, e.g. the fully phase desynchronised state, the chimera state with a purely phase synchronised subgroup or that with a phase synchronised subgroup having defects, then the corresponding fraction of laminar sites $m^{*}_{\sigma}$ for the group $\sigma$ should be calculable from the probabilities which we have obtained in the previous section. Once the system stabilises into an attractor, the numerically calculated value of $m^{*}_{\sigma}$ should also match the corresponding value calculated using the probabilities. In such a case $m^{*}$ also denotes the probability that the state variable $s_i$  at any randomly chosen site $i$ in any group at any time step has the value one. 
\par In the previous section we have calculated $P(x_1, x_2)$ which is the probability for a certain combination $x_1, x_2$ to occur. If the state variable $s_{n}^{\sigma}(i)$ at site $i$ in the group $\sigma$ at the time step $n$ has a value one, then either $s_{n - 1}^{\sigma}(i) = 0$ or $s_{n - 1}^{\sigma}(i) = 1$. The probabilities are given by $P^{x_1, x_2}(1|0)$ for the $0 \rightarrow 1$ transition and $P^{x_1, x_2}(1|1)$ for the $1 \rightarrow 1$ transition, as defined previously and have been calculated from the CML(see Figs. \ref{fig: prob_full_desync}, \ref{fig: prob_chim_defect} and \ref{fig: prob_chim_pure} (b)). If there are $x_1$ laminar sites in group one, at time step $n$, then it is easy to see that the probability that $s^{1}_{n}(i) = 1$ in group one is given by $x_1/N$ and that for sites having value zero is $1 - x_1/N$. The same argument can be given for the sites in group two with a total of $x_2$  laminar sites. Also we note that if all the sites in a group are laminar sites at time step $n$, then the probability of finding a laminar site is one at that time step. For example, when the system settles into a chimera state with purely phase synchronised group then the probability for a site in the synchronised group, to be one at any time step is exactly one which we have already seen in the previous section. Based on these arguments we can write $m^{*}_{\sigma}$ for the group $\sigma$ as, 

\begin{equation}
\begin{split}
 m^{*}_{\sigma} = \sum\limits_{x_\sigma^{'} = 0}^{N} \Bigg[ P(0, x_{\sigma^{'}})P^{0, x_{\sigma^{'}}}(1|0) + P(N, x_{\sigma^{'}})P^{N, x_{\sigma^{'}}}(1|1)
&+ \sum\limits_{x_\sigma = 1}^{N - 1}\bigg( P(x_{\sigma}, x_{\sigma^{'}})P^{x_\sigma, x_{\sigma^{'}}}(1|0)\frac{N - x_\sigma}{N}\\
&+ P(x_{\sigma}, x_{\sigma^{'}})P^{x_\sigma, x_{\sigma^{'}}}(1|1)\frac{x_{\sigma}}{N} \bigg) \Bigg]
 \label{den}
 \end{split}
 \end{equation}
 
It is easy to see that a similar equation can be written for $1 - m^{*}_{\sigma}$ which is the fraction of burst sites. We calculate $m^{*}_{\sigma}$ for fully phase desynchronised state, chimera states with defects in the phase synchronised group and the chimera with the purely phase synchronised group in Table \ref{table_1}. These values match with the fraction of laminar sites when they are directly calculated from the variation of the phases when the CML settles into different attractors. 

\begin{table}[ht]
\caption{We use the parameter values $\Omega = 0.27$, $N = 150$ and $K = 10^{-5}$ for each of the phase configurations below.}
\centering
%\begin{tabular}{ | m{5.5cm} | m{1.5cm}| m{1.5cm} | m{1.5cm} | m{1.5cm} | m{1.5cm} | m{1.5cm} |}
\begin{tabular}{c c c c c c c}
\hline\hline
\centering Attractor &$m^{*}_{1}$&$m^{*}_{2}$ &$m^{*}_{1}$ from CML& $m^{*}_{2}$ from CML\\ 
\hline
\centering Fully phase desynchronised state ($\epsilon_1 = 0.75$)& 0.565 & 0.544 & 0.565 & 0.544 \\ 
\centering Chimera state with defects in synchronised group ($\epsilon_1 = 0.93$) & 0.989 & 0.347 & 0.982 & 0.344 \\ 
\centering Chimera state with purely synchronised group ($\epsilon_1 = 0.82$) & 1.00 & 0.340 & 1.0 & 0.337 \\ 
\hline
\end{tabular}
\label{table_1}
\end{table}
In the next section we show that we can also find the density of the laminar sites from a mean field approximation of the cellular automata. 

\subsubsection*{\textbf{A mean field equation of the CA model}}\label{mf}
In this section we follow the prescription given by Mikkelsen et.al. in \cite{bohr2003} to calculate fraction of laminar sites for any attractor of the CML. We obtain a mean field equation for the cellular automata for the fraction of laminar sites in each of the groups when the coupled map lattice settles into any one of the attracting states (i.e. cases 1 - 3 as defined earlier). We calculate the probabilities, $P(x_1, x_2)$, $P^{x_1, x_2}(s_{n+ 1}(i)^{\sigma}|s_{n}^{\sigma}(i))$ when the dynamics of the system becomes stationary in any one of the attractors. By our definition the transition probability $P^{x_1, x_2}(s_{n+ 1}(i)^{\sigma}|s_{n}^{\sigma}(i))$ is identical irrespective of the choice of $i$ at a time step $n$ and can be considered as a mean field which is same at all sites in the group $\sigma$ at that time step. This also implies that we have two mean fields for the CA for each of the values of $\sigma'$. Now let us assume that $m_{\sigma}(t)$ be an arbitrary initial value of the fraction of laminar sites for the given stationary attractor dynamics. A linear difference equation in terms of the mean fields or the transition probabilities and the fraction of laminar sites, $m_{\sigma}(t)$, can be obtained using the same arguments as before (see Eq. \ref{den}). Finally we write the mean field equation as, 

\begin{equation}
\begin{split}
 m_{\sigma}(t + 1) = \sum\limits_{x_\sigma^{'} = 0}^{N} \Bigg[ P(0, x_{\sigma^{'}})&P^{0, x_{\sigma^{'}}}(1|0) + 
 P(N, x_{\sigma^{'}})P^{N, x_{\sigma^{'}}}(1|1)\\
&+ \sum\limits_{x_\sigma = 1}^{N - 1}\bigg( P(x_{\sigma}, x_{\sigma^{'}})P^{x_\sigma, x_{\sigma^{'}}}(1|0)(1 - m_{\sigma}(t))\\
&+ P(x_{\sigma}, x_{\sigma^{'}})P^{x_\sigma, x_{\sigma^{'}}}(1|1)m_{\sigma}(t) \bigg) \Bigg]\\
\end{split}
\label{mf_equation}
\end{equation}

\begin{figure}[H]
\centering \begin{tabular}{ccc}
\hspace{-0.2cm}
\includegraphics[width = 5.5cm, height = 4.5cm]{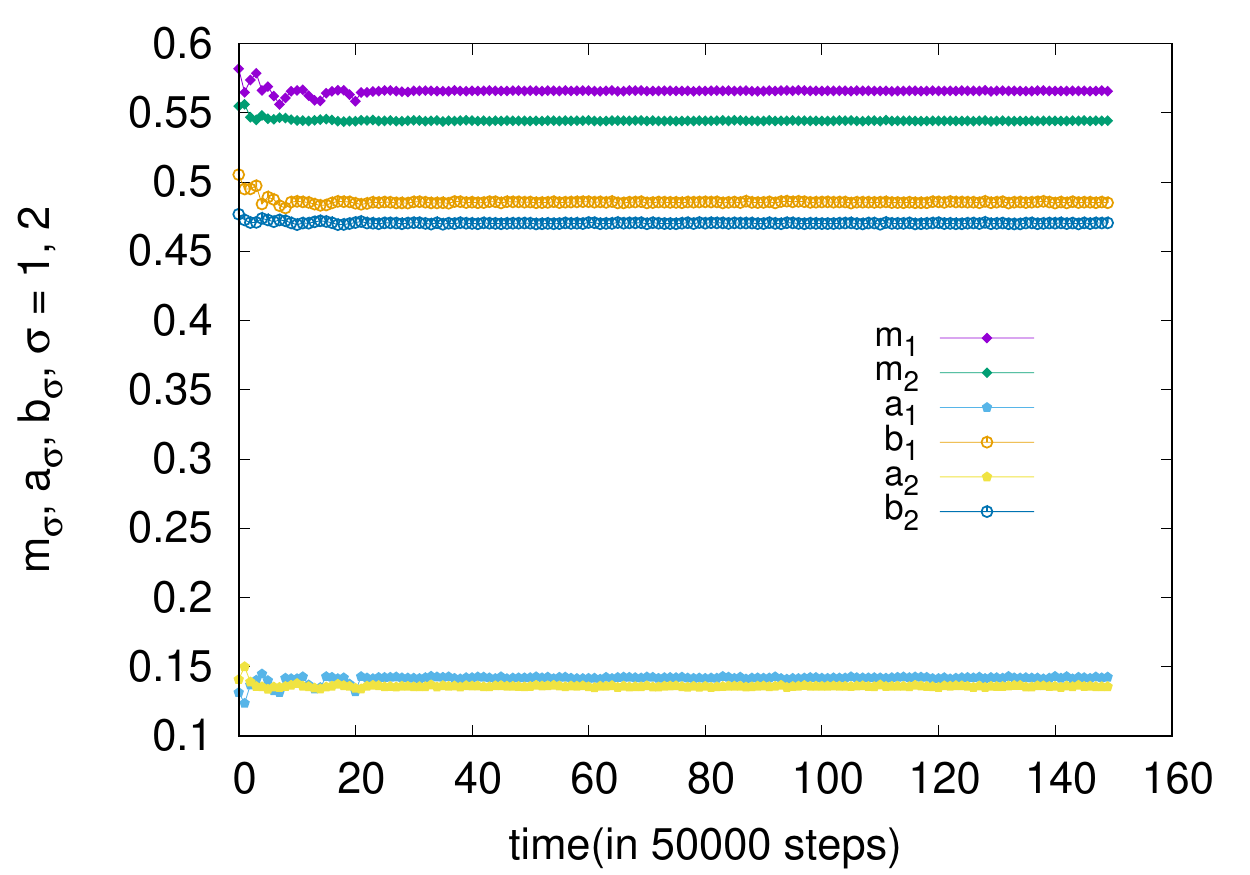}&
\includegraphics[width = 5.5cm, height = 4.5cm]{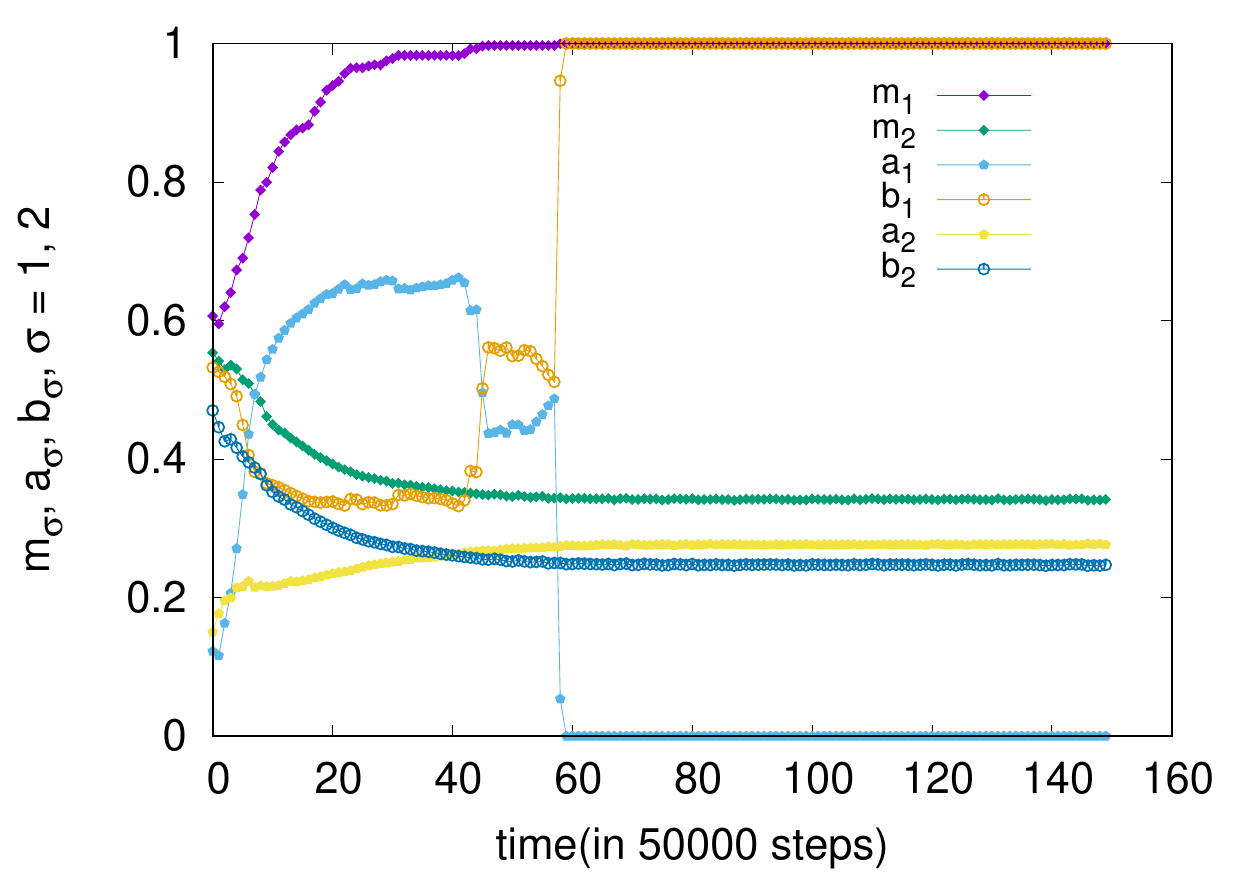}&
\includegraphics[width = 5.5cm, height = 4.5cm]{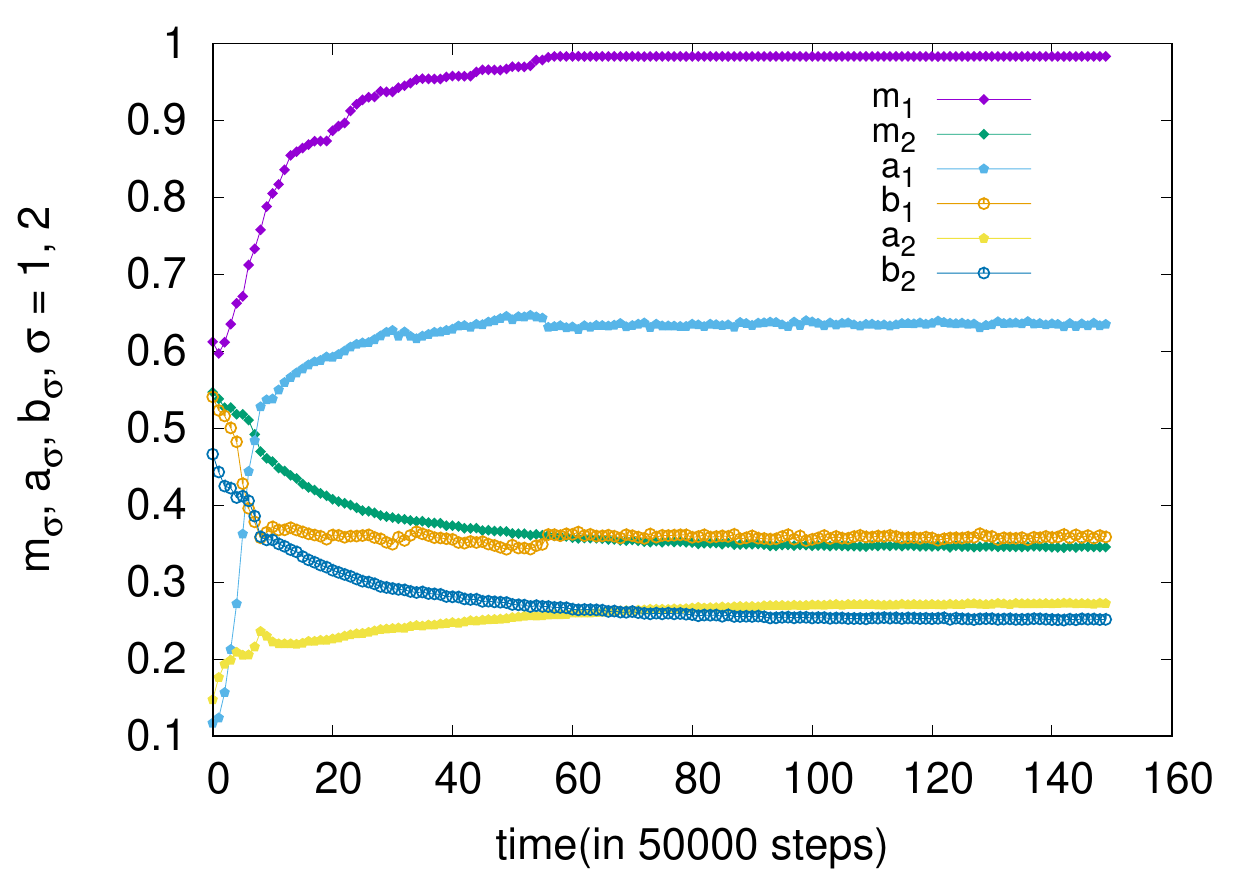}\\
(a) & (b) & (c)\\
\end{tabular}
\caption{\label{fig: m_a_b}\footnotesize (color online) The variation of $m_{\sigma}, a_{\sigma}$ and $b_{\sigma}$ for (a) fully phase desynchronised state $(\epsilon_1 = 0.75)$ (b) chimera state with purely synchronised subgroup $(\epsilon_1 = 0.82)$ and (c) chimera state with defects in the synchronised subgroup $(\epsilon_1 = 0.93)$. Other parameters are fixed at $K = 10^{-5}, \Omega = 0.27, N = 150$. Starting from the initial condition the conditional probabilities are calculated for the variation of the phases at the lattice site in CML for every 50000 time steps.}
\end{figure}

This is a linear equation of the form $m_{\sigma}(t + 1) = f(m_{\sigma}(t)) = a_{\sigma} m_{\sigma}(t) + b_{\sigma}$ where, 
\begin{equation}
\begin{split}
a_{\sigma} &= \sum\limits_{x_{\sigma'} = 0}^{N} \sum\limits_{x_\sigma = 1}^{N - 1} \bigg( P(x_{\sigma}, x_{\sigma{'}})P^{x_\sigma, x_{\sigma{'}}}(1|1) - 
P(x_{\sigma}, x_{\sigma{'}})P^{x_\sigma, x_{\sigma{'}}}(1|0) \bigg)\\
b_{\sigma} &= \sum\limits_{x_{\sigma'} = 0}^{N} \Bigg[ P(0, x_{\sigma{'}}P^{0, x_{\sigma{'}}}(1|0) +
P(N, x_{\sigma{'}})P^{N, x_{\sigma{'}}}(1|1) + \sum\limits_{x_\sigma = 1}^{N - 1} P(x_{\sigma}, x_{\sigma{'}})P^{x_\sigma, x_{\sigma{'}}}(1|0) \Bigg]\\
\end{split}
\label{ab}
\end{equation}
When the system settles into one of the attractors, the quantities $a_{\sigma}$ and $b_{\sigma}$ also settle to fixed values since the transition probabilities which are calculated from the variation of the phases of the CML as they settle to steady values. We calculate the transition probabilities from the CML and calculate $a_\sigma, b_\sigma$ using Eq. \ref{ab} and thereby find $m_\sigma$. Fig. \ref{fig: m_a_b} shows the transient in the evolution of $a_\sigma, b_\sigma \text m_\sigma$ and the steady values are shown in table \ref{table_2}.

The fixed points of the equation \ref{mf_equation} in the $m_1, m_2$ space are given by, 
\begin{equation}
 \begin{split}
  \widetilde{m}_1 &= \frac{b_1}{1 - a_1}\\ 
  \widetilde{m}_2 &= \frac{b_2}{1 - a_2}\\
 \end{split}
 \label{fixed_point}
\end{equation}

Since the fixed points in the $m_1, m_2$ space must lie in the interval [0:1], we find from the fixed point equation  (Eqs. \ref{fixed_point}) that $a_{\sigma}$ and $b_{\sigma}$ must satisfy the conditions, $a_1 + b_1 \leq 1$, $a_2 + b_2 \leq 1,  a_1,a_2 \neq 1$ and $b_1, b_2 \geq 0$. The conditions are satisfied by the final steady value of the $a_1, a_2, b_1, b_2$ values shown in the Table \ref{table_2}

\begin{figure}[H]
  \begin{tabular}{ccc}
  \hspace{-.2cm}
 \includegraphics[scale = 0.45]{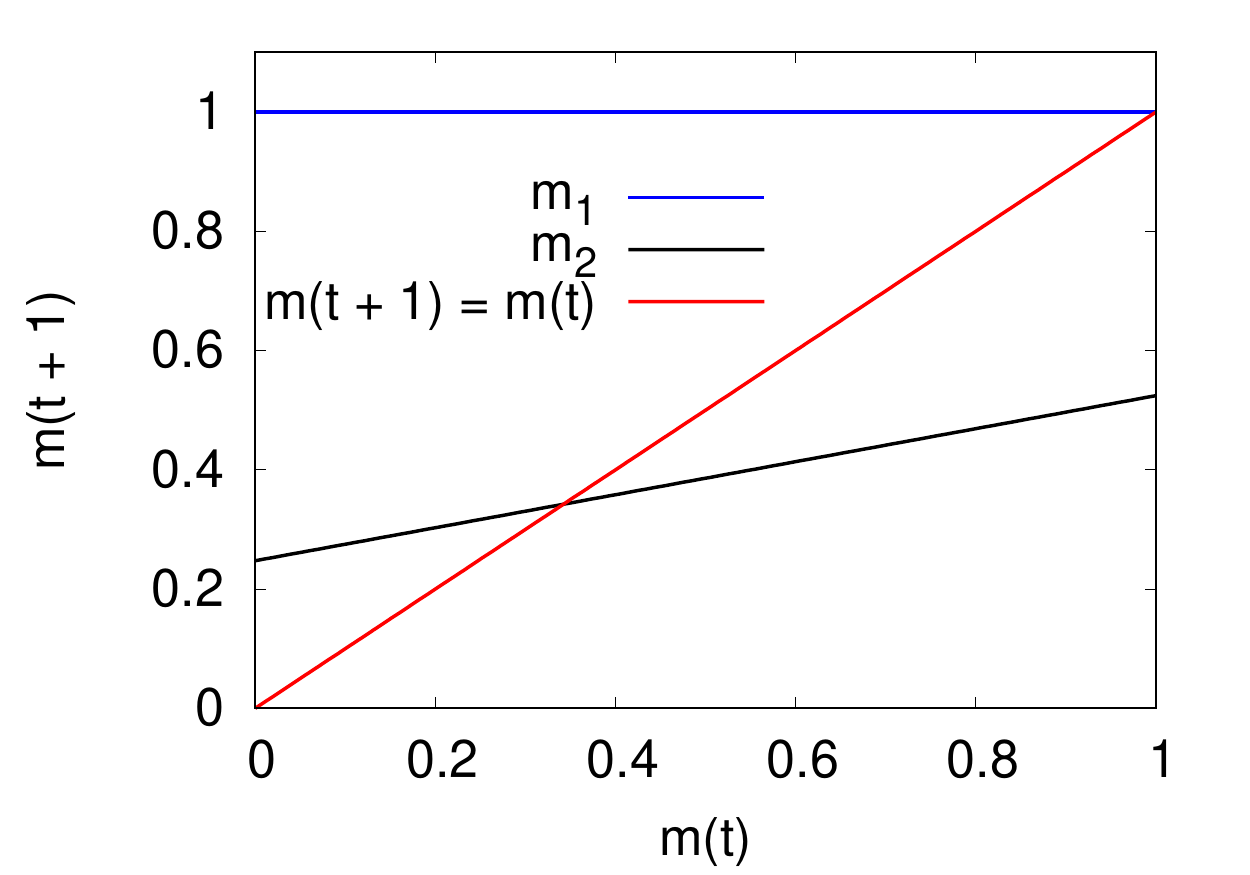}&
 \hspace{-.5cm}
 \includegraphics[scale = 0.45]{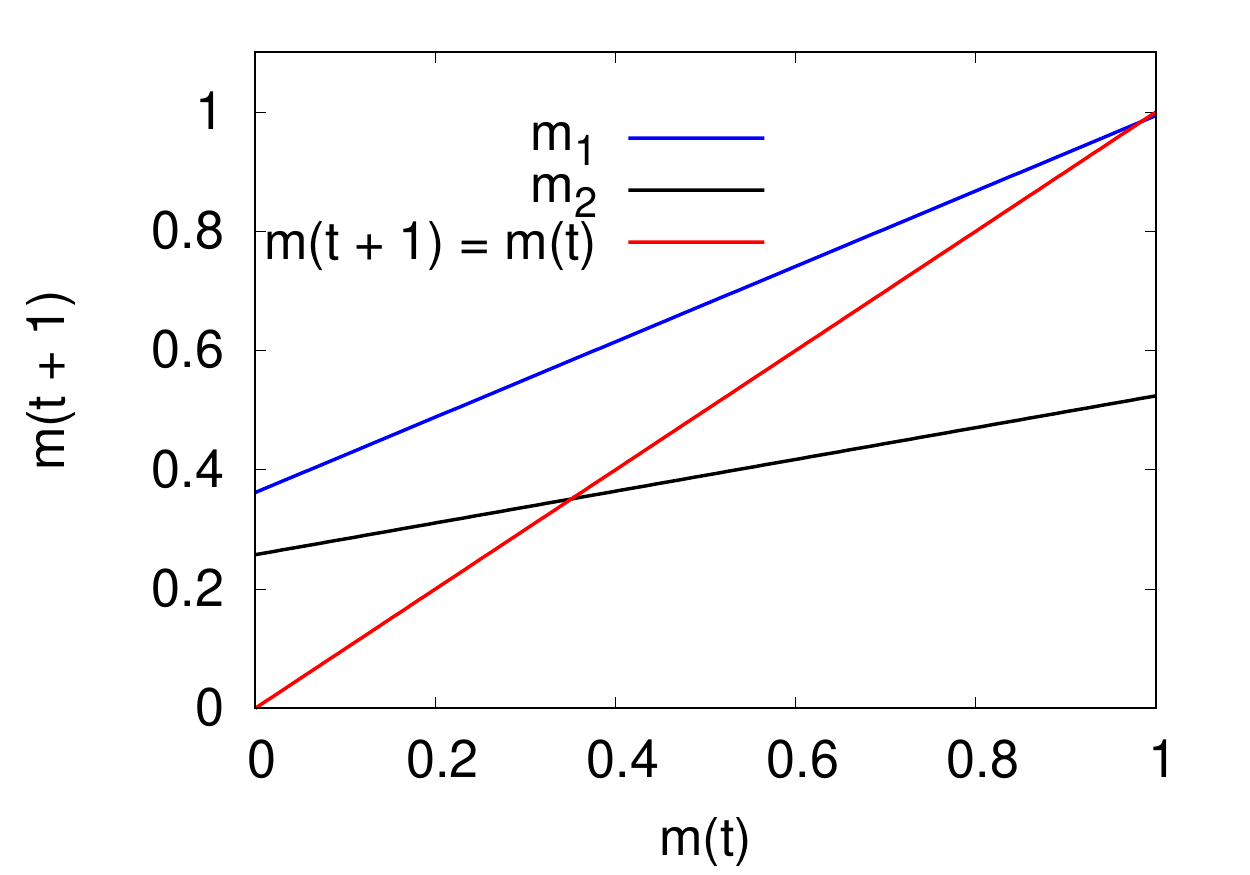}&
 \hspace{-.5cm}
 \includegraphics[scale = 0.45]{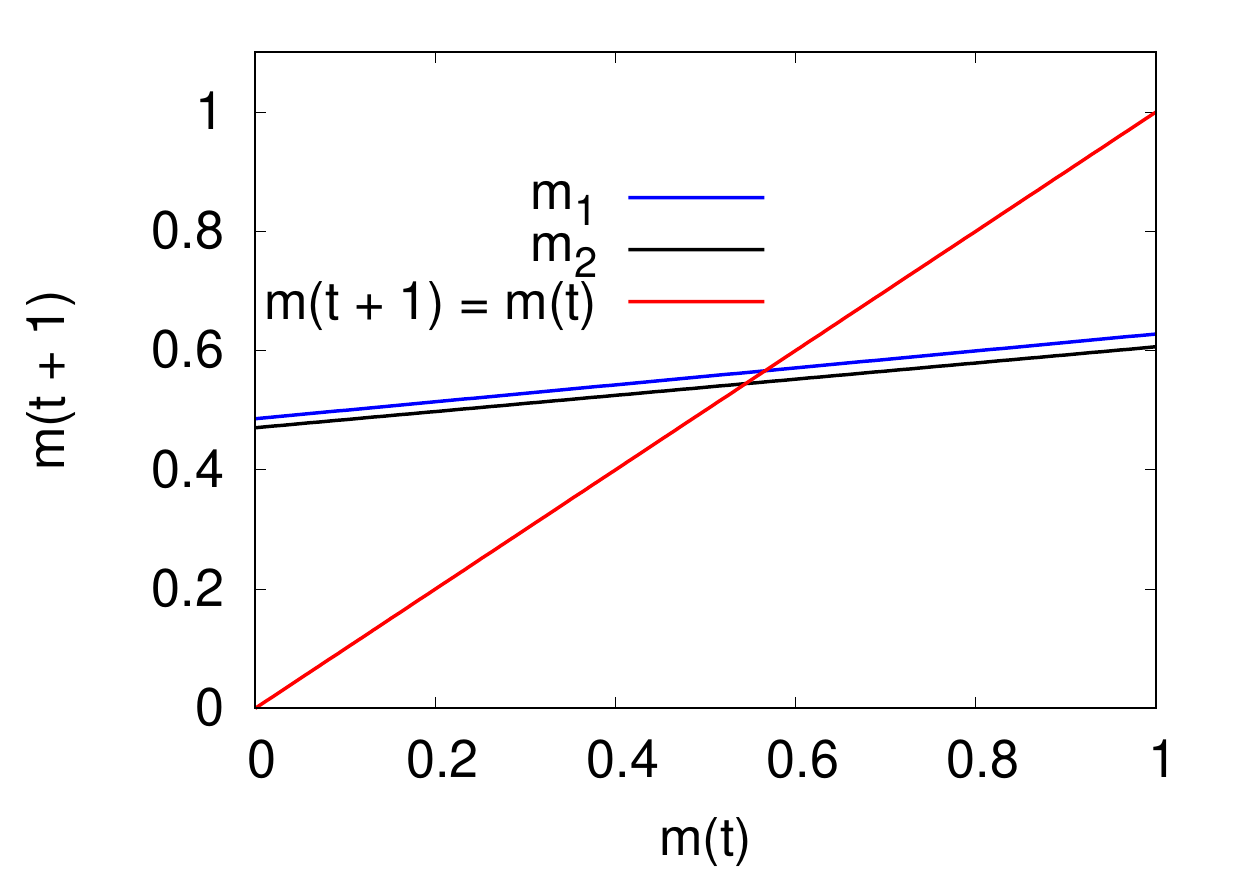}\\
 (a) & (b) & (c)\\
\hspace{-.2cm}
 \includegraphics[scale = 0.45]{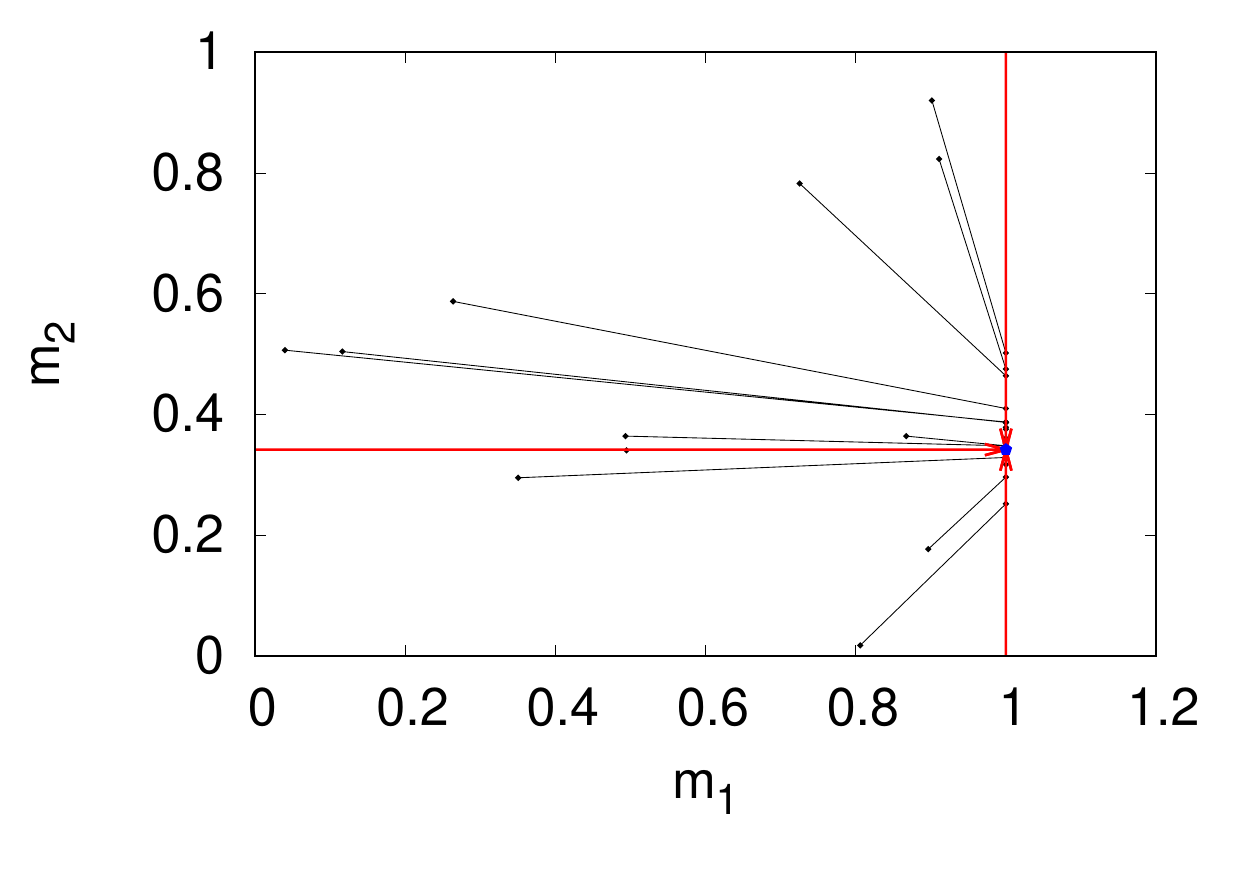}&
 \hspace{-.5cm}
 \includegraphics[scale = 0.45]{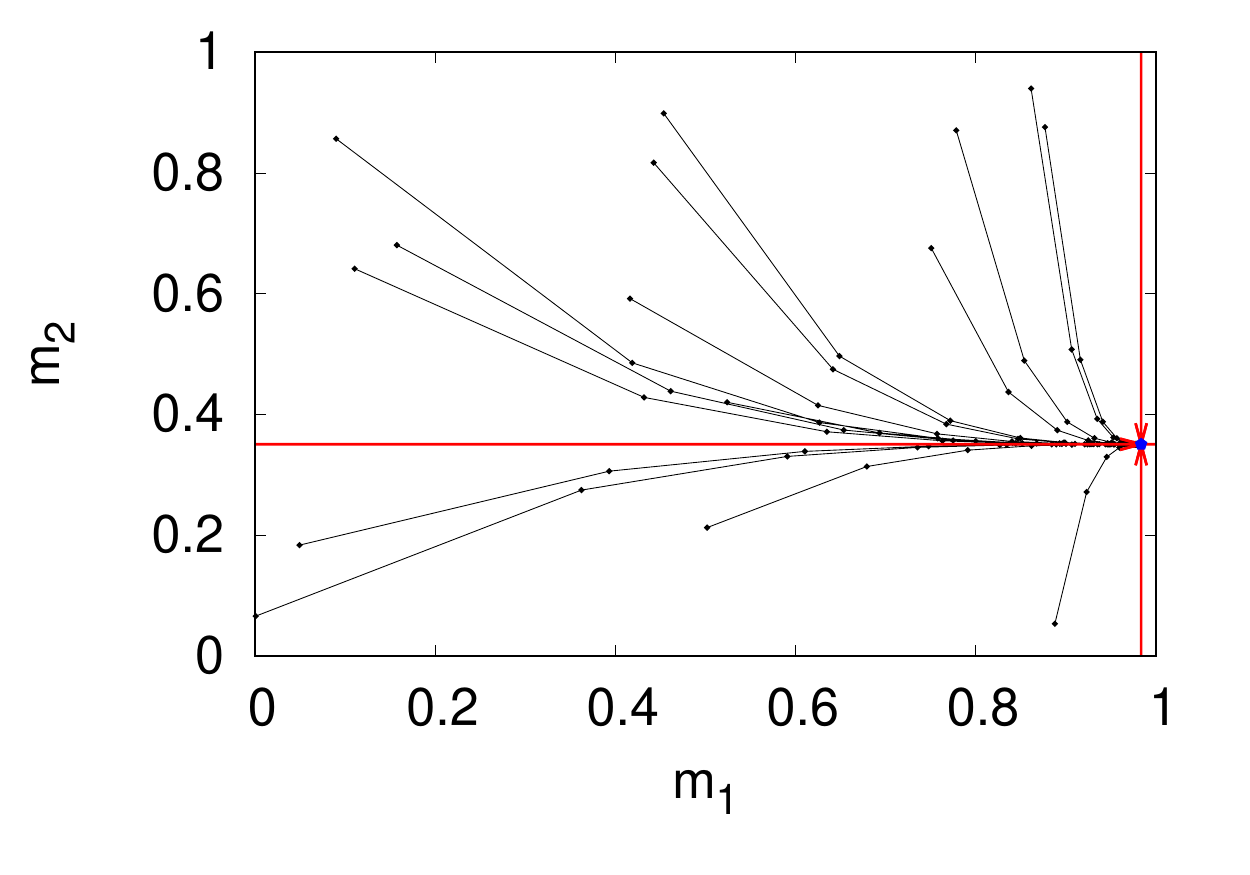}&
 \hspace{-.5cm}
 \includegraphics[scale = 0.45]{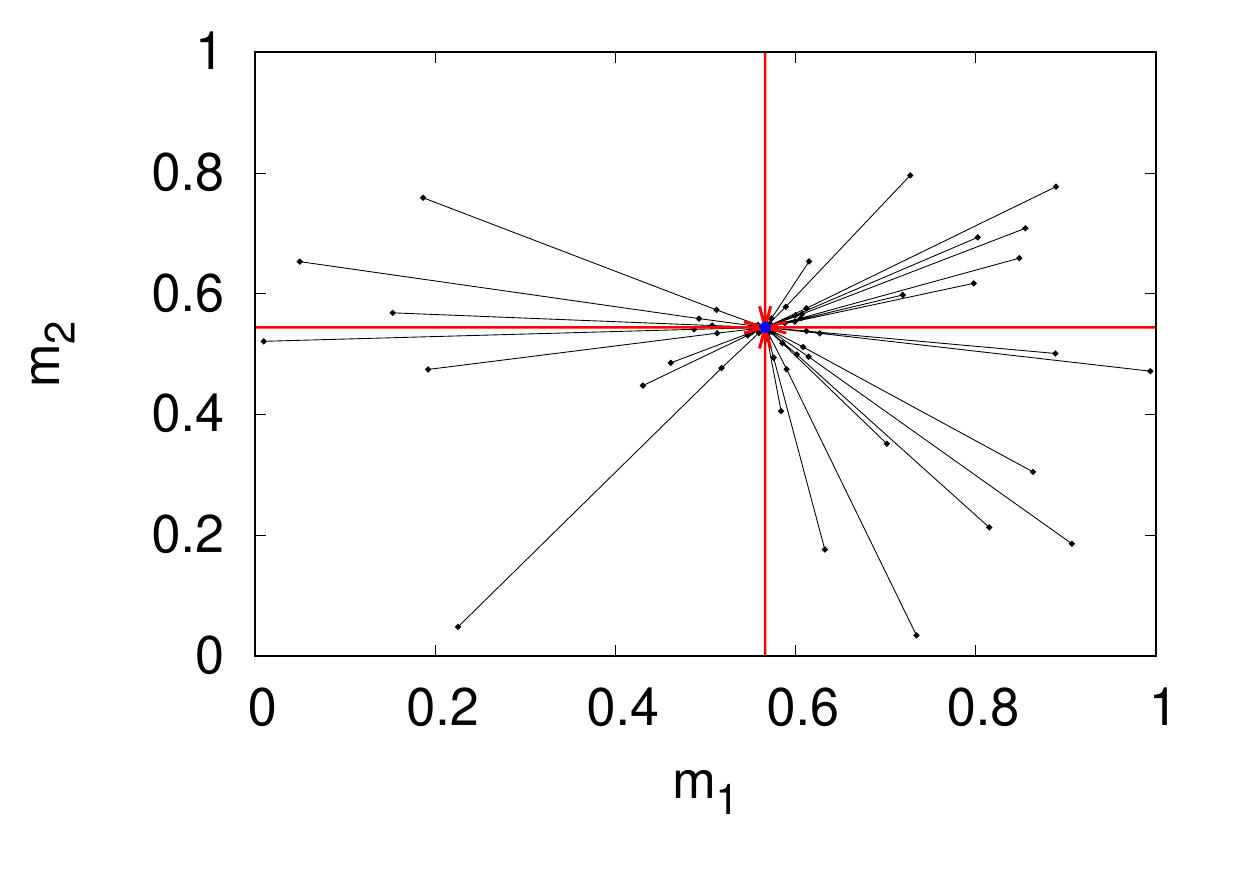}\\
 (d) & (e) & (f)\\
\end{tabular}
\caption{\label{fig: m_1_m_2} \footnotesize (color online) Return maps of the mean field equation for (a)Chimera state with purely phase synchronised subgroup $\epsilon_1 = 0.82$ (b) Chimera state with defects in the phase synchronised group $\epsilon_1 = 0.93$ with $K = 10^{-5}, \Omega = 0.27, N = 150$ (c) fully phase desynchronised state ($\epsilon_1 = 0.75$). We use random initial points between zero and one. The evolution of $m_1$ and $m_2$ are shown in Fig. \ref{fig: m_1_m_2}.(d), (e) and (f). We use the values of $a_{\sigma}$ and $b_{\sigma}$ from table \ref{table_2} and iterate them for (d)chimera with purely synchronised subgroup, (e) chimera with defects in the synchronised subgroup and (f) fully phase desynchronised state. In each of these three figures the consecutive points in each of the trajectories are joined by a line. The phase trajectories are shown in black and fixed points are shown in blue in each. The eigenvectors are shown in red arrows. }
\end{figure}

\begin{table}[ht]
\caption{\footnotesize{We use the parameter values $\Omega =0.27, K = 10^{-5}, N = 150$ to obtain $a_\sigma$, $b_\sigma$ and $m_\sigma$ for each attracting state.}}
\begin{tabular}{c c c c c c c}
\hline\hline
\centering Attractor &$a_{1}$&$b_{1}$&$a_{2}$&$b_{2}$&$\widetilde{m}_1$&$\widetilde{m}_2$\\ 
\hline
 Fully phase desynchronised state $(\epsilon_1 = 0.75)$ & 0.1421 & 0.4856 & 0.1357 & 0.4705 & 0.5659 & 0.5443\\ 
 Chimera state with defects in synchronised group $(\epsilon_1 = 0.93)$ & 0.6321 & 0.3618 & 0.2664 & 0.2575 & 0.9834 & 0.3511 \\ 
 Chimera state with purely synchronised group $(\epsilon_1 = 0.82)$ & 0.0 & 1.0 & 0.2768 & 0.2474 & 1.0 & 0.3422 \\ 
\hline
\end{tabular}
\label{table_2}
\end{table}
The Jacobian for the set of equations given by Eq.\ref{mf_equation} is written as, 
\begin{equation}
 J = \begin{bmatrix}
      a_1 & 0\\
      0 & a_2\\
\end{bmatrix}_{\widetilde{m}_1, \widetilde{m}_2}
\label{jacobian}
\end{equation}
We find from table \ref{table_2} that $\widetilde{m}_1$ and $\widetilde{m}_2$ are stable fixed points with eigenvalues $\lambda_1 = a_1$ and $\lambda_2 = a_2$ having the eigenvectors $m_2 = \widetilde{m}_2$ and $m_1 = \widetilde{m}_1$ respectively for the Jacobian in Eq. \ref{jacobian}. These fixed points are globally stable in the $m_1, m_2$ space as it is clear from the linear form of the equation \ref{mf_equation} for $m_{\sigma}(t)$ and the corresponding return maps (see Figs. \ref{fig: m_1_m_2}.(a), (b), (c)) for each specific values of $a_1$ and $b_1$ for the three types of attractors mentioned in table \ref{table_2}. We also see in Figs. \ref{fig: m_1_m_2}(d), (e) and (f) that randomly chosen initial conditions in the $m_1$ and $m_2$ space converge to fixed points, corresponding to each of the attractors. In the case of the chimera state with pure synchronisation in group one (case 1) we find that $a_1 = 0$, $0 < a_2 < 1$. Accordingly Fig. \ref{fig: m_1_m_2} (d) also shows that all randomly chosen initial conditions evolve rapidly along the eigenvector $m_2 = \widetilde{m}_2 $ and then converges towards the fixed point along the eigenvector $m_1 = \widetilde{m}_1$. This point in the $\widetilde{m}_1, \widetilde{m}_2$ space becomes unstable when chimera state with defects in the synchronised groups are seen in CML. In this case we find $a_1 > a_2$ for the chimera phase state with defects in the synchronised group (case 2). As a result all the trajectories evolves at a faster rate along the eigenvector $m_1 = \widetilde{m}_1$ as compared to the other eigenvector, towards the stable fixed point (see Fig. \ref{fig: m_1_m_2}.(e)). We find from Table \ref{table_2} that the eigenvalues of the Jacobian \ref{jacobian} have approximately equal values $a_1 \approx a_2$ for the fully phase desynchronised state (case 3). Hence all the trajectories have nearly identical rate of convergence to the fixed point which can be verified from the phase space trajectories in Fig. \ref{fig: m_1_m_2}.(f). In this range of parameters where globally desynchronised states are seen, the $\widetilde{m}_1, \widetilde{m}_2$ values for both type chimera states become unstable and all trajectories converge to the attractor corresponding to the fully desynchronised state  in the $\widetilde{m}_1, \widetilde{m}_2$ space.

\section{\label{transition}Transition from the chimera state and reconstruction of the phase diagram}
We have seen previously that the global order parameter and group-wise order parameters, $(R_n, R^{1}_{n}, R^{2}_{n})$, are indicators that can differentiate between fully phase synchronised configurations, partially phase synchronised configurations (e.g. chimera states) and fully phase desynchronised configurations. The phase diagram in Fig. \ref{fig: order_zoom} as well as the cross section taken at $\log_{10}K = -5.5$ (see Fig. \ref{fig: density_1}.a) show that at $\epsilon_1 \approx 0.8$ there is a transition from the fully desynchronised state to chimera states as $\epsilon_1$ increases to one. The signatures of these transitions can also be seen in the variation of the quantity $\widetilde{m}_\sigma$ which we have defined in the previous section. We now calculate $P^{x_1, x_2}(s_{n + 1}^{\sigma}(i)|(s_{n}^{\sigma}(i))$, $P(x_1, x_2)$ use them to find out $\widetilde{m}_{\sigma}$ using equation \ref{den} as $\epsilon_1$ varies between $0.65$ and one with $K = 10^{-5}, \Omega = 0.27$. 
\par
\begin{figure}[H]
 \centering \begin{tabular}{cc}
 \includegraphics[scale = 0.55]{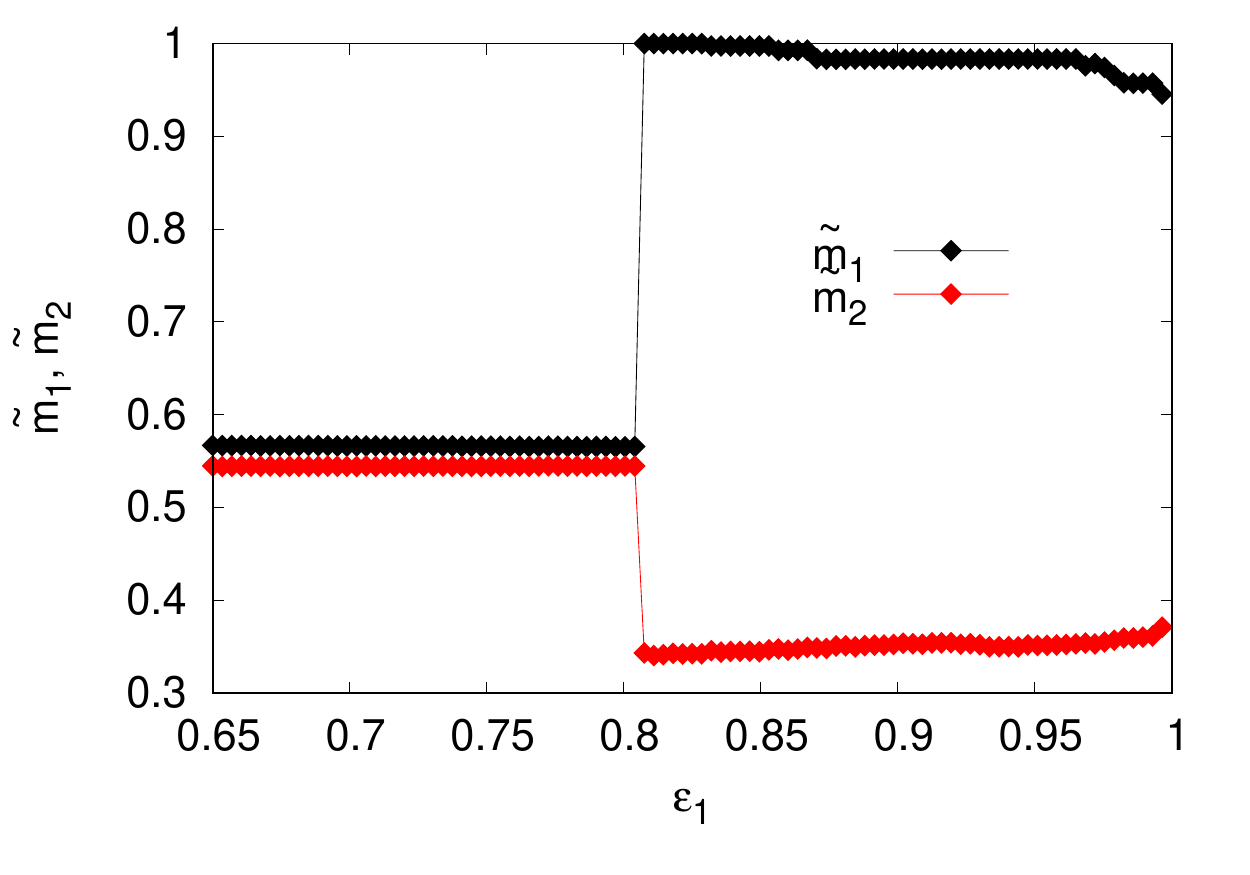}&
 \includegraphics[scale = 0.55]{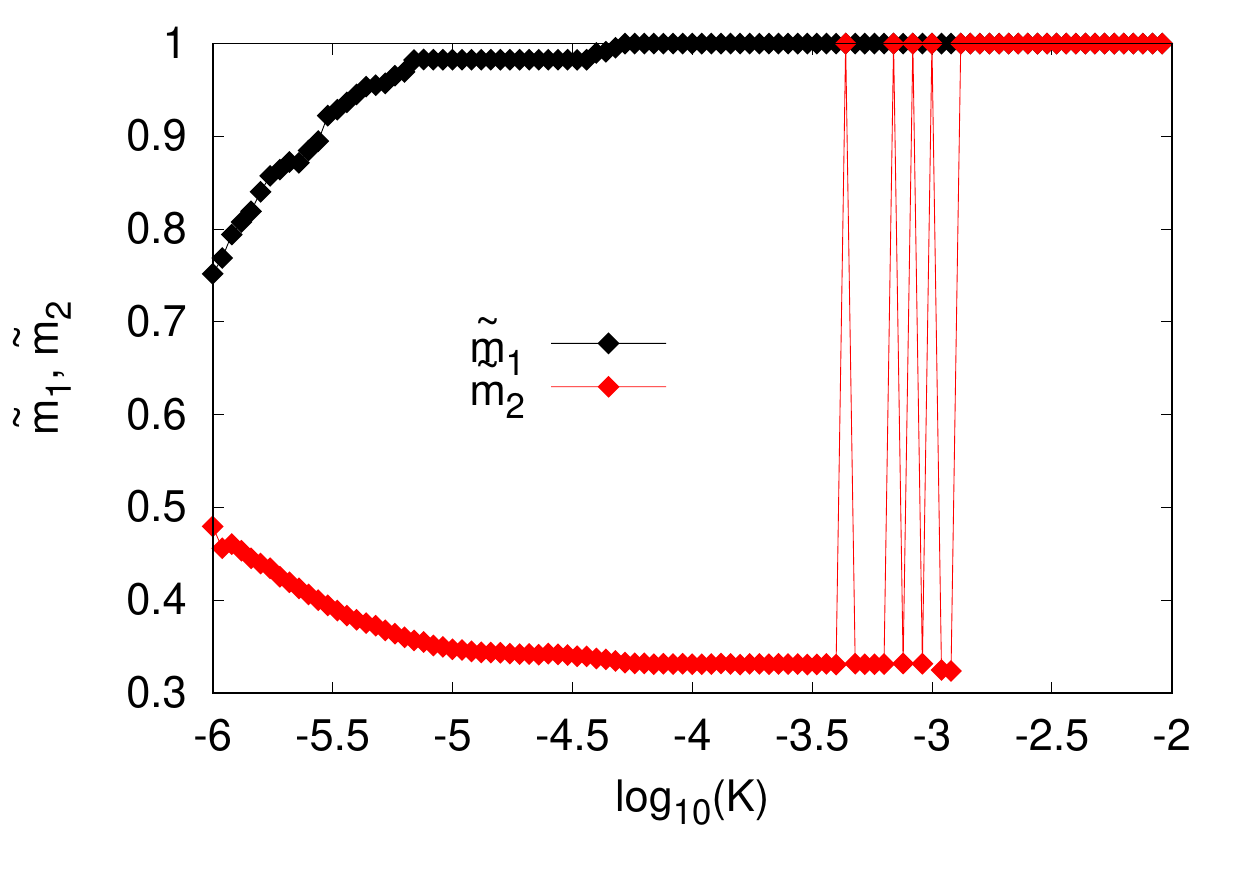}\\
 (a) & (b)\\
 \end{tabular}
 \caption{\label{fig: density_2} \footnotesize (color online) (a) The fractions $\widetilde{m}_1, \widetilde{m}_2$ are calculated at parameters $N = 150, K = 10^{-5}, \Omega = 0.27$. The signature of full phase de-synchronisation to chimera phase state can be seen at $\epsilon_1 = 0.8$. At this value of the $\widetilde{m}_{1}$ become one while $\widetilde{m}_{2}$ remain less than one which it should be for a chimera phase state in the CML. (b) The variation of $\widetilde{m}_{1}$ and $\widetilde{m}_{2}$ are plotted as $\log_{10}K$ varies between $-6$ and $-2$ keeping $\epsilon_1$ fixed at $0.93$ with $\Omega = 0.27$. The chimera states with defects in the synchronised group appears when $\log_{10}K < -4.27$ while the chimera with purely phase synchronised group appear between $-4.3 < \log_{10}K < -3.4$. When $\log_{10}K$ is between $-3.36$ and $-2.9$ we observe that the system changes between purely phase synchronised chimera state and two clustered state with small variations of $K$. The system settles to the two clustered state when $\log_{10}K > -2.9$. }
\end{figure}

 The variation of $\widetilde{m}_1, \widetilde{m}_2$ with parameter is clearly indicative of the transition from the fully phase desynchronised state to the chimera phase state in the CML (see Fig. \ref{fig: density_2}.(a)) with increasing values of $\epsilon_1$ as $\widetilde{m}_{1}, \widetilde{m}_{2} \approx 0.55$ when $\epsilon_1 < 0.81$ and $\widetilde{m}_{1} \approx 1, \widetilde{m}_{2} \approx 0$ when $\epsilon_1 > 0.81$. In fact between $\epsilon_1 = 0.808$ and $0.828$ we observe that $\widetilde{m}_{1} = 1$ and $\widetilde{m}_{2} \approx 0$ which confirms  the existence of a chimera state with a purely phase synchronised subgroup. When $\epsilon_{1} > 0.828$ we find that $\widetilde{m}_{1} \lesssim 1$ indicating that there are defects in synchronised group. The number of  defects slowly increase as $\epsilon_1$ increases to one for this fixed value of $K$. Comparing Figs. \ref{fig: density_1}.(a) and \ref{fig: density_2}(a) we can see that $\widetilde{m}_{\sigma}$ can differentiate correctly between the chimera with a purely synchronised subgroup and the chimera state with defects in the synchronised subgroup. The variation of the order parameters in Fig. \ref{fig: density_1}.b show another cross section taken at $\epsilon_1 = 0.93$ in the phase diagram where similar behaviour is found in the variation of $\widetilde{m}_1$ and $\widetilde{m}_2$ (see Fig. \ref{fig: density_2}.(b)). In particular we find that when $-5.2 < \log_{10}K < -4.3$ chimeras with defects in the phase synchronised cluster appear as $\widetilde{m}_1 \lesssim 1, \widetilde{m}_2 \approx 0$ in this range while the chimera states with a purely synchronised subgroup appear when $-4.3 < \log_{10}K < -3.4$, since in this range of $K$, $\widetilde{m}_1 = 1, \widetilde{m}_2 \approx 0$. 

\begin{figure}[H]
 \centering \begin{tabular}{cc}
 \hspace{-0.7cm}
 \includegraphics[scale = 0.7]{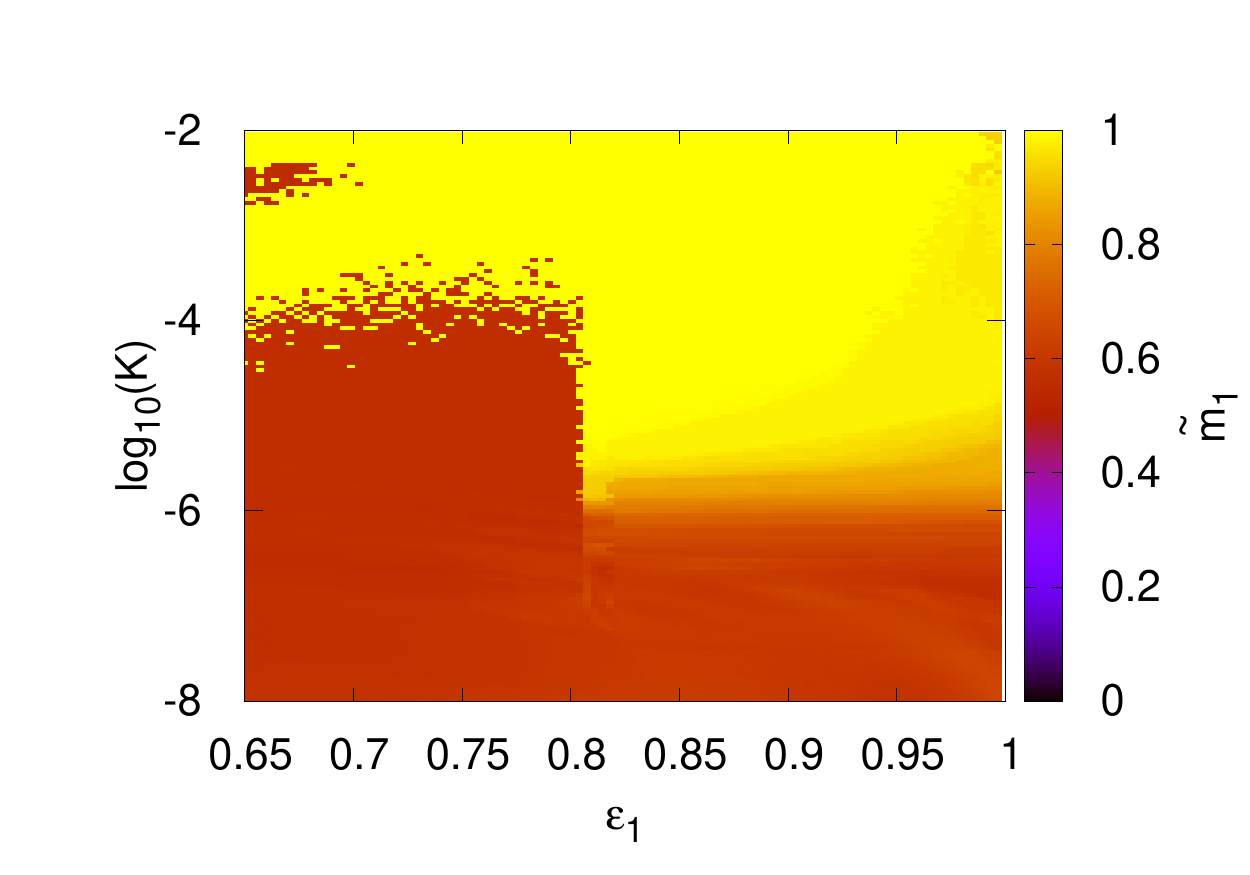}&
 \hspace{-0.7cm}
 \includegraphics[scale = 0.7]{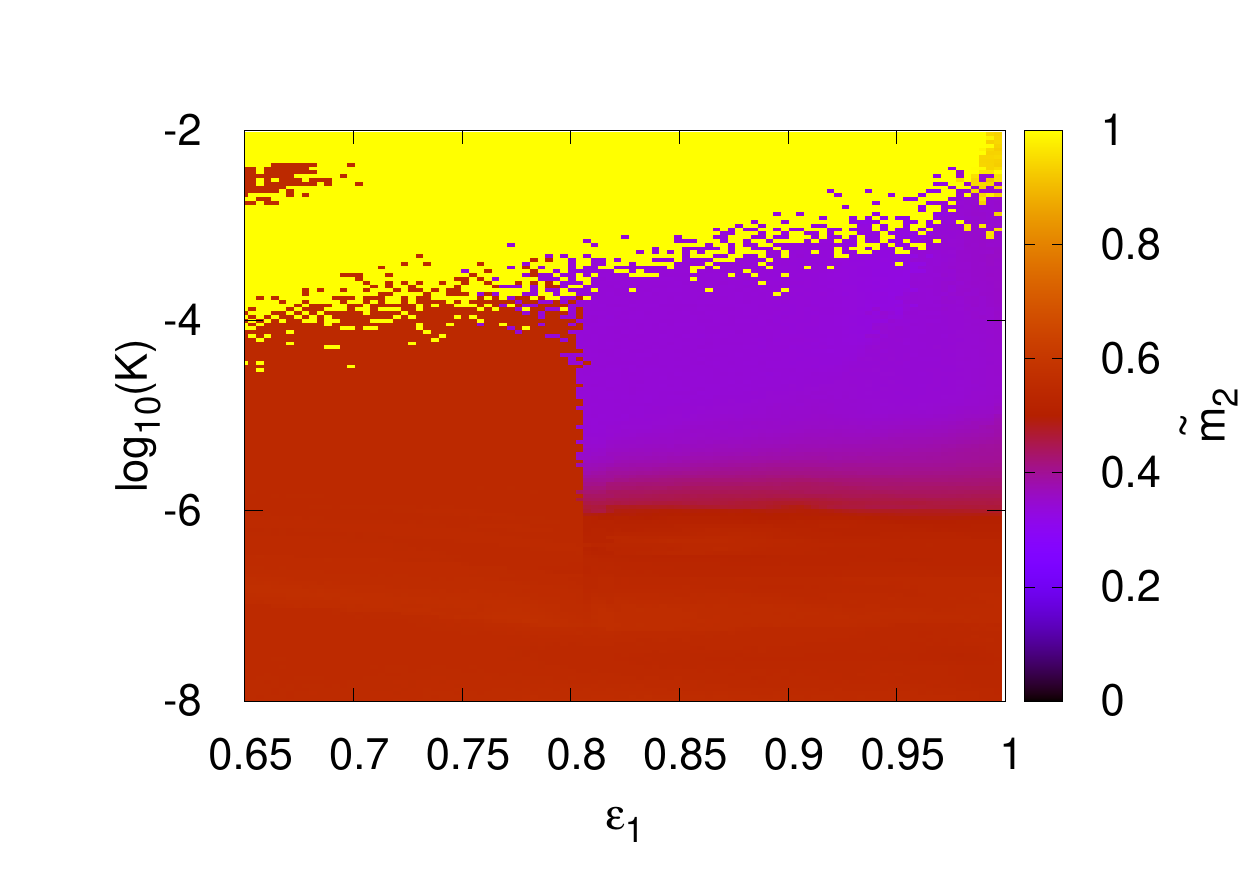}\\
 (a) & (b)\\
 \end{tabular}
\caption{\label{fig: mean_phase_diag} \footnotesize (color online) (a)$\widetilde{m}_1$ and (b) $\widetilde{m}_2$ are calculated from after calculating $a_1, b_1, a_2, b_2$ from the knowledge of the transition probability calculated from the space time variation of the phases of maps in CML using the parameters $10^{-8} < K < 10^{-2}, 0.65 < \epsilon_1 < 1, \Omega = 0.27, N = 150$.}
\end{figure}

We calculate $\widetilde{m}_{1}$ and $\widetilde{m}_2$ for group one and two, for the range of parameters given by $-8 < \log_{10}K < -2$ and $0.65 < \epsilon_1 < 1$ for $\Omega = 0.27$. Figures \ref{fig: mean_phase_diag}.(a) and (b) show that the cellular automaton results in a chimera configuration in the region approximately given by $0.8 < \epsilon_1 < 1$ and $-5.5 < \log_{10}K < -4$. We see that other types of configurations are seen near the boundary of this region. At $\epsilon_1 = 0.8$ chimera configurations are seen on one side whereas  fully desynchronised configurations are seen on the other. At the boundary of the parameter region of the chimera states two clustered state are found  between $-4 < \log_{10}K < -3$ for $0.8 < \epsilon_1 < 1$. We see that for this range of $\epsilon_1$ and for  $\log_{10}K > -3$, both $m_1$ and $m_2$ are one, signalling  the region where two clustered states are found. Within the same range of $\epsilon_1$ if we decrease $K$ we see that defects start to appear in  group one, as $m_1$ decreases from one. As $\log_{10}K$ nears $-6$ the number of defects increases in this range of $\epsilon_1$. As $\log_{10}K$ is close to $-6$ the defects in group one cause $m_1$ to be be comparable to $m_2$ implying that the chimera configuration is lost. Similarly, the fraction $m_1 \lesssim 1$ as $\epsilon_1$ increases from $0.8$ to one when $\log_{10}K$ is between the range $-5.5$ and $-4$ implying the appearance of defects in the synchronised group. The fully desynchronised phase configuration with $0.5 < m_1, m_2 < 0.6$ is seen at the parameters $-5.5 < \log_{10}K < -4$ and $\epsilon_1 < 0.8$ in Fig. \ref{fig: density_2}. Thus our mean field analysis accurately reproduces the phase diagram of the CML in the region of interest and  verifies the construction of the cellular automaton. 

\subsection*{Signatures  of transitions in spatiotemporal behaviour in the equivalent CA model}
The key to the construction of the CA is the correct choice of transition probabilities. The identity between the phase diagrams obtained from the solution of the mean field equation Eq. \ref{mf_equation} and that obtained from the complex order parameters of the CML validates the choice of transition probabilities used in the CA. We note that while several possible probabilities can be extracted from the numerical evolution of the CML, an incorrect identification of transition probabilities for the CA would have resulted in a nonidentical phase diagram. Since the transition probabilities extracted here reflect the global coupling structure of the CML, we have successfully constructed a cellular automaton which is equivalent to the CML in the region of interest. We note that the CA models of Refs. \cite{chate1988, bohr2003, zahera2010}, all have local interaction and hence our CA belong to a class distinct from them. 

\par
\begin{figure*}
\centering\begin{tabular}{cc}
\includegraphics[scale = 0.58]{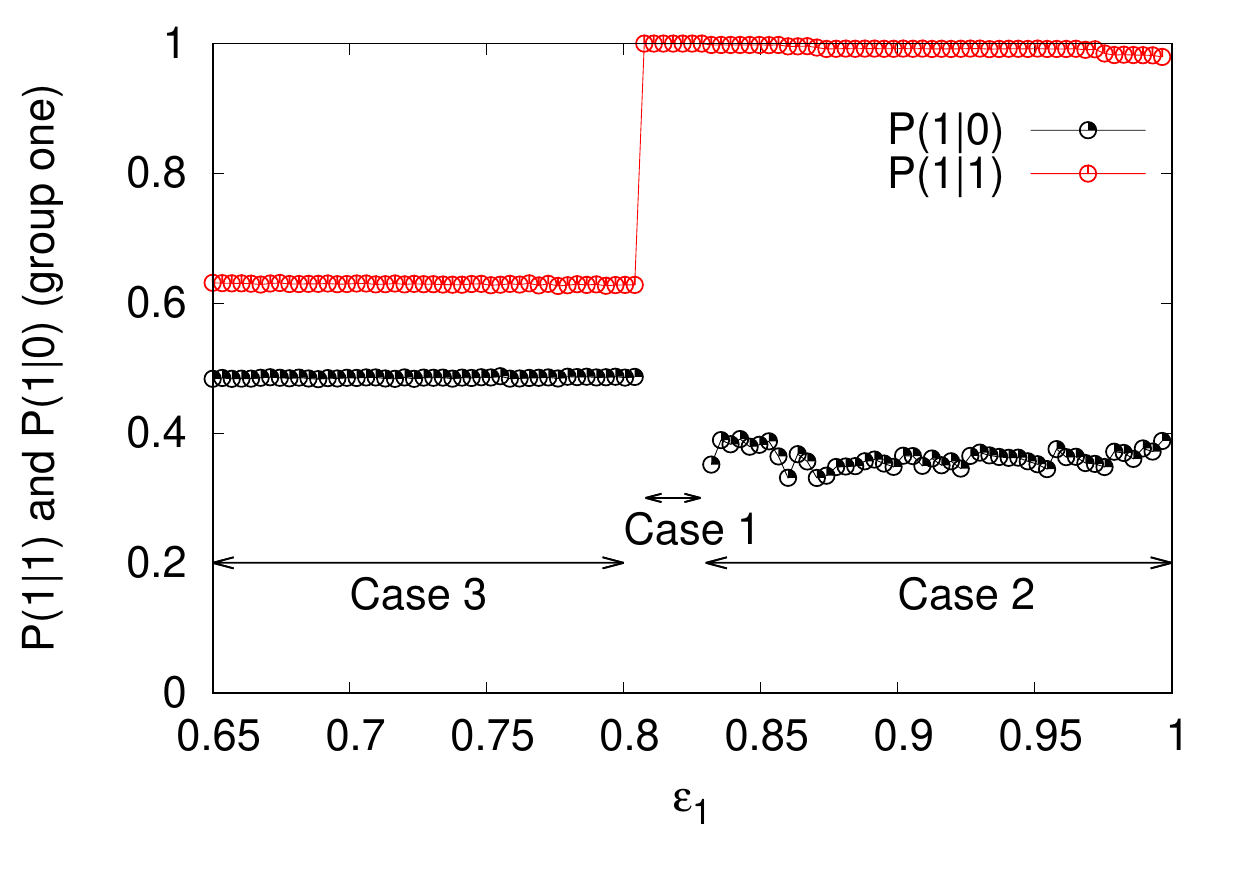}&
\includegraphics[scale = 0.58]{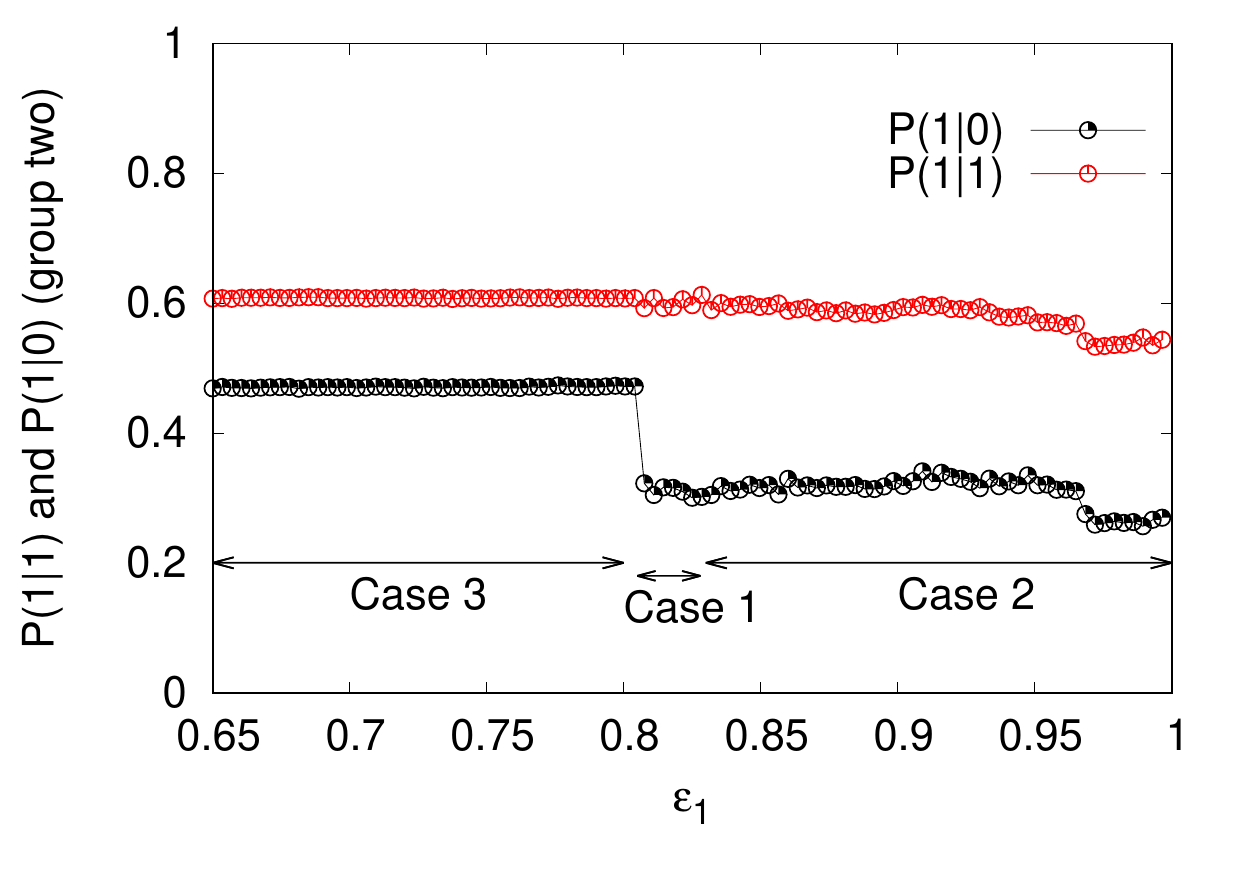}\\
(a) & (b)\\
\end{tabular}
\caption{\label{fig: p_eps_long_trans}\footnotesize (color online) The variation of the mean values of the transition probabilities with respect to $\epsilon_1$ are plotted. Mean values of $P(1|0)$, $P(1|1)$ for (a) group one and (b) two are shown. Other parameters are kept fixed at $K = 10^{-5}, \Omega = 0.27, N = 150$. We discard initial $3 \times 10^{6}$ time steps as transient while evolving the CML and use the next $10^{6}$ time steps to calculate the above quantities.}
\end{figure*}

\begin{figure*}
\centering\begin{tabular}{cc}
\includegraphics[scale = 0.58]{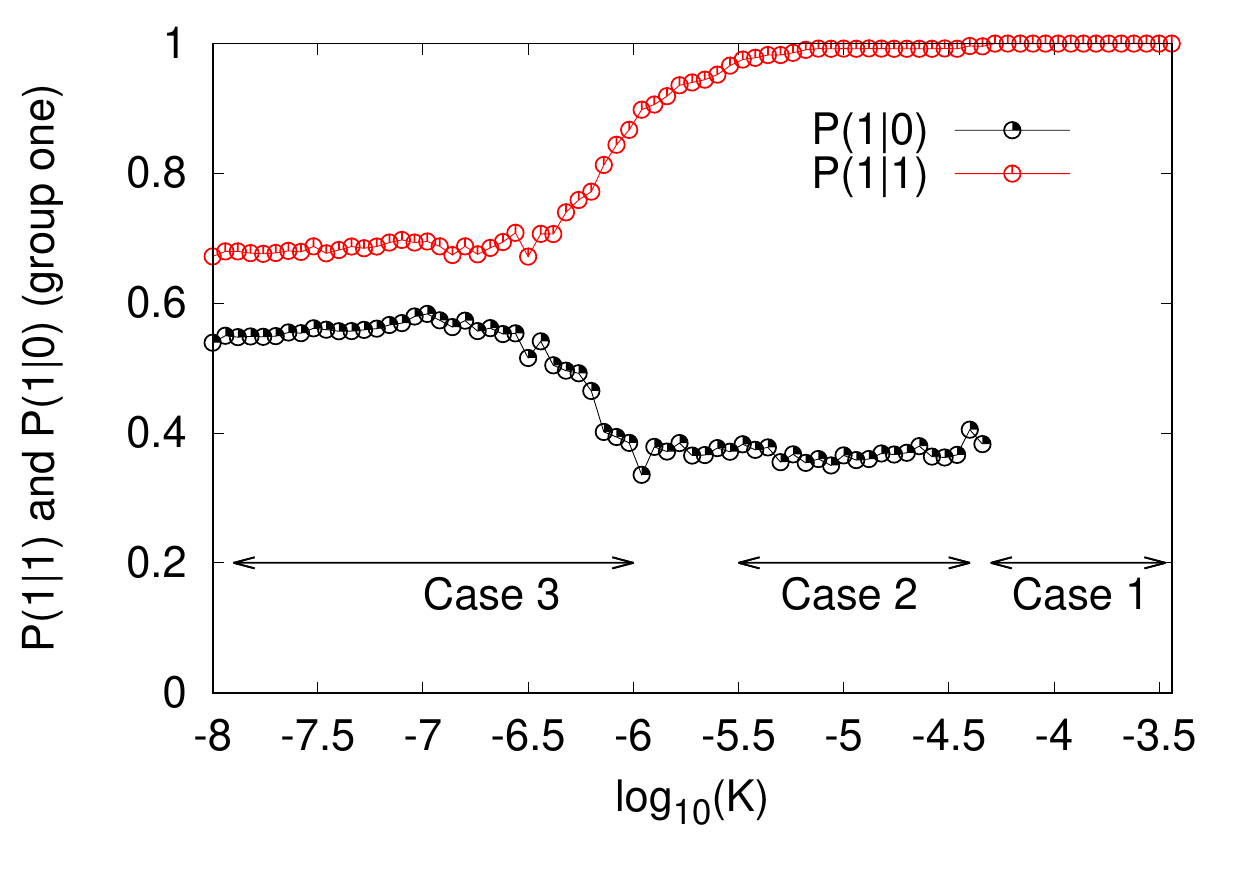}&
\includegraphics[scale = 0.58]{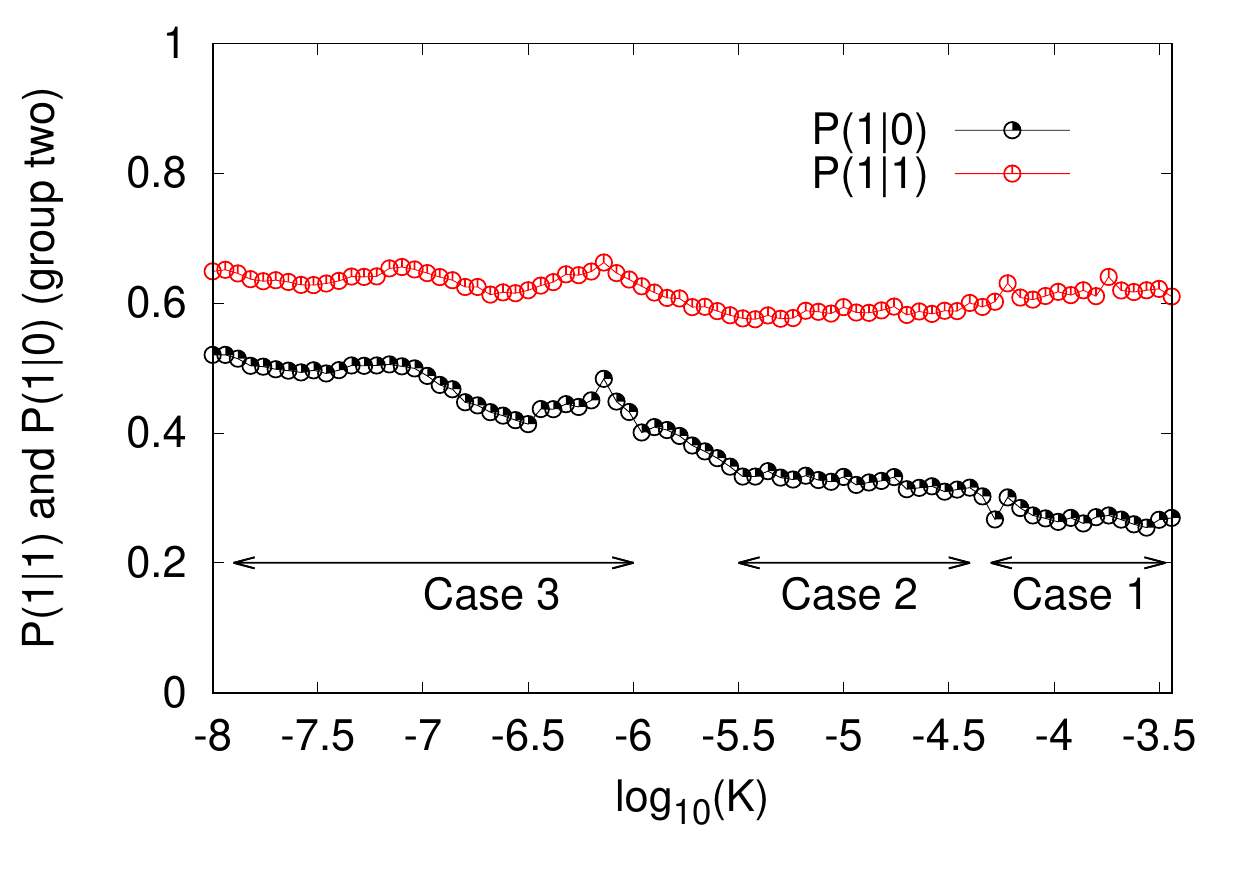}\\
(a) & (b)\\
\end{tabular}
\caption{\label{fig: p_K_long_trans}\footnotesize (color online) The variation of the mean values of the transition probabilities with respect to $K$ are plotted. Mean values of $P(1|0)$ and $P(1|1)$ for (a) group one and (b) group two is shown. Other parameters are kept fixed at $\epsilon_1 = 0.93, \Omega = 0.27, N = 150$.  We discard initial $3 \times 10^{6}$ time steps as transient while evolving the CML and use the next $10^{6}$ time steps to calculate the above quantities.}
\end{figure*}

 We now discuss the spatiotemporal behaviour of this cellular automaton and compare it with the CML. Figures \ref{fig: order_zoom} and \ref{fig: mean_phase_diag} show a region of the parameter space ($0.8 \lesssim \epsilon_1 < 1$ and $-5.5 < \log_{10}K \lesssim -2$) that supports chimera phase states (case 1 and case 2 as discussed in the section \ref{PD}). We calculate the transition probabilities calculated for the $\epsilon_1$ and $K$ values used in Figs. \ref{fig: density_1} and \ref{fig: density_2} to see how the transition probabilities reflect transitions between chimera states and other phase configurations. Figures \ref{fig: p_eps_long_trans}(a) and (b) show the variation of the average independent transition probabilities, $P(1|0)$ and $P(1|1)$ which are $P^{x_1, x_2}(1|0)$ and $P^{x_1, x_2}(1|1)$ averaged over all possible combinations of $x_1$ and $x_2$ such that the transition probabilities are nonzero. These are shown for both the groups separately for $0.65 < \epsilon_1 < 1, \log_{10}K = -5$ while Figs. \ref{fig: p_K_long_trans}.(a) and (b) show the same quantities for each of the groups for the values $-8 < \log_{10}K < -3.5, \epsilon_1 = 0.93$. For the chimera states with a purely synchronised group one shown as case 1 in Figs. \ref{fig: p_eps_long_trans}(a) and \ref{fig: p_K_long_trans}(a), the dynamics of the laminar sites is deterministic as $P(1|1)$ is exactly one. As only laminar sites survive, $P(1|0)$ cannot be computed numerically for this group. When there are defects in the synchronised group of the chimera states (shown as case 2 in Figs. \ref{fig: p_eps_long_trans}.(a) and \ref{fig: p_K_long_trans}.(a)), $P(1|1)$ takes values which are near one $(\approx 0.98)$ while $P(1|0)$ takes values much less than one. This implies that for the chimera states in case 2, the dynamics of the laminar sites is approximately deterministic while the dynamics of the burst sites is probabilistic in group one. In the case of fully phase desynchronised states (case 3 in Figs. \ref{fig: p_eps_long_trans}.(a) and \ref{fig: p_K_long_trans}.(a)), the dynamics of sites in group one remains probabilistic as both $P(1|0)$ and $P(1|1)$ have fractional values which are less than one. However, Figs. \ref{fig: p_eps_long_trans}(b) and \ref{fig: p_K_long_trans}(b) show that the dynamics of group two is probabilistic for the all the three cases above, since both $P(1|1)$ and $P(1|0)$ are less than one for each case. Thus the CA shows a transition from a fully probabilistic CA to a partly deterministic CA at the chimera bifurcation boundary of the CML. We note that this kind of transition is completely characteristic of a bifurcation from a chimera state where one subgroup is synchronized, and will be seen in other contexts as well. 

\par
\begin{figure*}
\centering\begin{tabular}{cc}
\includegraphics[scale = 0.67]{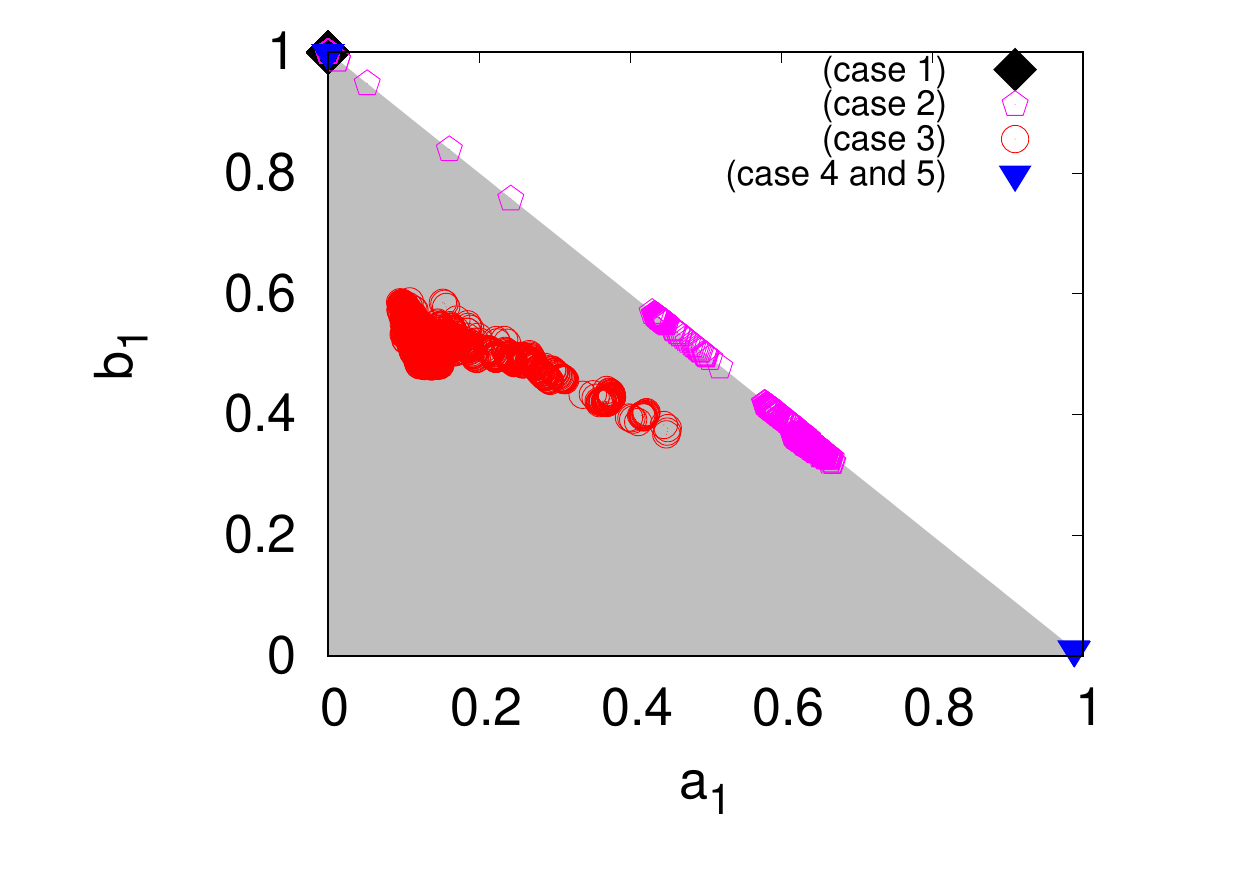}&
\includegraphics[scale = 0.67]{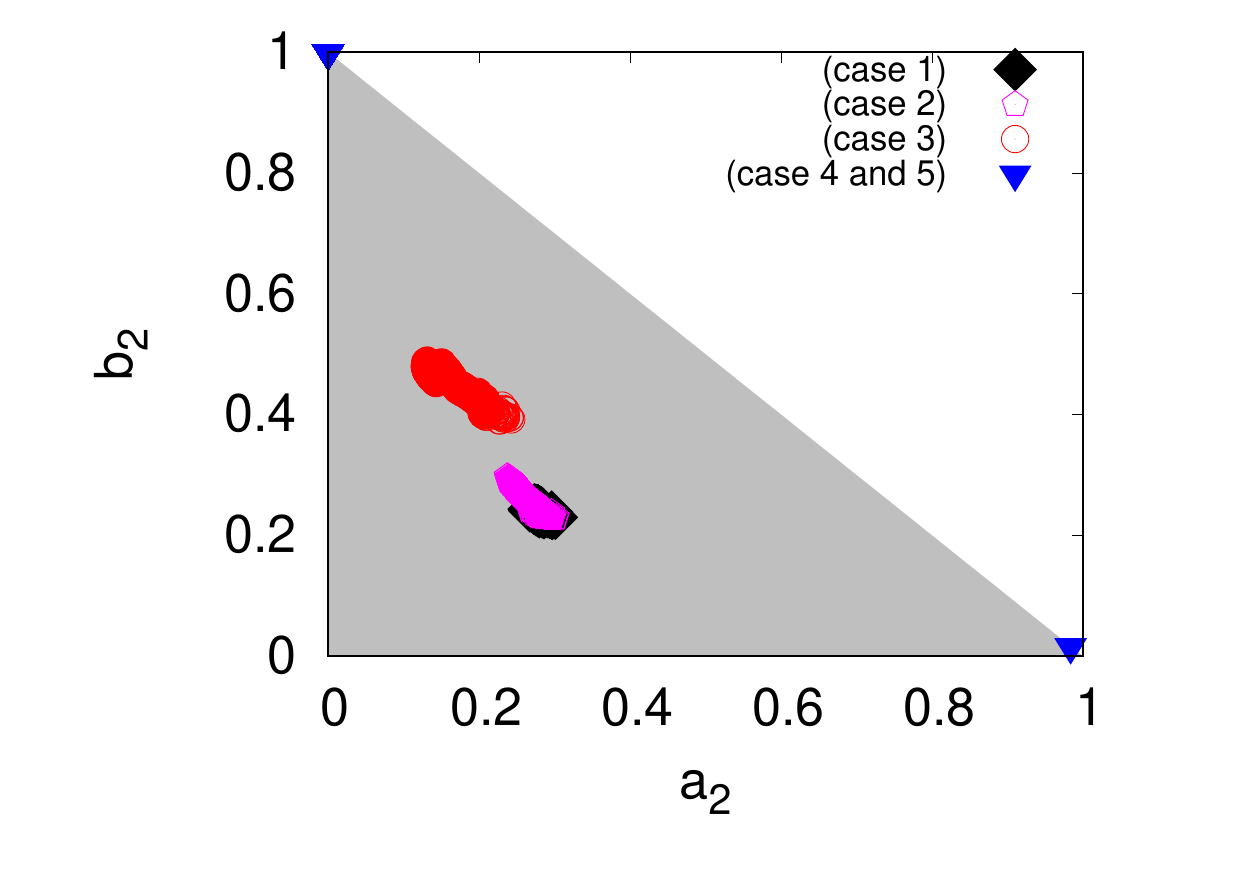}\\
\end{tabular}
\caption{\label{fig: m_a_b_2d}\footnotesize (color online) The region  for the existence of the a mean field fixed point which lies between $0$ and $1$ (left) $\widetilde{m}_1$ for group one and (right) $\widetilde{m}_2$ for group two is shown in grey. In both the figures, the black $\rhombus$ denotes a stable chimera phase with pure synchronisation (case 1) in group one, magenta $\pentago$ denotes the case of chimera states with defects (case 2), red $\circlet$ denotes fully phase desynchronised phases (case 3) and the blue $\trianglepb$ denotes the case of clustered states (case 4 and case 5). In each of these cases the values of $a_\sigma$ and $b_\sigma$ $(\sigma = 1, 2)$ satisfy the  conditions given in Table 3.}
\end{figure*}

 Since we have verified in Section V that the identification of the CA and the mean field equation is accurate, quantitative inferences regarding the emergence of the chimera state and the other states can also be made via the fixed points of the mean field equations (i.e. Eqs \ref{fixed_point}). We have seen that since the quantities $\widetilde{m}_1, \widetilde{m}_2$ must be between zero and one, we must have $a_\sigma + b_\sigma \leq 1$. This allowed region is indicated in grey in Figs. \ref{fig: m_a_b_2d}(a) and (b) for $\sigma = 1, 2$. 

We also indicate in this region, the actual  fixed points, calculated numerically using the transition probabilities for the spatiotemporal behaviours found in our CML (cases  1 - 5 in section \ref{PD}) for both groups 1 and 2. For the chimera state in case 1, the fixed points of the mean field equation for group one are found at the parameter values $b_1 = 1$ and $a_1 = 0$ (indicated by the black $\rhombus$ in Fig. \ref{fig: m_a_b_2d}.(a)). This satisfies  the condition $a_1 + b_1 = 1$ since $\widetilde{m}_1 = 1$ when all sites in group one are laminar. For the chimera state of case 2 the fixed points for group one (indicated by the magenta $\pentago$ in Fig. \ref{fig: m_a_b_2d}(a)) lie along the edge of the allowed $a_1, b_1$ space denoted by $a_1 + b_1 \lesssim 1$.  This is reasonable  since we have the fraction  $\widetilde{m}_1 \lesssim 1$ due to the presence of a few burst sites in the phase synchronised cluster. For both these chimera states, the fixed points of the mean field equation for the phase desynchronised group i.e. $\widetilde{m}_2$ are found in the allowed region $a_2 + b_2 < 1$, and the location of the fixed points for the two cases overlaps with each other (see Fig. \ref{fig: m_a_b_2d}.(b)). In the case of the fully phase desynchronised state (case 3 in section \ref{PD}), the fixed points lie in a region inside the solution space of both the groups (indicated by the red $\circlet$ in Figs. \ref{fig: m_a_b_2d}(a) and (b)) in the allowed region. The phase clustered states where $\widetilde{m}_1 = 1$ and $\widetilde{m}_2 = 1$ (cases 4 and 5) are seen at $a_\sigma = 0, b_\sigma = 1$ and $a_\sigma = 1, b_\sigma = 0$ for $\sigma = 1, 2$ (shown by blue $\trianglepb$s) in both the Figs. \ref{fig: m_a_b_2d}(a) and (b). In this case, the conditions $a_1 + b_1 = 1$ and $a_2 + b_2 = 1$ are satisfied due to the absence of burst sites in both the groups, and thus the solutions lie on the diagonal. 

As  the parameters in the CML vary, (e.g. as shown in Figs. \ref{fig: p_eps_long_trans} and \ref{fig: p_K_long_trans}) the position of the stable fixed points of the mean field equation for the CA in the $a_\sigma, b_\sigma$ space moves as fixed points corresponding to different cases become stable, as shown in Figs. \ref{fig: m_a_b_2d}.(a) and (b). For example, in Fig. \ref{fig: m_a_b_2d}, as $\epsilon_1$ increases from 0.65 to one with $K$ fixed at $10^{-5}$, the fixed points (Eqs. \ref{fixed_point}) jump from the region corresponding to full phase desynchronised states (case 3, denoted by $\circlet$) to that for chimera states with a purely phase synchronised group one (case 1, denoted by $\rhombus$) and then to the chimera state with defects in the phase synchronised group (case 2, shown by $\pentago$). Similarly, when $K$ is increased from $10^{-8}$ to one with fixed $\epsilon_1 = 0.93$, the fixed points hop from the fully desynchronised state (case 3, denoted by $\circlet$) to chimera phase states with defects in the synchronised group (case 2, indicated by \pentago) to chimera states without defects (case 1, denoted by $\rhombus$) and then to phase clustered states (case 4 and 5, shown by $\trianglepb$) (see Fig. \ref{fig: m_a_b_2d}). Table \ref{table_3} lists the stability conditions on $a_\sigma, b_\sigma (\sigma = 1, 2)$ for each of the cases seen here. 

\begin{table}[ht]
\caption{\label{table_3}\footnotesize Specific criteria for the existence of chimera states and other phase configuration in terms of the transition probabilities of the CA obtained from the fixed point solution of the mean field equation given in Eq. \ref{fixed_point}.}
\begin{tabular}{ | l | l | }
\hline\hline
\centering Attractor &Condition for existence\\ 
\hline\hline
 \makecell{Chimera state with purely\\ phase synchronised group (case 1 )\\$(a_1 + b_1 = 1$, $a_2 + b_2 < 1)$} & \makecell{$\sum\limits_{x_{2} = 0}^{N} \left(\sum\limits_{x_1 = 1}^{N}P(x_1, x_2)P^{x_1, x_2}(1|1) + P(0, x_2)P^{0, x_2}(1|0)\right) = 1 $,\\ $\sum\limits_{x_{1} = 0}^{N} \left(\sum\limits_{x_2 = 1}^{N}P(x_1, x_2)P^{x_1, x_2}(1|1) + P(x_1, 0)P^{x_1, 0}(1|0)\right) < 1$}\\
 \hline
 \makecell{Chimera state with defects in\\ phase synchronised group (case 2 )\\$(a_1 + b_1 \lesssim 1$, $a_2 + b_2 < 1)$} & \makecell{$\sum\limits_{x_{2} = 0}^{N} \left(\sum\limits_{x_1 = 1}^{N}P(x_1, x_2)P^{x_1, x_2}(1|1) + P(0, x_2)P^{0, x_2}(1|0) \right) \lesssim 1 $,\\ $\sum\limits_{x_{1} = 0}^{N} \left(\sum\limits_{x_2 = 1}^{N}P(x_1, x_2)P^{x_1, x_2}(1|1) + P(x_1, 0)P^{x_1, 0}(1|0)\right) < 1$}\\
 \hline
 \makecell{Fully phase synchronised\\ state (case 3 )\\$(a_1 + b_1 < 1$, $a_2 + b_2 < 1)$} & \makecell{$\sum\limits_{x_{2} = 0}^{N} \left(\sum\limits_{x_1 = 1}^{N}P(x_1, x_2)P^{x_1, x_2}(1|1) + P(0, x_2)P^{0, x_2}(1|0)\right) < 1 $,\\ $\sum\limits_{x_1 = 0}^{N} \left(\sum\limits_{x_2 = 1}^{N}P(x_1, x_2)P^{x_1, x_2}(1|1) + P(x_1, 0)P^{x_1, 0}(1|0)\right) < 1$}\\
 \hline
\makecell{Phase clustered state (case 4)\\$(a_1 + b_1 = 1$, $a_2 + b_2 = 1)$} & \makecell{$\sum\limits_{x_{2} = 0}^{N}\left( \sum\limits_{x_1 = 1}^{N}P(x_1, x_2)P^{x_1, x_2}(1|1) + P(0, x_2)P^{0, x_2}(1|0)\right) = 1 $,\\ $\sum\limits_{x_1 = 0}^{N} \left( \sum\limits_{x_2 = 1}^{N}P(x_1, x_2)P^{x_1, x_2}(1|1) + P(x_1, 0)P^{x_1, 0}(1|0)\right) = 1$}\\
 \hline
\end{tabular}
\end{table}

\section{\label{conclusion}Conclusion}
To summarise, we have analysed a system which shows novel chimera behaviour, viz. a mixed state with a synchronised part and a spatiotemporally intermittent part. This behaviour is seen in a coupled map lattice consisting of two groups of globally coupled sine circle maps with different values of intergroup coupling and intra-group coupling. The system shows a variety  of solutions in different regions of the parameter space. A phase diagram is obtained using the complex order parameter. Our analysis focusses on the spatiotemporally intermittent chimeras where coherent and incoherent regions coexist. We note here that the distribution of laminar lengths (i.e. the distribution of the number of consecutive sites which show laminar behaviour), seen here does not show power law behaviour, as seen in Refs. \cite{zahera2005, zahera2006}, but falls off exponentially\cite{footnote}. This is due to the global nature of the coupling used here, unlike the diffusive coupling used in Ref. \cite{zahera2005, zahera2006}.

We set up  a procedure to identify the laminar and burst sites of the system from the values of the phases of the maps during their evolution. This method is general and is capable of identifying the laminar and burst stages of the lattice sites for space time variations of other extended dynamical systems and experimentally realisable systems such as coupled laser models and Josephson junction oscillator arrays. Further analysis of the system is carried out by constructing an equivalent cellular automaton for the CML on the lines of Ref. \cite{bohr2003}. The equivalent cellular automaton is constructed by defining transition probabilities appropriate  for the global coupling topology of the system. The  nature of the transition seen in the CA probabilities at the point where the CML shows a transition to spatiotemporally intermittent behavior is analyzed. We find that  the subgroup probabilities corresponding to the synchronized part of the chimera shows a transition from deterministic to probabilistic behavior, in consistence with the spatiotemporal behavior of the solutions. Thus the transition from a probabilistic CA to a partially deterministic CA signals the bifurcation to the chimera state. It is expected that this will be a general feature for the transition in the CA for all cases where the chimera contains a synchronized subgroup.

We  also derive  mean field equations for the fraction  of  laminar/turbulent sites using the transition probabilities. The fixed point of these mean field equations correctly gives the fraction of the laminar sites in each of the groups which is confirmed by the numerical results for the CML. The nature of the fixed point is discussed using linear stability analysis. Further, a phase digram is constructed using our fixed point analysis, which confirms the phase diagram using the order parameters obtained earlier. 

\par Thus the cellular automaton, which we obtain, possesses a unique global interaction structure. The transition probabilities which we have identified and validated via our analysis, accurately represent the global coupling of the CA. We obtain conditions involving these transition probabilities to identify chimera phase states along with other phase configurations which are seen in the CML. Thus, the equivalent cellular automaton proves to be an effective tool in the analysis of the chimera states for this system. Being a constructive model \cite{kaneko1996}, this CA not only represents the CML under consideration but can serve as an independent construct to analyze the variety of spatiotemporal behaviours seen in globally coupled oscillator models. We note that the CA shows a characteristic transition to partially deterministic behavior on the transition to the chimera state.  We hope that our methods will provide pointers for the analysis of chimera states in other extended dynamical systems as well. 

\bibliographystyle{plainnat}

\end{document}